\documentclass[a4paper,11pt]{article}
\pdfoutput=1 
\usepackage{jheppub} 
\usepackage[T1]{fontenc} 
\usepackage{multirow}
\usepackage{enumitem}
\usepackage{mathtools}

\renewcommand{\d}{\mathrm{d}}
\newcommand{\dv}{d_{\mathrm{v}}}
\newcommand{\ds}{d_{\mathrm{s}}}
\newcommand{\dbar}{\bar{d}}
\renewcommand{\u}{\mathrm{u}}
\newcommand{\uv}{u_{\mathrm{v}}}
\newcommand{\us}{u_{\mathrm{s}}}
\newcommand{\ubar}{\bar{u}}
\newcommand{\g}{{\mathrm g}}
\newcommand{\q}{{\mathrm q}}
\newcommand{\qbar}{\bar{\mathrm q}}

\newcommand{\W}{{\mathrm W}}
\newcommand{\Z}{{\mathrm Z}}

\newcommand{\as}{\alpha_{\mathrm{s}}}

\newcommand{\aew}{\alpha_{\mathrm{w}}}
\newcommand{\pT}{p_{\perp}}
\newcommand{\pTs}{p^2_{\perp}}

\newcommand{\dP}{\d\mathcal{P}}
\newcommand{\y}{\boldsymbol{y}}
\newcommand{\qtis}{\tilde{q}^2}

\newcommand{\qv}{\boldsymbol{q}}

\newcommand{\Herwig}{\textsf{Herwig}}

\newcommand{\GeV}{\mathrm{GeV}}
\newcommand{\PySix}{\textsc{Pythia 6}}
\newcommand{\PyEight}{\textsc{Pythia 8}}

\newcommand{\Eq}{Equation\,}
\newcommand{\Eqs}{Equations\,}
\newcommand{\Sec}{Section\,}
\newcommand{\Secs}{Sections\,}

\newcommand{\Ansatz}{Ansatz\,}
\newcommand{\Fig}{Figure\,}
\newcommand{\Figs}{Figures\,}
\newcommand{\Tab}{Table\,}
\newcommand{\Prop}{Property\,}
\newcommand{\Props}{Properties\,}
\newcommand{\App}{Appendix\,}

\subheader{\begin{flushright} CERN-TH-2019-072 \\ MAN/HEP/2019/005 \\ MCnet-19-12 \end{flushright}}

\title{\boldmath A Monte-Carlo simulation of double parton scattering}

\author[a]{Baptiste Cabouat,}
\author[b]{Jonathan R. Gaunt}
\author[a]{and Kiran Ostrolenk}

\affiliation[a]{University of Manchester, School of Physics and Astronomy,\\Schuster Building, Oxford Road, Manchester M13 9PL, United Kingdom}
\affiliation[b]{CERN Theory Division, 1211 Geneva 23, Switzerland}

\emailAdd{baptiste.cabouat@manchester.ac.uk}
\emailAdd{jonathan.richard.gaunt@cern.ch}
\emailAdd{kiran.ostrolenk@manchester.ac.uk}


\arxivnumber{1906.04669}

\abstract{In this work, a new Monte-Carlo simulation of double parton scattering (DPS) at parton level is presented. The simulation is based on the QCD framework developed recently by M. Diehl, J. R. Gaunt and K. Sch\"{o}nwald. With this framework, the dynamics of the $1\to2$ perturbative splittings is consistently included inside the simulation, with the impact-parameter dependence taken into account. The simulation evolves simultaneously two hard systems from a common hard scale down to the hadronic scale. The evolution is performed using an angular-ordered parton shower which is combined with a set of double parton distributions that depend explicitly on the inter-parton distance. An illustrative study is performed in the context of same-sign WW production at the LHC, with the quark content of the proton being limited to three flavours. In several distributions we see differences compared to DPS models in Herwig, Pythia, and the DPS ``pocket formula''.}

\begin{document} 
\maketitle
\flushbottom

\section{Introduction}

In high-energy proton-proton collisions such as the ones that occur at the Large Hadron Collider (LHC), the underlying event can be an important background to a variety of signals. Therefore, in order to accurately describe experimental data, it is necessary to develop a simulation of the underlying event \cite{Olive:2016xmw, Buckley:2011ms}. Current event generators such as \Herwig\, \cite{Bahr:2008pv, Bellm:2015jjp, Bellm:2017bvx}, \textsc{Pythia} \cite{Sjostrand:2006za, Sjostrand:2014zea} and \textsc{Sherpa} \cite{Schumann:2007mg,Gleisberg:2008ta,Bothmann:2019yzt}, model the underlying event with a good agreement with experimental data. However, these models can be improved in order to reach an even higher accuracy, which is required for the next generation of proton-proton colliders and for beyond-the-standard-model searches.

At parton level, the underlying event is generated by two major components: multiple parton interactions (MPI) and parton showers. Most of the parton showers currently implemented are at a leading-logarithm (LL) accuracy, which means that the leading logarithms are fully resummed within a so-called Sudakov form factor. In fact, they include also some next-to-leading-logarithm (NLL) effects, mostly via colour coherence, four-momentum conservation, running of the strong coupling, etc... \cite{Buckley:2011ms} In order to improve the simulation of the underlying event, one can try to implement new parton showers that would include more aspects such as spin correlations, subleading-colour effects, new choices of kinematics, next-to-leading-order (NLO) splitting kernels, transverse-momentum-dependent parton distributions (TMD), amplitude-level evolutions, etc... Many efforts have been made in those directions recently \cite{Hoche:2017iem, Hoche:2017hno, Hoeche:2017jsi, Bury:2017jxo, Cabouat:2017rzi, Martinez:2018ffw, Dasgupta:2018nvj, Richardson:2018pvo, Hoang:2018zrp, Platzer:2018pmd, Cormier:2018tog, Nagy:2019pjp, Bewick:2019rbu, Forshaw:2019ver}.

Another way of improving the simulation of the underlying event is to develop new models of MPI. At cross-section level, a scattering with $n$ parton-parton interactions ($n$PS) is usually suppressed compared to a scattering with a single parton-parton interaction (SPS). More precisely, the ratio between the total cross sections scales as \cite{Gaunt:2012, Diehl:2017wew}

\begin{equation}
\frac{\sigma_{n\mathrm{PS}}}{\sigma_\mathrm{SPS}}\sim\left(\frac{\Lambda^2}{Q_h^2}\right)^{n-1},
\end{equation}

\noindent with $Q_h$ the energy scale at which the proton is probed and $\Lambda\sim 1\,\GeV$, the characteristic scale of the proton. One can see that for high energy processes where $Q_h\gg\Lambda$, the $n$PS total cross section is strongly suppressed compared to the SPS one. However, for scales $Q_h$ of the order of $\Lambda$, $n$PS is unsuppressed. In fact, most of the underlying event is composed of these so-called ``soft'' MPI. Moreover, $n$PS constitutes a systematic background of SPS signals. For high energy scales such as at the LHC, one can expect the ratio $n$PS/SPS to be enhanced. Indeed, the protons are probed at lower momentum fractions, where the population of partons is greater \cite{Gaunt:2012}. Therefore, MPI models must be included inside the event generators in order to accurately describe experimental data.

In Quantum Chromodynamics (QCD), a correct description of MPI would require multi-parton distribution functions (mPDFs) $F_{i_1,\dots,i_n}\left(x_1,\dots,x_n,\{\y_{kl}\},\mu_1^2,\dots,\mu_n^2\right)$ which give the joint probability of finding $n$ partons of flavours $i_1,\dots,i_n$ within the same proton with longitudinal momentum fractions $x_1,\dots,x_n$ when those partons participate in $n$ different interactions characterised by the scales $\mu_1^2,\dots,\mu_n^2$ \cite{Diehl:2011yj}. The set of impact parameters $\{\y_{kl}\}$ parametrises the relative distances between the $n(n-1)/2$ pairs of partons involved. An mPDF is a complicated object which takes into account all the correlations between the $n$ partons belonging to the same proton. Those correlations originate for example from kinematic constraints, quantum-number conservation rules (i.e. the sum rules) or from dynamical effects such as the $1\to2$ parton splittings that occur inside the proton \cite{Gaunt:2009re}. The mPDFs are non-perturbative quantities, and their calculation for arbitrary $n$ is far beyond the current state of the art (for progress in the case $n=2$, see \cite{Bali:2018nde}). Theoretically, they can be extracted from the ``light-cone''  wavefunction of the proton \cite{Blok:2010ge, Chang:2012nw, Rinaldi:2014zoa, Broniowski:2016trx}, but this wavefunction contains Fock states with an arbitrary number of particles, which makes its modelling presently impossible. Furthermore, unlike for the single parton distribution functions (sPDFs), one cannot constrain the mPDFs using current experimental data.

The usual event generators thus adopt the following strategy to model MPI \cite{Bahr:2008dy, Bahr:2008spa, Corke:2011yy, Sjostrand:2004pf, Sjostrand:2017cdm}. First, a hard process and its kinematics are selected, without taking into account the possible presence of secondary parton-parton interactions. After this, secondary parton-parton interactions are added. Each system is then evolved by using a parton shower. In the current event generators, the mPDFs are typically written as a product of sPDFs. This ansatz is then modified in order to take into account the kinematic constraints and the quantum-number conservation rules \cite{Sjostrand:2004pf, Sjostrand:2017cdm}. 

This ansatz leads to a reasonably good description of MPI. However, it fails to take into account some parton-parton correlations which are not necessarily negligible. For example, models based on the light-cone wavefunction of the proton with only three quarks showed strong correlations at large momentum fractions, especially in spin and in colour \cite{Chang:2012nw, Rinaldi:2014zoa}, and these correlations should be implemented. Also, the dynamical correlations due to the $1\to2$ splittings should be included as well. These dynamical correlations are important \cite{Korotkikh:2004bz, Gaunt:2009re, Gaunt:2012dd}, especially for small momentum fractions and large energy scales, where the other correlations tend in general to be reduced \cite{Manohar:2012jr, Diehl:2014vaa}. Therefore, there is a need for a realistic set of mPDFs. This set must be based on theoretical works, since very little experimental data have been gathered so far.

Unfortunately, no set of mPDFs is available for the moment. However, the case $n = 2$, referred to as double parton scattering (DPS), has been widely studied \cite{Kirschner:1979im, Shelest:1982325, Snigirev:2003cq, Gaunt:2009re, Ryskin:2011kk, Blok:2011bu, Gaunt:2011xd, Diehl:2011yj, Ryskin:2012qx, Gaunt:2012dd, Manohar:2012jr, Manohar:2012pe, Blok:2013bpa, Diehl:2014vaa,  Diehl:2015bca, Buffing:2017mqm,  Diehl:2017wew,  Diehl:2017kgu, Diehl:2018kgr, Gaunt:2018eix, Diehl:2019rdh}. Despite the lack of experimental data, several groups managed to produce sets of double parton distribution functions (dPDFs) that represent an improvement over the simple product ansatz. These sets were produced by solving the double DGLAP equations (dDGLAP), an extension of the usual DGLAP equations. The first analytical solutions of the dDGLAP equations were derived in \cite{Shelest:1982325, Kirschner:1979im, Snigirev:2003cq}, but those solutions cannot be used as such for phenomenological studies where numerical solutions are needed instead. Fortunately, the first numerical set of leading-order (LO) dPDFs was generated by JG and W. J. Stirling and is referred to as GS09 \cite{Gaunt:2009re}. With this set, the contributions to the dPDFs from the $1\to2$ splittings are included, although without any dependence on the inter-parton distance. This set showed some differences with the usual ansatz made for dPDFs, especially considering the longitudinal correlations between partons \cite{Gaunt:2009re}. Later, significant progress was made in the theoretical description of DPS, in particular with regards to incorporating the \mbox{$1\to2$} splittings in a consistent way, with the impact-parameter dependence fully taken into account. This new approach to DPS led to new QCD frameworks which were developed by four different groups, namely B. Blok et. al. \cite{Blok:2011bu, Blok:2013bpa},  M. G. Ryskin and A. M. Snigirev \cite{Ryskin:2011kk, Ryskin:2012qx}, A. V. Manohar and W. J. Waalewijn \cite{Manohar:2012pe} and M. Diehl, JG and K. Sch\"{o}nwald \cite{Diehl:2017kgu}.

As mentioned above, the total cross section for DPS is suppressed compared to the SPS one for high energy scales. However, for differential cross sections, the situation can be rather different. For example, consider the final state $A+B$ characterised by the scale $Q_h$. This final state can be obtained with the SPS \mbox{pp $\to A+B$} or with the DPS composed of the two subprocesses pp $\to A$ and \mbox{pp $\to B$}. Let\footnote{Bold symbols are used for two-dimensional vectors in the plane perpendicular to the beam axis.} $\qv_A$ and $\qv_B$ be the transverse momenta of the final states $A$ and $B$ with respect to the beam axis in the centre-of-mass frame of the pp system. For small momenta i.e. $|\qv_A|\sim |\qv_B|\sim \Lambda$ for DPS, and $|\qv_A+\qv_B|\sim \Lambda$ for SPS, it can be shown that \cite{Diehl:2011yj, Diehl:2017wew}

\begin{equation}
\frac{\d\sigma^\mathrm{DPS}_{(A,B)}}{\d^2\qv_A\d^2\qv_B}\sim\frac{\d\sigma^\mathrm{SPS}_{A+B}}{\d^2\qv_A\d^2\qv_B}\sim\frac{1}{\Lambda^2Q_h^4}.
\end{equation}

\noindent Thus, at differential level, DPS and SPS contribute with the same strength in some regions\footnote{Examples of such regions are the ones where the transverse momenta are small or where the separation between the rapidities of the final states $A$ and $B$ is large.} of phase space.

\begin{figure}[t!] 
\centering
\includegraphics[width=0.45\textwidth]{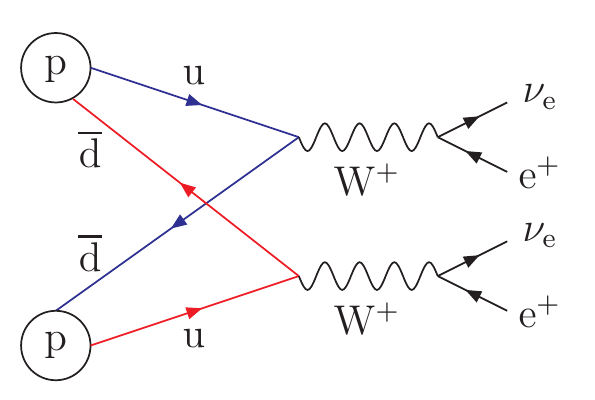}
\includegraphics[width=0.45\textwidth]{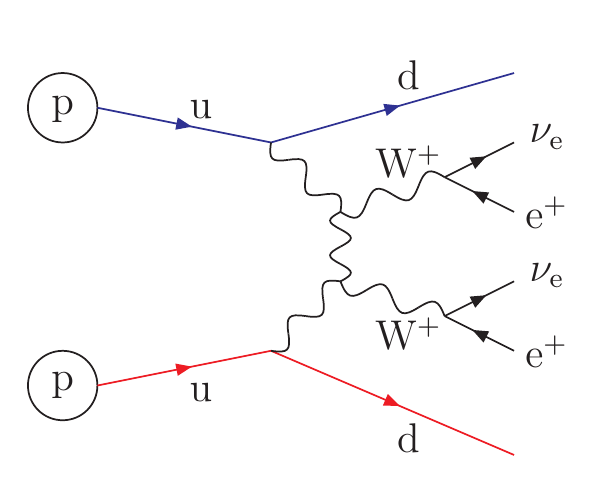} \\
(a) \hspace{200pt} (b)
\caption{Production of a $\W^+\W^+$ pair at a pp collider: (a) via DPS, (b) via SPS. More Feynman diagrams participate in the SPS process but they are not represented here. Both SPS and DPS contribute to the total cross section for same-sign WW production.}
\label{FigWW}
\end{figure}

For some processes, SPS can even be suppressed by a higher multiplicity of couplings. An example of such a process is same-sign WW production (i.e. $\W^+\W^+$ or $\W^-\W^-$) represented in \Fig \ref{FigWW}. Both diagrams include vertices whose strength is given by the electroweak coupling $\aew$. If one ignores the decays of the W bosons, it can be seen that the cross section for WW production via DPS scales as $\mathcal{O}(\aew^2)$, whereas the one for WW production via SPS is of order\footnote{Other diagrams participate in the SPS process and lead to contributions of order $\mathcal{O}(\aew^2\as^2)$.} $\mathcal{O}(\aew^4)$. Since in the perturbative regime (which is our framework) the couplings are relatively small, it turns out that the SPS and DPS cross sections are comparable in the instance of same-sign WW production. For these reasons, DPS cannot be neglected and needs to be accurately modelled. The DPS contribution to this process has been extensively studied from the theoretical side \cite{Kulesza:1999zh, Cattaruzza:2005nu, Maina:2009sj, dEnterria:2012jam, Gaunt:2010pi, Ceccopieri:2017oqe, Cao:2017bcb, Cotogno:2018mfv} and the first experimental evidence for DPS-initiated same-sign WW has recently been obtained by the CMS collaboration \cite{CMS:2019jog}.

The aim of this paper is thus to present a simulation of DPS which is based on the QCD framework developed in \cite{Diehl:2017kgu}, referred to later as the DGS framework. This framework has many advantages which makes it well-suited to a Monte-Carlo implementation. For example, it involves dPDFs which are defined for an individual hadron and which have a probabilistic interpretation. The parton-level simulation which is proposed in this work consists in a phase-space generator and an angular-ordered parton shower. First, two hard systems are generated using the full DPS cross section. Thereafter, the two systems are showered simultaneously in order to give a set of final-state partons. In both steps of the simulation, the dynamics of the $1\to2$ splittings as well as the impact-parameter dependence are fully taken into account, with dPDFs depending explicitly on this parameter. 

It is not the first time that the $1\to2$ splittings are included within a simulation of the underlying event. Indeed, B. Blok and P. Gunnellini developed an approach to include $1\to2$ splitting effects inside the MPI machinery of \mbox{\PyEight} \cite{Blok:2015rka, Blok:2015afa}. However, in this approach, the idea is to reweight the effective cross section\footnote{See \Sec \ref{sec:ReviewDPS} for a description of the effective cross section.} of the DPS process on an event-by-event basis with a factor that takes into account the $1\to2$ splitting effects. A similar strategy was used in \cite{Cotogno:2018mfv} to study the impact of the spin correlations between partons in same-sign WW production, where here the effects of $1\to2$ splittings were not considered. The approach of \cite{Blok:2015rka, Blok:2015afa} utilises a reweighting factor for the effective cross section that was calculated for a particular set of dPDFs, together with some appropriate choice of scale for the chosen processes. These dPDFs were calculated relying on particular factorisation assumptions. It is not clear how straightforward it would be for a general end-user to adapt this approach to incorporate different dPDFs and use it for arbitrary processes and kinematics. In the simulation presented in this work, the approach is process independent and does not rely on a particular set of dPDFs, which makes it more flexible and user-friendly. A first-principles implementation is also convenient for future developments, such as a consistent combination of SPS and DPS processes. Note that this simulation constitutes the first implementation of the full DGS framework; neither the work from \cite{Blok:2015rka, Blok:2015afa}, nor the one from \cite{Cotogno:2018mfv} implements all the features of the framework. It is also the first simulation which gives a geometrical picture of the evolution which is consistent with the $1\to2$ splitting mechanism.\footnote{See \Sec \ref{Sec:Mrg}.}

The first objective of this simulation is to allow a phenomenological study of any set of dPDFs that suits the DGS framework as well as the sum rules that those dPDFs must satisfy \cite{Gaunt:2009re}. It is important to emphasise here that the simulation is not bound to a specific set of dPDFs. Other dPDF sets can be plugged into this simulation without having to modify the whole approach, as long as they follow the prescription given in \cite{Diehl:2017kgu}. The second objective is to test some features related to the overall DGS framework, such as the dynamics of the $1\to2$ splittings or the impact-parameter dependence. The new aspects of this simulation might improve the current MPI models in the future, which is the underlying goal of this work.

The plan of this paper is the following. First, reviews of parton showers and DPS are given in \Secs \ref{Sec:PS} and \ref{Sec:DPS} respectively. Then, the simulation of DPS which has been developed is introduced in \Sec \ref{Sec:dShower} and the results of the simulation in the context of $\W^+\W^+$ pair production are presented in \Sec \ref{Sec:results}. Finally, the main ideas and the conclusions are gathered in the summary.

\section{Current parton shower algorithms}
\label{Sec:PS}

\subsection{Selection of the hard process}

Let us consider a proton-proton collision that happens at a centre-of-mass energy $\sqrt{s}$. The collision leads to the production of a final state $A$. In an event generator, the collision is described as the interaction between a parton $i$ coming from one proton with a parton $j$ coming from the other. In the centre-of-mass frame of the pp system, partons $i$ and $j$ have four-momenta\footnote{Here, the transverse momenta of partons $i$ and $j$ with respect to the beam axis (called ``primordial'' or ``intrinsic'' transverse momenta) are neglected i.e. partons $i$ and $j$ are assumed to be collinear with the incoming protons. This framework is called ``collinear factorisation''. In the PDFs, those transverse momenta are integrated over.} $p_{i,j}=x_{1,2}(\sqrt{s}/2)(1;0,0,\pm1)$, which results in a squared invariant mass of \mbox{$\hat{s}=(p_i+p_j)^2=x_1x_2s$}. The total cross section for the process pp $\to A$ can be written using the factorisation formula \cite{Collins:1989gx}

\begin{equation}
\sigma_A(s)=\sum_{i,j}\iint \d x_1\,\d x_2 \;f_i(x_1,\mu^2)\,f_j(x_2,\mu^2)\,\hat{\sigma}_{ij\to A}(\hat{s}=x_1x_2s,\mu^2).
\label{eq:Fact}
\end{equation}

\noindent The integrated parton-level cross section $\hat{\sigma}_{ij\to A}$ describes the short-range physics (matrix element + phase space) of the interaction between the two partons $i$ and $j$ coming from the incoming protons. This partonic cross section may be calculated using perturbation theory, in practice up to a few terms. The factorisation scale $\mu^2$ can be seen as an artificial delineation between short-range ($\hat{\sigma}_{ij\to A}$) and long-range (PDFs) physics. Below $\mu^2$, the physics is absorbed into the PDFs, which are universal (i.e. independent of the process $ij\to A$).  One should have \mbox{$\mu^2\gg \Lambda^2\sim1\,\mathrm{GeV}^2$} in order for perturbation theory to be applicable for $\hat{\sigma}_{ij\to A}$. 

In an event generator, Monte-Carlo techniques \cite{Buckley:2011ms, Sjostrand:2006za} are used to select a hard process $ij\to A$ as well as its kinematics according to \Eq (\ref{eq:Fact}). This equation should be kept in mind since it will be compared later to its equivalent for DPS.

\subsection{QCD radiation}

Once a hard process has been selected, it needs to be evolved in order to take into account the extra emissions of colour-charged particles \cite{Buckley:2011ms}. First, an evolution variable $Q^2$ is defined. The evolution then brings the system from the hard scale $Q_h^2$ down to $Q_0^2\sim\Lambda^2$ by going downwards in $Q^2$ i.e. from the hard process down to the non-perturbative regime. Within the parton-shower framework, this is done by defining a branching probability for each QCD branching. For final-state radiation (FSR) i.e. radiation coming from the final-state legs of the system, these branching probabilities are calculated by approximating matrix elements in the limit where the emitted parton is collinear to the radiating one, referred to as the collinear limit. The expression of the branching probabilities at LO for FSR can be found in \cite{Buckley:2011ms, Sjostrand:2009ad}.

Radiation associated with the partons which initiate the hard process is called initial-state radiation (ISR). In the case of proton-proton collisions, these partons actually come from protons. Hence, the inner structure of the proton needs to be taken into account in order to give an accurate description of ISR. This can be achieved using PDFs.

The main idea in the case of ISR is to work with the ensemble of partons described by the PDFs and to relate the branching probability to the variation of this set under a change in the evolution variable $Q^2$. This variation is calculated with the DGLAP equations. The system of partons is evolved by using the so-called ``backward evolution'' \cite{Sjostrand:1985xi, Bengtsson:1986gz}. The aim is thus to reconstruct the past history of the parton which initiates the hard process by using a conditional branching probability. This probability is defined as the rate at which the number of partons of flavour $i$ changes during a variation $\d Q^2$ of the scale. More precisely \cite{Sjostrand:1985xi, Bengtsson:1986gz}:

\begin{equation}
\begin{split}
\dP_{i}^\mathrm{ISR}=& \frac{\d Q^2}{Q^2}\sum_k\frac{\as}{2\pi}\,\frac{x/zf_k(x/z,Q^2)}{xf_i(x,Q^2)}\,P_{k\to i}(z)\,\d z \\
&\times\exp\left(-\sum_k\int_{Q^2}^{Q_h^2}\frac{\d {Q'}^2}{{Q'}^2}\int_{x}^{1}\frac{\as}{2\pi}\,\frac{x/z'f_k(x/z',{Q'}^2)}{xf_i(x,{Q'}^2)}\,P_{k\to i}(z')\,\d z'\right).
\end{split}
\label{eq:dPISR}
\end{equation}

\noindent This quantity is the probability that a given parton of flavour $i$, with longitudinal momentum fraction $x$, is resolved during an evolution from a scale $Q_h^2$ down to $Q^2$ and then appears as coming from a parton of flavour $k$ with momentum fraction $x/z$ \cite{Sjostrand:1985xi, Bengtsson:1986gz}. The functions $P_{k\to i}(z)$ are the LO unregularised\footnote{These functions introduce the singularity $z=1$ inside \Eq (\ref{eq:dPISR}). In practice, this singularity is avoided by limiting the range for the values of $z$: $z<z_\mathrm{max}(Q^2)<1$, where the boundary may depend on $Q^2$.} splitting kernels and $\as$ is the strong coupling. The exponential factor is the no-emission probability. It ensures unitarity, which means that the probability density defined by Equation (\ref{eq:dPISR}) is normalised to unity \cite{Sjostrand:2009ad}. This no-emission probability is closely related to the Sudakov form factor defined in perturbative calculations; the main difference being the presence of a non-perturbative PDF ratio. In both cases, the role of the exponential factor is to resum some large logarithms that can spoil the perturbative expansion. For this reason it is customary to refer to the no-emission probability as the Sudakov form factor and this latter term will be used in the following. The inner structure of the proton is correctly taken into account in \Eq (\ref{eq:dPISR}) with the presence of a ratio of PDFs; the evolution is ``guided'' by the PDFs. This expression is valid in the collinear limit only, since the splitting kernels in the DGLAP equations are derived with this approximation. 

The evolution variable $Q^2$ should be chosen to be proportional to the virtuality of the off-shell parton which is radiating. In \PySix, the evolution variable was chosen to be exactly the virtuality \cite{Sjostrand:2004ef}, whereas in \PyEight, \textsc{Sherpa} \cite{Schumann:2007mg}, \textsc{Dire} \cite{Hoche:2015sya}, \textsc{Vincia}\cite{Fischer:2016vfv} and \textsc{Ariadne} \cite{Lonnblad:1992tz} it is the transverse momentum squared $\pTs$ of the emitted parton with respect to the initial direction of the radiating one. In \Herwig, the evolution variable for FSR is defined to be \cite{Gieseke:2003rz}

\begin{equation}
\qtis_\mathrm{FSR}=\frac{p_k^2-m_k^2}{z(1-z)},
\label{eq:qtisFSR}
\end{equation} 

\noindent where $p_k^2$ is the four momentum squared of the parton $k$ which is currently radiating and $m_k$, its rest mass. In the case of ISR, the initial-state parton $i$ gets a space-like virtuality during the backward evolution which is given by $p_i^2-m_i^2$. The evolution variable is thus defined as

\begin{equation}
\qtis_\mathrm{ISR}=\frac{-(p_i^2-m_i^2)}{1-z}.
\label{eq:qtisISR}
\end{equation}

For the branching $k\to i+j$ of a final-state parton $k$, if $E_{k}$ is its energy, then $\qtis\simeq E_{k}^2\,\theta_{ij}^2$ in the limit $\theta_{ij}\to 0$, where $\theta_{ij}$ is the opening angle between the three-momenta of partons $i$ and $j$. Therefore, an evolution downwards in $\qtis$ implies that the angle $\theta_{ij}$ must decrease through the evolution. This property is referred to as angular ordering \cite{Bahr:2008pv, Gieseke:2003rz}. It has an important consequence, namely that the shower is coherent. This means that in an event in which the branching $k\to i+j$ happens, partons $i$ and $j$ radiate soft gluons coherently, so that at angles greater than $\theta_{ij}$, it is as if the gluons are radiated by parton $k$ \cite{Webber:1986mc}. A parton shower which is not coherent radiates too much and does not reproduce the results that one can obtain with matrix-element calculations in the limit where the emitted gluons are soft (referred to as the soft region). 

In the simulation of DPS which will be presented later, the evolution variable will be $Q^2=\qtis$, as in \Herwig.

\subsection{Kinematics}
\label{Sec:PSkin}

The branching probabilities are used to select probabilistically a set of phase-space points $\{(Q_n^2,z_n)\}$, which represents all the extra emissions that have been added to the original hard process. The algorithm which is employed to achieve that is called the ``veto algorithm'' \cite{Sjostrand:2006za, Bahr:2008pv, Buckley:2011ms}. Each new iteration of the algorithm uses the scale $Q^2$ of the previous one as its $Q_h^2$ in the expression of the branching probability. Once the shower has been generated, it is necessary to set up kinematics that preserve four-momentum conservation and ensure that all final-state partons are on-mass-shell. Multiple strategies exist. In the following, the kinematics defined in \Herwig\, \cite{Bahr:2008pv} will be presented since the evolution variable $Q^2=\qtis$ is used.

Let us consider the branching $k \to i+j$. Within the parton-shower framework, this branching is a phase-space point $(\qtis,z)$. The four-momenta of the two new partons coming from the branching need to be constructed from these two shower variables. The first quantity which needs to be defined is the virtuality of the radiating parton. Its value can be extracted from \Eqs (\ref{eq:qtisFSR}) and (\ref{eq:qtisISR}) for FSR and ISR respectively. The virtualities are necessary to compute the magnitude $\pT$ of the transverse momentum of partons $i$ and $j$ with respect to the direction of the momentum of their mother $k$. This one is given by \cite{Bahr:2008pv}

\begin{equation}
\pTs=z(1-z)p_{k}^2-(1-z)p_i^2-zp_j^2.
\end{equation}

\noindent For FSR, partons $i$ and $j$ are set on-mass-shell so $p_{i,j}^2=m_{i,j}^2$. Therefore

\begin{equation}
\label{eq:pTFSR}
\pTs = z^2(1-z)^2\qtis + z(1-z)m_{k}^2-(1-z)m_i^2-zm_j^2.
\end{equation}

\noindent In the case of ISR, it is partons $k$ and $j$ that are set on-mass-shell. Moreover, partons $k$ and $i$ are assumed to be massless. Thus

\begin{equation}
\pTs = (1-z)^2\qtis-zm_j^2.
\end{equation}

\noindent The magnitude $\pT$ defined above is not enough to fully specify the relative transverse momentum of partons $i$ and $j$. Indeed, one needs to define the orientation of this momentum in the plane perpendicular to the direction of parton $k$. This is achieved by selecting an azimuthal angle $\varphi$. This angle is typically uniformly distributed between 0 and $2\pi$. However, this flat distribution can be biased to take into account the azimuthal correlations between the partons which are due to the fact that the spins and polarisations carried by the partons lead to additional interferences \cite{Webber:1986mc}.

The variables $\pT$ and $\varphi$ are enough to describe the transverse motion of partons $i$ and $j$ with respect to the direction of parton $k$. The longitudinal motion is fixed by the shower variable $z$. Parton $i$ carries a fraction $z$ of the longitudinal component of the momentum of parton $k$ whereas parton $j$ carries a fraction $1-z$.

The kinematics of a branching $k\to i + j$ can be iterated in order to construct the kinematics of the whole shower. In the case of FSR, each colour-charged final-state leg of the hard process generates its own shower and is therefore called the progenitor. The result of the shower is a jet of partons which are distributed around the direction of this progenitor. To each progenitor is associated another leg of the hard process which carries the matching colour/anticolour. This other leg is referred to as colour partner. The shower is generated in the rest frame of the progenitor--partner pair, referred to as a dipole. The $+z$ direction is defined to be along the direction of the momentum of the progenitor. This frame is most of the time different from the centre-of-mass frame of the pp system (laboratory frame). In particular, the $z$-axis of the dipole may not coincide with the beam axis of the laboratory frame. The relative transverse momenta of the subsequent branchings are then computed iteratively using \Eq (\ref{eq:pTFSR}). The longitudinal parts are calculated by applying the definitions of the longitudinal fractions $z_n$. At the end of the procedure, the whole shower generated by the progenitor is boosted back to the laboratory frame. After the kinematics has been constructed for FSR, all the progenitors now have a time-like virtuality, since they radiated. This is to be expected, but the problem is that the kinematics of the hard process was selected by considering the progenitors on-mass-shell. Therefore, the kinematics which has been newly established breaks four-momentum conservation. A way to recover momentum conservation is to boost the generated jets along the direction of their respective progenitor such that the invariant mass $\sqrt{\hat{s}}$ of the hard process remains the same as the one which was selected before\footnote{Recall that the kinematics of the hard process is selected using \Eq (\ref{eq:Fact}).} the shower \cite{Bahr:2008pv}. 

For ISR, the procedure is similar. The hard process is initiated by two partons which have been extracted from the proton beams and with momenta \mbox {$p_{i,j}=x_{1,2}(\sqrt{s}/2)(1;0,0,\pm1)$} in the laboratory frame. A colour partner is assigned to each one of them and the showers are generated in the dipole rest frames, as for FSR. In the case where the two initial-state partons form a colour singlet (e.g. W or $\Z^0$ productions), the $z$-axis of the dipole is aligned with the beam axis. However, in the case where an initial state is colour connected to a final state (e.g. in gg $\to$ gg scattering) then the $z$-axis of the dipole is not aligned with the beam axis and one needs to boost the resulting shower. After the backward evolution has been performed, the new partons that are extracted from the beams have momenta $p'_{i,j}=x'_{1,2}(\sqrt{s}/2)(1;0,0,\pm1)$ in the laboratory frame. The fractions $x'_{1,2}$ can be related to the fractions $x_{1,2}$ selected before the shower by using the definitions of the $z_n$: $x'_{1,2}=x_{1,2}/\prod_{n}z_n$. The transverse part of the kinematics is calculated iteratively with respect to the $z$-axis in the dipole rest frame. Each shower is then boosted back to the laboratory frame. The two partons which are initiating the hard process now have acquired a transverse momentum and a space-like virtuality. This is because some emissions were attached to those two partons, which turned them into virtual particles. However, the kinematics of the hard process before the shower was established with the momenta $p_{i,j}$. Therefore, momentum conservation is also lost in the ISR case. One solution here is to rescale the momenta of the partons that are initiating the hard process such that the invariant mass squared $\hat{s}=x_{1}x_{2}s$ and the rapidity $Y=(1/2)\ln(x_{1}/x_{2})$ of the hard process remain equal to their original values (i.e. before ISR). The rescaling factors are found by solving two algebraic equations and they define the longitudinal boosts that must be applied to the two partons extracted from the beams, as well as to the emissions which have been attached to them. The transverse kick has to be absorbed globally by the whole final state by applying a transverse boost to this latter. Indeed, no transverse kick can be given to the momenta $p'_{i,j}$ since it is convenient to keep those momenta along the $z$-axis (i.e. collinear to the incoming protons) \cite{Bahr:2008pv}. 

All the details regarding the construction of the kinematics for FSR and ISR can be found in \cite{Bahr:2008pv}. Before closing this section, let us come back to two technical aspects which must be mentioned. First, it has not been specified yet from which scale $Q_h^2$ the evolution should start. There are multiple answers to this question. In \Herwig, each progenitor starts its evolution with its own starting scale. The starting scale $\qtis_{h,i}$ of a progenitor $i$ is related to the starting scale $\qtis_{h,j}$ of its colour partner $j$ via the relationship \cite{Bahr:2008pv, Buckley:2011ms}

\begin{equation}
\label{eq:startScale}
\qtis_{h,i}\,\qtis_{h,j}=m_{ij}^4,
\end{equation}

\noindent where $m_{ij}^2$ is the dipole mass squared. In the case where $i$ and $j$ are both initial-state partons, referred to as an initial-initial (II) dipole, $m_{ij}^2$ is simply the invariant mass squared $\hat{s}$ of the hard process. However, if the dipole is stretched between an initial-state parton and a final-state parton (IF/FI dipole), then $m_{ij}^2$ is not $\hat{s}$ anymore and is instead equal to the Mandelstam variable $-\hat{t}$ (or $-\hat{u}$). The condition given by \Eq (\ref{eq:startScale}) ensures that the soft region of the dipole phase space is correctly covered by the respective showers of partons $i$ and $j$, without overlap. One can see that there is a certain degree of freedom in \Eq (\ref{eq:startScale}). Indeed, any combination of $(\qtis_{h,i},\qtis_{h,j})$ that satisfies this condition will ensure that the soft region is covered. However, those combinations will give different results outside the soft region. This is because an angular-ordered shower describes correctly the emission pattern of a dipole in the soft and collinear regions only. Outside these regions, matrix elements must be used. In particular, the hard non-collinear region of the phase space (referred to as the ``dead cone'') is not populated by an angular-ordered shower and must be filled with matrix elements  \cite{Gieseke:2003rz}. The issues related to the dead cone in the case of DPS will not be addressed in this work. In the default \Herwig, the choices $\qtis_{h,i}=\qtis_{h,j}=\hat{s}$ and $\qtis_{h,i}=\qtis_{h,j}=-\hat{t}$ are made for the II and IF/FI dipoles respectively \cite{Bahr:2008pv}. It will be seen later that this choice should be modified for the DPS case.

The second technical aspect is the argument of the running strong coupling $\as$ that is used in the parton shower. One possible choice is the evolution variable $\qtis$. However, it is argued in \cite{Amati:1980ch,Ciafaloni:1981nm,Catani:1989ne,Catani:1990rr} that, in the context of an angular-ordered shower, using the transverse momentum squared $\pTs$ takes into account some NLO effects, which makes this choice better than $\qtis$ itself. This strategy is referred to as the ``Monte-Carlo scheme''. Deriving the same result in the case of DPS is beyond the scope of this work. Nevertheless, we decide to use the same scheme in the following and some arguments for making such a choice will be given in \Sec \ref{Sec:dShowerSh}. Note that considering $\qtis$ or the virtuality as the argument of the strong coupling are valid choices too in the context of a LO shower, since the differences between those choices lead to contributions which are beyond the accuracy of the shower.

\section{Double vs. single parton scattering}
\label{Sec:DPS}

In the following, a review of multiple parton interactions and, in particular, double parton scattering is given.

\subsection{Current MPI models}
\label{Sec:currentMPI}

\Eq (\ref{eq:Fact}) describes a proton-proton collision as a single parton-parton collision. However, many partons can be extracted from the same proton and those partons may also initiate other parton-parton collisions referred to as secondary interactions. There are several models for MPI. Within \Herwig, the number of secondary interactions is selected according to some distribution derived from the eikonal model \cite{Durand:1987, Bahr:2008dy, Bahr:2008spa}. Each subsystem is thereafter showered independently. At the end of the procedure, if four-momentum conservation is violated,\footnote{For example, the partons have extracted more energy than is available inside the proton.} the event is regenerated. This is repeated until all kinematic constraints are fulfilled \cite{Bahr:2008dy}. In \PyEight, the strategy is different. The generation of secondary interactions is combined with ISR and FSR in a unique sequence of decreasing transverse-momentum values. More specifically, a probability, differential in $\pTs$, to have a secondary interaction is defined \cite{Corke:2011yy, Sjostrand:2004pf, Sjostrand:2017cdm}. As for the eikonal model used in \Herwig, this differential probability is derived from the cross sections of the QCD $2 \to 2$ processes.\footnote{Processes such as $\q\g\to\q\g$, $\g\g\to\g\g$, $\q\q\to\q\q$, ...} During a common evolution which is performed by going downwards in $\pTs$, three scales are generated with the veto algorithm: one for ISR, one for FSR and one for MPI. The highest scale determines what actually happens. This procedure is called ``interleaved evolution'' \cite{Corke:2010yf, Sjostrand:2004ef}.

The common evolution of the different subsystems involves mPDFs. As mentioned in the introduction, a typical ansatz for the mPDFs is to assume that they can be expressed as a product of sPDFs. In the instance of DPS, this means that the dPDFs are written as \cite{Gaunt:2009re}

\begin{equation}
F_{ij}(x_1,x_2,\y,\mu^2_1,\mu^2_2)\simeq f_i(x_1,\mu^2_1)\,f_j(x_2,\mu^2_2)\,F(\y),
\label{eq:AnsatzProd}
\end{equation}

\noindent where $\y$ is the distance between the two partons in the plane transverse to the momentum of the proton. The $\y$-dependence of the dPDFs has been factorised out into some distribution $F(\y)$ which one assumes to be flavour and scale independent. This latter is usually modelled by using the electromagnetic form factor of the proton with a characteristic size of the order of the radius of the proton \cite{Bahr:2008dy, Corke:2010yf}. 

\Eq (\ref{eq:AnsatzProd}) is convenient, but it fails to take into account the correlations between the partons belonging to the same proton. Some of these correlations are a consequence of the number and momentum sum rules that the mPDFs must verify in order to give a realistic description of the proton. The number sum rules state that the proton has two valence u quarks and one valence d quark. In the case of SPS, their expressions can be found in a textbook and read

\begin{equation}
\int_0^1f_{\u_\mathrm{v}}(x,\mu^2)\,\d x = 2,
\hspace{70pt}
\int_0^1f_{\d_\mathrm{v}}(x,\mu^2)\,\d x = 1,
\end{equation}

\noindent where the valence components of the sPDFs are defined\footnote{For the sPDF sector, it is assumed that the sea component is defined as $\us=\ubar$.} as $f_{\u_\mathrm{v}}(x,\mu^2) = f_{\u}(x,\mu^2)-f_{\bar{\u}}(x,\mu^2)$ (and similarly for $\d_{\mathrm{v}}$). In the following, this kind of relation will be symbolically written as $\uv = u - \ubar$. The momentum sum rule imposes that all the partons probed at a given scale must carry the full momentum of the proton they belong to. More specifically, the sPDFs must satisfy the following equation

\begin{equation}
\int_0^1\left(\sum_{i=1}^{n_f}[xq_i(x,\mu^2)+x\bar{q}_i(x,\mu^2)] +xg(x,\mu^2)\right)\d x=1,
\end{equation}

\noindent with $n_f$ the number of quark flavours considered. An extension of these sum rules for the DPS case has been proposed in \cite{Gaunt:2009re}. They are referred to as the Gaunt-Stirling (GS) sum rules and have been extensively studied \cite{Gaunt:2009re,Ceccopieri:2014ufa,Diehl:2018kgr}. It is explained in \cite{Diehl:2018kgr} that it is the integrals\footnote{Those integrals need to be regularised in order to get a finite result, as it will be seen later.} over $\y$ of the dPDFs that satisfy the GS sum rules.

The GS sum rules strongly constrain the dPDFs. For example, if one defines the dPDF $\dv\dv$ as the probability density to extract two valence d quarks from the same proton, then this dPDF must be identically zero at all scales. Also, finding a parton with momentum fraction $x_1$ reduces the probability to find a second parton with a fraction $x_2$ which is close to the value $1-x_1$. In order to approximately include such effects, the usual sPDFs are rescaled and renormalised inside \PyEight. For example, the probability to extract a valence d quark with momentum fraction $x_2$, knowing that a parton of flavour $i$ has been extracted beforehand at a scale $\mu_1^2$ with a fraction $x_1$, is given by \cite{Sjostrand:2004pf}

\begin{equation}
f_{\d_{\mathrm{v}}}^{(i)}(x_2,\mu_2^2\,|\,x_1,\mu_1^2)=\frac{N_{\d_{\mathrm{v}}}^{(i)}}{1-x_1}\,f_{\d_{\mathrm{v}}}\left(\frac{x_2}{1-x_1},\mu_2^2\right),
\label{eq:rescale}
\end{equation}

\noindent where $N_{\d_{\mathrm{v}}}^{(i)}$ is equal to zero if $i=\d_{\mathrm{v}}$ and unity otherwise. This ensures that at most one valence d quark can be extracted from the proton. The rescaling $x_2\leftarrow x_2/(1-x_1)$ ensures that the kinematic constraint $x_1+x_2\leq 1$ is fulfilled. It can be checked that such a distribution satisfies the following sum rule 

\begin{equation}
\int_0^{1-x_1}f_{\d_{\mathrm{v}}}^{(i)}(x_2,\mu_2^2\,|\,x_1,\mu_1^2)\,\d x_2 = N_{\d_{\mathrm{v}}}^{(i)}.
\end{equation}

Similar distributions are defined for the other flavours. With such a scheme, the factorised form of the dPDFs given by \Eq (\ref{eq:AnsatzProd}) is replaced by the following expression inside \PyEight\, \cite{Fedkevych:2018}

\begin{equation}
\begin{split}
F^{\mathrm{Py}}_{ij}(x_1,x_2,\y,\mu^2_1,\mu^2_2)=\frac{F(\y)}{2}&\left(f_i(x_1,\mu_1^2)\,f_j^{(i)}(x_2,\mu_2^2\,|\,x_1,\mu_1^2) \right . \\
& \left . +f_j(x_2,\mu_2^2)\,f_i^{(j)}(x_1,\mu_1^2\,|\,x_2,\mu_2^2)\right).
\end{split}
\end{equation}

\noindent These dPDFs approximately satisfy the GS sum rules \cite{Fedkevych:2018} and are clearly symmetric in the sense that $F^{\mathrm{Py}}_{ij}(x_1,x_2,\y,\mu^2_1,\mu^2_2)=F^{\mathrm{Py}}_{ji}(x_2,x_1,\y,\mu^2_2,\mu^2_1)$. The rescaling introduced in \Eq (\ref{eq:rescale}) generates a suppression of the dPDFs close to the kinematic limit. More precisely, the distribution given by \Eq (\ref{eq:rescale}) tends smoothly towards zero whenever the kinematic boundary $x_1+x_2=1$ is approached. In QCD studies, it is customary to implement this kinematic suppression by adding a phase-space factor to \Eq (\ref{eq:AnsatzProd}) which is usually of the form $(1-x_1-x_2)^p\,\Theta(1-x_1-x_2)$, with $p\geq 1$ and $\Theta$, the Heaviside function \cite{Korotkikh:2004bz}. 

More features are implemented inside the MPI model of \PyEight \,\cite{Sjostrand:2004pf}. For example, the concept of ``companion'' quark is introduced to take into account the fact that sea quarks are always produced in pairs. More precisely, each time a sea quark is extracted from a proton, it leaves behind its corresponding antiquark which is included inside the structure of the beam remnant. This latter quark is referred to as the ``companion''.

In \Herwig, the factorised form (\ref{eq:AnsatzProd}) is preserved, without adding any phase-space factor. As mentioned above, the kinematic constraint is enforced by vetoing the events that do not satisfy it. Moreover, the MPI machinery cannot extract too many valence quarks since the backward evolutions of the secondary interactions are forced to terminate on a gluon, whereas the one of the hard process finishes necessarily on a valence quark. In practice, this is achieved by evolving the secondary interactions using sPDFs with the valence contributions subtracted out \cite{Bahr:2008dy}.

In the following, unless explicitly mentioned, the $F_{ij}$ will refer to the $\y$-dependent dPDFs as defined in \cite{Diehl:2017kgu} i.e. the factorised form given by \Eq (\ref{eq:AnsatzProd}) will not be used and has been recalled here for historical reasons only.

\subsection{Review of double parton scattering}
\label{sec:ReviewDPS}

It has been seen that a factorisation formula can be written for SPS, recall \Eq (\ref{eq:Fact}). In the same way, a factorisation formula can also be derived for DPS  \cite{Diehl:2011yj, Diehl:2015bca, Diehl:2017kgu, Vladimirov:2017ksc, Diehl:2018wfy}.  Consider a final state $A+B$ produced during a pp collision at a centre-of-mass energy of $\sqrt{s}$. It is assumed that the production of such a final state can be described with the subprocesses pp $\to A$ and \mbox{pp $\to B$} i.e. a DPS. It is possible to define two different factorisation scales $\mu_A^2$ and $\mu_B^2$, one for each subprocess. The total cross section for the process pp $\to A+B$ via a DPS process only is \cite{Gaunt:2009re,Diehl:2017kgu}

\begin{equation}
\begin{split}
\sigma^\mathrm{DPS}_{(A,B)}(s)=\frac{1}{1+\delta_{AB}}\sum_{i,j,k,l}&\int\d x_1\,\d x_2\,\d x_3\,\d x_4\,\hat{\sigma}_{ij\to A}(\hat{s}_{12}=x_1x_2s,\mu_A^2)\,\hat{\sigma}_{kl\to B}(\hat{s}_{34}=x_3x_4s,\mu_B^2) \\
&\times\int \d^2\y\,\Phi^2(y\nu)\,F_{ik}(x_1,x_3,\y,\mu_A^2,\mu_B^2)\,F_{jl}(x_2,x_4,\y,\mu_A^2,\mu_B^2).
\end{split}
\label{eq:FactDPS}
\end{equation}

\begin{figure}[t!] 
\centering
\includegraphics[width=0.4\textwidth]{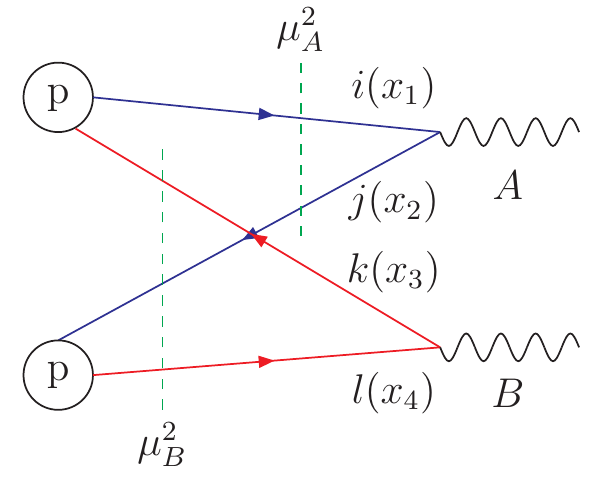}
\includegraphics[width=0.5\textwidth]{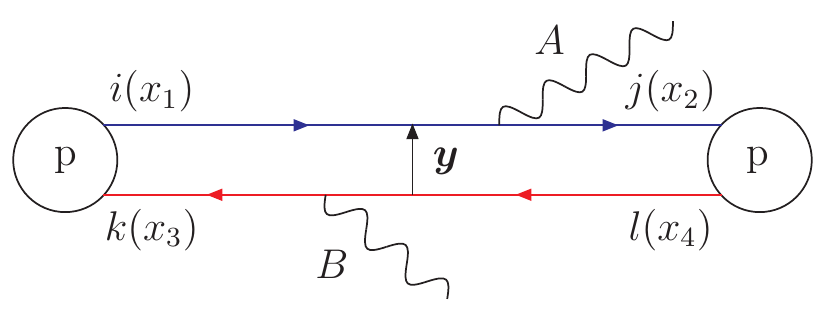} \\
(a) \hspace{200pt} (b)
\caption{Sketch of a DPS at a pp collider leading to the production of the final state $A+B$: (a) in terms of Feynman diagrams. The green dash-lines symbolise the choice of factorisation scale for each subprocess. The \Fig (b) offers another view on the collision. Here, the transverse distance $\y$ between the partons is represented.}
\label{FigFactDPS}
\end{figure}

\noindent This formula can be seen as a product of two \Eqs (\ref{eq:Fact}), especially for the short-range part. Indeed, the short-range part of \Eq (\ref{eq:FactDPS}) is the product of the parton-level cross sections for the subprocesses $ij\to A$ and $kl\to B$, where these have associated squared invariant masses of $\hat{s}_{12}=x_1x_2s$ and $\hat{s}_{34}=x_3x_4s$ respectively. The long-range part is more complicated. It involves the dPDFs $F_{ij}(x_1,x_2,\y,\mu_A^2,\mu_B^2)$. It can be intuitively understood that the transverse distance $\y$ between the two partons has to be the same for the two incoming protons in order for the two pairs of partons to actually collide in two separate hard interactions \cite{Gaunt:2012}. Since $\y$ is not a measurable quantity, this degree of freedom must be integrated over in order to give a physical cross section. The function $\Phi$ is a cut-off at small $y=|\y|$ which will be discussed later. The quantity in front of the sum in \Eq (\ref{eq:FactDPS}) is a symmetry factor which is equal to one half if $A=B$ and to unity otherwise \cite{Gaunt:2009re}. It comes from phase-space integration. It is worth recalling that \Eqs (\ref{eq:Fact}) and (\ref{eq:FactDPS}) are valid for partons that are collinear with the incoming protons (i.e. the primordial transverse momenta are neglected in the hard processes and in measurements). Illustrations of this formula are given in \Fig \ref{FigFactDPS}.

For historical purposes, it is interesting to see what \Eq (\ref{eq:FactDPS}) gives in the case where the factorisation ansatz given by \Eq (\ref{eq:AnsatzProd}) is used. With the factorised form, the integral over $\y$ can be performed. This leads to the definition of an effective cross section\footnote{Under the approximation of \Eq (\ref{eq:AnsatzProd}), the effect of including $\Phi$ is power suppressed and $\Phi$ can thus be dropped.} \cite{Bahr:2008spa, Gaunt:2009re}

\begin{equation}
\label{eq:sigmaEff}
\sigma_\mathrm{eff}=\left(\int F^2(\y)\,\d^2\y\right)^{-1},
\end{equation}

\noindent and \Eq (\ref{eq:FactDPS}) can be recast as the so-called ``DPS pocket formula''

\begin{equation}
\label{eq:pocket}
\sigma^\mathrm{DPS}_{(A,B)}(s)=\frac{\sigma^\mathrm{SPS}_A(s)\,\sigma^\mathrm{SPS}_B(s)}{(1+\delta_{AB})\,\sigma_\mathrm{eff}},
\end{equation}

\noindent where $\sigma^\mathrm{SPS}_A$ and $\sigma^\mathrm{SPS}_B$ are the SPS cross sections for the processes pp $\to A$ and pp $\to B$ given by \Eq (\ref{eq:Fact}). This formula is particularly simple to use in practice to estimate the size of the DPS contribution. Historically, this formula has been used to experimentally measure the effective cross section $\sigma_\mathrm{eff}$ \cite{Ryskin:2011kk,Treleani:2007gi, Bahr:2013gkj}. More precisely, the D0 and CDF collaborations have measured $\sigma_\mathrm{eff}$ at a p$\bar{\mathrm p}$ collider by using the DPS contribution to \mbox{p$\bar{\mathrm p}\to \gamma +3$ jets}, with $A=\gamma +1$ jet and $B=2$ jets \cite{Abazov:2009gc, Abe:1997xk, Abe:1997bp}. The D0 collaboration found $\sigma_\mathrm{eff}=16.4$ mb whereas the CDF collaboration measured $\sigma_\mathrm{eff}=14.5$ mb. In  \cite{Treleani:2007gi, Bahr:2013gkj}, it is explained that including some theoretical considerations leads to a lower value than the one extracted by the D0 and CDF collaborations. More measurements have been performed recently at the LHC by the ATLAS, CMS and LHCb collaborations \cite{Aaij:2012dz, Aad:2013bjm, ChatrChyan:2013xxa, Aaij:2015wpa, Aaij:2016bqq, Aaboud:2016fzt, Aaboud:2016dea, Sirunyan:2017hlu}. What can be remembered from these measurements is that $\sigma_\mathrm{eff}\sim 1/\Lambda^2$ i.e. it is of the order of the transverse area of the proton \cite{Gaunt:2012}. Note that $\sigma_\mathrm{eff}$ is process independent according to \Eq (\ref{eq:AnsatzProd}). However, if one uses \Eq (\ref{eq:pocket}) to define $\sigma_\mathrm{eff}$ (as experimentalists do), then one may find that it depends on process, scale, etc...

Let us now present the main features of the DGS framework introduced in \cite{Diehl:2017kgu}. This framework involves $\y$-dependent dPDFs i.e. no factorisation ansatz is used. These dPDFs satisfy the following two properties:

\begin{enumerate}[label=\roman*.]
\item \label{enu:propi} These dPDFs satisfy the homogeneous dDGLAP evolution equations \cite{Diehl:2011yj, Diehl:2017kgu}. These equations can be derived by considering the renormalisation of the $\y$-dependent dPDF operator. In the case where the two factorisation scales are set to be equal i.e. \mbox{$\mu_A^2=\mu_B^2=\mu^2$}, they read 

\begin{equation}
\begin{split}
\mu^2\frac{\partial}{\partial\mu^2} \,F_{ij}(x_1,x_2,\y,\mu^2)=&\sum_{i'}\int_{x_1}^{1-x_2}\frac{\d x_1'}{x_1'}\,\frac{\as(\mu^2)}{2\pi}\,\hat{P}_{i'\to i}\left(\frac{x_1}{x_1'}\right)\,F_{i'j}(x_1',x_2,\y,\mu^2) \\
& + \sum_{j'}\int_{x_2}^{1-x_1}\frac{\d x_2'}{x_2'}\,\frac{\as(\mu^2)}{2\pi}\,\hat{P}_{j'\to j}\left(\frac{x_2}{x_2'}\right)\,F_{ij'}(x_1,x_2',\y,\mu^2),
\end{split}
\label{eq:homdDGLAP}
\end{equation}

\noindent where $\hat{P}_{i'\to i}(z)$ are the usual regularised splitting kernels. This equation is basically the sum of two usual DGLAP terms. The only differences are the presence of the dPDFs and the fact that the upper boundary of the integral is not unity anymore but is now determined by the kinematic condition $x_1+x_2\leq 1$. The evolution of the dPDF $F_{ij}(x_1,x_2,\y,\mu^2)$ with the scale $\mu^2$ therefore turns out to be simply the sum of the contributions from the evolution of each one of the two partons $i$ and $j$. It is important to note that the impact parameter $\y$ does not contribute to the evolution at all.

\item \label{enu:propii} At small $y$ and for scales $\mu\sim 1/y$, the dPDFs should be given, up to formally power-suppressed corrections, by a perturbative splitting expression involving the sPDFs. This can be derived by considering the operator product expansion of the dPDFs at small $y$. At LO, the expression is given by \cite{Diehl:2011yj}

\begin{equation}
F_{ij}^\mathrm{spl,pt}(x_1,x_2,\y,\mu^2)=\frac{1}{\pi y^2}\frac{f_k(x_1+x_2,\mu^2)}{x_1+x_2}\,\frac{\as(\mu^2)}{2\pi}\,P_{k\to i+j}\left(\frac{x_1}{x_1+x_2}\right).
\label{eq:SplPT}
\end{equation}

This term takes into account the fact that the pair of partons $ij$ can originate from the perturbative splitting of a parton $k$ with longitudinal momentum fraction $x_1+x_2$. The flavour $k$ is uniquely determined by the flavours $i$ and $j$ for LO QCD splittings so no sum is needed. If there is no flavour $k$ such that the branching $k \to i+j$ is allowed, because of colour or flavour considerations, then the perturbative splitting expression for the pair $(i,j)$ is equal to zero. This small-$y$ expression involves the unregularised splitting kernel $P_{k\to i+j}(z)$ and the sPDF of parton $k$, which gives the probability of probing such a flavour $k$ at the scale $\mu$. There is no need to regularise the splitting kernel since virtual loops cannot lead to a $1\to 2$ splitting \cite{Gaunt:2009re}. 
\end{enumerate}

In \cite{Diehl:2017kgu} a model set of dPDFs was constructed satisfying these constraints, and it is this set of dPDFs that is used in our numerical studies. We slightly adjusted this set to approximately take account of momentum and number sum-rule constraints, as discussed in \Sec \ref{Sec:Scheme} and \App \ref{app:inputs}. However, it is important to remind the reader that the framework which will be presented in the next section is not tied to this particular dPDF set. One can use any dPDFs that are consistent with the DGS framework (namely satisfying \Props (\ref{enu:propi}) and (\ref{enu:propii}) above) and that approximately satisfy the momentum and number sum-rule constraints.

Let us now briefly review the model set of dPDFs used in \cite{Diehl:2017kgu}. The authors modelled the dPDFs as follows

\begin{equation}
F_{ij}(x_1,x_2,\y,\mu^2)=F_{ij}^\mathrm{int}(x_1,x_2,\y,\mu^2)+F_{ij}^\mathrm{spl}(x_1,x_2,\y,\mu^2),
\end{equation}

\noindent where $F_{ij}^\mathrm{int}$ is called the intrinsic component of the dPDFs and it contains the non-perturbative contributions to the dPDFs. The term $F_{ij}^\mathrm{spl}$ is referred to as the splitting component and corresponds to the contribution from the perturbative $1\to 2$ splittings. Both $F_{ij}^\mathrm{int}$  and $F_{ij}^\mathrm{spl}$ separately satisfy the homogeneous dDGLAP equations \cite{Diehl:2011yj, Diehl:2017wew, Diehl:2017kgu}. Each component has its own starting scale for the evolution. The intrinsic component is initialised at the scale $\mu_0=1$ GeV by

\begin{equation}
\begin{split}
F_{ij}^\mathrm{int}(x_1,x_2,\y,\mu_0^2)=&\frac{1}{4\pi h_{ij}(x_1,x_2)}\exp\left(-\frac{y^2}{4h_{ij}(x_1,x_2)}\right) \\
& \times f_i(x_1,\mu_0^2)\,f_j(x_2,\mu_0^2)\,\frac{(1-x_1-x_2)^2}{(1-x_1)^2\,(1-x_2)^2},
\end{split}
\label{eq:StartInt}
\end{equation}

\noindent which is basically \Ansatz (\ref{eq:AnsatzProd}) with a phase-space factor times a Gaussian in $y$ with a width which depends on $x_{1,2}$ and on the flavours $i$ and $j$. The $y$ shape comes from the link between dPDFs and generalised parton distributions (GPDs) in the approximation of uncorrelated partons \cite{Diehl:2011yj, Diehl:2014vaa, Diehl:2004cx}. The widths of the Gaussians $h_{ij}(x_1,x_2)$ are defined as \cite{Diehl:2014vaa}

\begin{equation}
h_{ij}(x_1,x_2)=\alpha_i'\ln\frac{1}{x_1}+\alpha_j'\ln\frac{1}{x_2} + B_i + B_j,
\label{eq:hCoeffs}
\end{equation}

\noindent where the $\alpha_i'$ and $B_i$ coefficients are obtained using GPD phenomenology and are usually scale-dependent. More precisely, at low scale, values for these coefficients are determined by fitting models for GPDs to data for electromagnetic form factors \cite{Diehl:2004cx} as well as for $\mathrm{J}/\psi$ photoproduction and deeply virtual Compton scattering \cite{Diehl:2007zu, Aktas:2005xu}. In the procedure sketched in \cite{Diehl:2017kgu}, the dependence on $x_{1,2}$ is actually removed for simplicity so the functions $h_{ij}(x_1,x_2)$ are brought back to flavour-dependent coefficients which are 

\begin{equation}
h_{\q_i\q_j}=7.06\,\GeV^{-2},
\hspace{40pt}
h_{\q_i\g}=5.86\,\GeV^{-2},
\hspace{40pt}
h_{\g\g}=4.66\,\GeV^{-2},
\label{eq:hVal}
\end{equation}

\noindent where $\q_i$ stands for any quark or antiquark. Those coefficients are obtained by setting \mbox{$x_1=x_2=10^{-3}$} in \Eq (\ref{eq:hCoeffs}) and using the values of the $\alpha_i'$ and $B_i$ coefficients given in \cite{Diehl:2014vaa}. The initial condition (\ref{eq:StartInt}) is then evolved according to the homogeneous dDGLAP equations from the scale $\mu_0^2$ up to $\mu^2$. For the splitting component, the starting scale\footnote{Including the coefficient $b_0$ inside the definition of the starting scale $\mu_y$ simplifies some expressions in \cite{Diehl:2017kgu}. The physics is not changed since $b_0$ is of the order of unity.} is $\mu_y=b_0/y^*$, with $y^*=y/\sqrt{1+y^2/y^2_\mathrm{max}}$, \mbox{$b_0=2e^{-\gamma_E}\simeq 1.12$} and $y_\mathrm{max}=0.5\,\mathrm{GeV}^{-1}$. The input is then

\begin{equation}
F_{ij}^\mathrm{spl}(x_1,x_2,\y,\mu_y^2)=\exp\left(-\frac{y^2}{4h_{ij}(x_1,x_2)}\right)F_{ij}^\mathrm{spl,pt}(x_1,x_2,\y,\mu_y^2).
\label{eq:StartSpl}
\end{equation}

\noindent This expression reduces for small $y$ to the perturbative splitting expression given in \Eq (\ref{eq:SplPT}), thereby ensuring that \Prop (\ref{enu:propii}) above is satisfied. At small $y$, the splitting component behaves like $1/y^2$ \cite{Diehl:2011yj, Diehl:2017wew, Diehl:2017kgu}. From a naive power counting, this would mean that the component $F_{ij}^\mathrm{int}$ is negligible compared to $F_{ij}^\mathrm{spl}$ for $y\to 0$. However, this might not be necessarily the case in practice.

The input for the splitting component is not only defined by the small-$y$ expression given by \Eq (\ref{eq:SplPT}). In particular, some modelling is added. First, a Gaussian factor is included in order to suppress the expression at large $y$ values, as for the intrinsic component. Second, the starting scale is defined to be $\mu_y$ and not $b_0/y$. This is to avoid the sPDF and the strong coupling being evaluated at a scale which is outside the perturbative regime. Indeed, with this choice, $\mu_y\to b_0/y_\mathrm{max}\simeq2.24$ GeV when $y\to +\infty$, which is still in the perturbative regime. The input (\ref{eq:StartSpl}) is also evolved by using the homogeneous dDGLAP equations, but starting from the scale $\mu_y$ \cite{Diehl:2017kgu}.

\begin{figure}[t!] 
\centering
\includegraphics[width=0.5\textwidth]{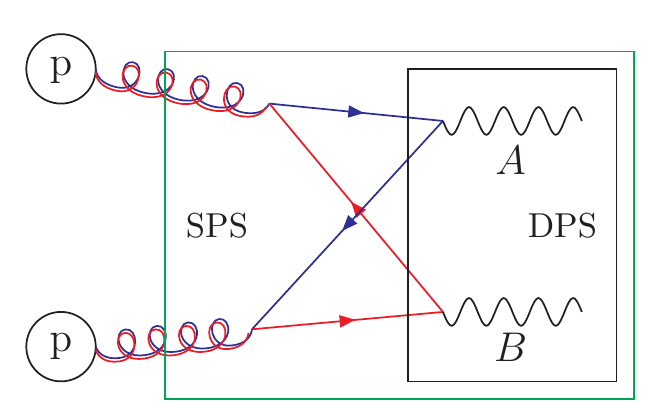}
\caption{Example of a process which can be seen either as a DPS or as a SPS. If the hard process is defined by the black box, then it is a DPS with the two subprocesses $\q\qbar\to A$ and $\q\qbar\to B$. In the case where the hard process is defined by the green box, then one has the SPS $\g\g\to A+B$. The pieces which are not included within the boxes are integrated out inside the PDFs.}
\label{FigyZero}
\end{figure}

Let us now come back to the general DGS framework. The $1/y^2$ behaviour of the splitting component is troublesome. Indeed, the expression diverges when $y\to 0$ and leads to an unphysical DPS cross section. As often in Feynman graph calculations, a divergence in the formulae is a manifestation of a problem of double counting. In that case, it is a double counting between DPS and SPS. Indeed, a $\q\qbar$ pair with separation $\y$ may\footnote{The colour configuration of the $\q\qbar$ pair must be in the octet representation.} be resolved as a single gluon at resolution scales smaller than $1/y$. This phenomenon is described by the $1\to2$ splitting mechanism. The double counting issue then appears since a DPS process with $1\to2$ splittings in both protons may also be regarded as a loop correction to the SPS process. The sketch given in \Fig \ref{FigyZero} shows how the same Feynman diagram can be seen either as an SPS or as a DPS.

The factorisation formula for DPS thus breaks down for small $y$ because of double counting between SPS and DPS. This issue can be solved by a two-step procedure explained in detail in \cite{Diehl:2011yj, Diehl:2017wew, Diehl:2017kgu}. The first step is to regulate the DPS total cross section at small $y$. This is done by including the function $\Phi$ inside the factorisation formula (\ref{eq:FactDPS}). The function $\Phi$ is chosen so that $\Phi(u)\to 1$ for $u\to+\infty$ and $\Phi(u)\to 0$ for $u\to 0$, which indeed regulates the integral. The cut-off scale $\nu$ separates DPS from SPS and can be seen as some new factorisation scale. In \cite{Diehl:2017kgu}, it is argued that one should choose $\nu\sim Q_h$, with $Q_h$ the hard scale which characterises the final-state $A+B$. In the following, the function $\Phi(u)$ is chosen to be the Heaviside function $\Theta(u-b_0)$. The second step is to make a subtraction that removes the double counting between DPS and SPS. The total cross section for the production of the final-state $A+B$ thus becomes\footnote{This expression is a simplified version of the one given in \cite{Diehl:2017kgu}. Some terms have been left aside.}

\begin{equation}
\label{eq:sub}
\sigma_{A+B}=\sigma_{A+B}^\mathrm{SPS}+\sigma_{(A,B)}^\mathrm{DPS}-\sigma_\mathrm{sub},
\end{equation}

\noindent where $\sigma_\mathrm{sub}$ is the integral over $\y$ of a quantity $\d\sigma_\mathrm{sub}/\d^2\y$ that is defined to satisfy $\d\sigma_\mathrm{sub}/\d^2\y\simeq\d\sigma_{(A,B)}^\mathrm{DPS}/\d^2\y$ for $y\lesssim 1/Q_h$ and $\d\sigma_\mathrm{sub}/\d^2\y\simeq\d\sigma_{A+B}^\mathrm{SPS}/\d^2\y$ for $y\gg 1/Q_h$. Thus, when the two partons are well separated, the production of $A+B$ is described as a DPS with the DPS cross section. In this region, the approximations used to derive the DPS cross section hold and  $\Phi(y\nu)\simeq 1$. In contrast, when the partons get really close ($y\sim 1/Q_h$), the DPS description is not valid anymore and the SPS cross section is used instead. It can be shown that the dependence of the total cross section $\sigma_{A+B}$ on the unphysical cut-off $\nu$ is removed via a cancellation between the $\nu$-dependent parts of $\sigma_{(A,B)}^\mathrm{DPS}$ and $\sigma_\mathrm{sub}$. More specifically, in \cite{Diehl:2017kgu}, the subtraction term is defined as the DPS cross section given by \Eq (\ref{eq:FactDPS}), but with the dPDFs replaced by the fixed order splitting expression (i.e. \Eq (\ref{eq:StartSpl}) without the Gaussian factor and with the scale $\mu_y$ replaced by the generic scale $\mu$). The function $\Phi(u)$ is also inserted inside the subtraction term. Since the dPDFs are dominated by the perturbative splitting expression at small $y$, one can understand that the $\nu$-dependences cancel, at least order by order in QCD.

\section{A parton-level simulation of DPS}
\label{Sec:dShower}

\subsection{The approach}

The main idea is to use \Eq (\ref{eq:FactDPS}) to generate two separate hard processes with their respective kinematics. After that, the two hard processes are showered simultaneously during a common evolution guided by the dPDFs. This procedure leads to the set of final-state partons. Let us compare with what is already done in the current event generators. The possibility of choosing two separate hard processes already exists. However, their kinematics are selected according to the usual SPS cross section given by \Eq (\ref{eq:Fact}). The kinematics are then rectified in order to take into account the kinematic constraints that link the two hard processes. The total DPS cross section is then calculated with the DPS pocket formula given by \Eq (\ref{eq:pocket}). Regarding the evolution of these two hard processes, the current models of MPI shower them almost independently in the sense that the dynamical correlations between the different subsystems are not taken into account. Using dPDFs should then catch some of these dynamical correlations. The impact parameter $\y$ is present in MPI models. However, the $\y$-dependent part of the dPDFs is factorised out into a function $F(\y)$, as in \Eq (\ref{eq:AnsatzProd}). The function $F(\y)$ is then modelled and used to calculate the average number $\left<n_\mathrm{MPI}\right>$ of secondary interactions. This is the eikonal model \cite{Durand:1987, Bahr:2008dy, Bahr:2008spa, Sjostrand:2017cdm}. In \Herwig, the number of secondary interactions is calculated straight from the eikonal model \cite{Bahr:2008dy, Bahr:2008spa}, whereas the function $F(\y)$ is used to weight the probability to have a new secondary interaction in \PyEight\, \cite{Sjostrand:2017cdm}. In this latter event generator, the factor $F(\y)$ is now dependent on the longitudinal fraction $x$ in order to take into account the longitudinal correlations between the partons \cite{Sjostrand:2017cdm}. In this work, the longitudinal correlations are included by using the $\y$-dependent dPDFs instead of the factorisation form given by \Eq (\ref{eq:AnsatzProd}).

\subsection{Selection of the two hard processes}
\label{Sec:2hard}

In order to select kinematic and flavour configurations for each one of the two hard processes, one needs to sample random variables according to \Eq (\ref{eq:FactDPS}). The generic method to achieve this is well known for the usual SPS cross section formula and is explained in great detail in \cite{Sjostrand:2006za}. In the case of DPS, the strategy is broadly the same. The major difference is the dimension of the phase space which jumps from three to seven. Indeed, each hard process is characterised by three non-trivial variables. The last variable is the impact parameter $y$. More specifically, \Eq (\ref{eq:FactDPS}) is rewritten as

\begin{equation}
\begin{split}
\sigma^\mathrm{DPS}_{(A,B)}(s)=\frac{1}{1+\delta_{AB}}\sum_{i,j,k,l}&\int\d \tau_A\,\d Y_A\,\d\hat{t}_A\,\d \tau_B\,\d Y_B\,\d\hat{t}_B\,\frac{\d \hat{\sigma}_{ij\to A}}{\d\hat{t}_A}\,\frac{\d\hat{\sigma}_{kl\to B}}{\d\hat{t}_B} \\
&\times \int 2\pi\, y\,\d y\,\Phi^2(y\nu)\,F_{ik}(x_1,x_3,\y,\mu^2)\,F_{jl}(x_2,x_4,\y,\mu^2),
\end{split}
\label{eq:FactDPSDiff}
\end{equation}

\noindent with $\tau_A=x_1x_2$, $Y_A=(1/2)\ln\left(x_1/x_2\right)$, $\tau_B=x_3x_4$ and $Y_B=(1/2)\ln\left(x_3/x_4\right)$. The Mandelstam variables $\hat{t}_A$ and $\hat{t}_B$ specify the transverse momenta of the outgoing particles in each one of the two $2\to2$ processes. The factor $2\pi$ comes from the fact that the dPDFs have been assumed to be independent of the azimuthal angle of the vector $\y$. In other terms, the dPDFs only depend on the magnitude $y$. In the following, the equal-scale case will be considered. This is because there is no set of unequal-scale $\y$-dependent dPDFs available yet. However, an extension of the algorithm presented below to the unequal-scale case is in principle achievable, see \Sec \ref{Sec:unequal}. The equal-scale case is suitable for processes such as same-sign WW production where one has $\mu_A^2\simeq \mu_B^2\simeq m_\W^2$, with $m_\W$, the mass of the W boson. More generally, the prescription which will be used in the following is to set the common factorisation scale $\mu^2$ to be equal to $\min(\mu_A^2,\mu_B^2)$. From a parton-shower point of view, this is the most realistic choice since the two hard processes should be resolved at the start of the evolution. In contrast, note that the arguments of the couplings used in the expressions of the differential parton-level cross sections do not have to be the same for the two hard processes.

The phase-space boundaries are mainly determined by the cuts (e.g. on the transverse momenta) and the kinematic constraints. The two main constraints are $x_1+x_3\leq 1$ and $x_2+x_4\leq 1$. If one selects first the variables for the subprocess pp $\to A$, then the variables $\tau_B$ and $Y_B$ are constrained by some limits which are functions of $\tau_A$ and $Y_A$. 

The constraints on the variable $y$ are less trivial. Theoretically, the integral goes up to $y=+\infty$. However, one expects the integrand to fall to zero quickly for values of $y$ larger than the radius of the proton. For this reason, one can in practice cut off the $y$ integral at some value $y_\mathrm{cut}$, if this value is much larger than the radius of the proton. The value $y_\mathrm{cut}=8\,\mathrm{GeV}^{-1}$ will be used here. This choice will be motivated in \Sec \ref{Sec:setup}. Note that even if $y > y_\mathrm{max}$, $\mu_y$ stays larger than $b_0/y_\mathrm{max}$ because of the $y^*$-prescription introduced in \Eq (\ref{eq:StartSpl}). The lower limit is given by the function $\Phi(y\nu)$. In our case $\Phi(y\nu)=\Theta(y\nu-b_0)$, which implies that the integral is non-zero for $y>b_0/\nu$ (or $\mu_y<\nu$). In the following, the choice $\nu=Q_h$ is made, with $Q_h$, the hard scale. The lower limit is thus $b_0/Q_h$. This is a natural requirement since the DPS description is not valid for $y\lesssim1/Q_h$. 

\subsection{Combining parton showers and dPDFs}
\label{Sec:dShowerSh}

The aim now is to use the dPDFs to guide the ISR evolution of the two hard processes. This idea is not completely new and has already been investigated in the past for \mbox{\PyEight} \cite{Sjostrand:2017cdm, Sjostrand:2004ef}. However, the authors were using the model of mPDFs presented in \Sec \ref{Sec:currentMPI} which is based on factorising the mPDFs as products of sPDFs. An evolution which uses the $\y$-dependent dPDFs is proposed here.

Let us consider two partons of flavours $i$ and $j$ belonging to the same incoming proton with momentum fractions $x_1$ and $x_2$ and participating in two different hard processes characterised by  the same hard scale $Q_h^2$. Simulating ISR for DPS requires to perform a simultaneous backward evolution of the two partons belonging to the same proton, starting from the scale $Q_h^2$. In order to do so, one needs to define a branching probability for the pair of flavours $ij$. This can be achieved by using the homogeneous dDGLAP equations (\ref{eq:homdDGLAP}) in the same spirit as what is usually done with the conventional parton showers \cite{Sjostrand:2017cdm, Sjostrand:2004ef}. One can write

\begin{equation}
\label{eq:dPij}
\begin{split}
\dP_{ij}=\frac{\d Q^2}{Q^2}&\left(\sum_{i'}\int_{x_1}^{1-x_2}\frac{\d x_1'}{x_1'}\,\frac{\as(\pTs)}{2\pi}\,P_{i'\to i}\left(\frac{x_1}{x_1'}\right)\,\frac{F_{i'j}(x_1',x_2,\y,Q^2)}{F_{ij}(x_1,x_2,\y,Q^2)} \right . \\
& \left . + \sum_{j'}\int_{x_2}^{1-x_1}\frac{\d x_2'}{x_2'}\,\frac{\as(\pTs)}{2\pi}\,P_{j'\to j}\left(\frac{x_2}{x_2'}\right)\,\frac{F_{ij'}(x_1,x_2',\y,Q^2)}{F_{ij}(x_1,x_2,\y,Q^2)} \right),
\end{split}
\end{equation}

\noindent where the fact that the argument of $\as$ should be the transverse momentum squared $\pTs$ of the branching has been anticipated, see \Sec \ref{Sec:PSkin}. Also, the factorisation scale $\mu^2$ has been replaced by the evolution variable $Q^2$, since the context of parton shower is now considered. The splitting kernels are now the unregularised ones, since the phase-space is regulated by a set of cut-offs. In order to get a well-defined probability which satisfies unitarity, one needs to add a Sudakov form factor as follows

\begin{equation}
\dP_{ij}^\mathrm{ISR}=\dP_{ij}\exp\left(-\int_{Q^2}^{Q_h^2}\dP_{ij}\right).
\label{eqdPijSud}
\end{equation}

\noindent $\dP_{ij}^\mathrm{ISR}$ is the ingredient needed to simulate ISR for DPS. The physical interpretation is the following. $\dP_{ij}^\mathrm{ISR}$ is the probability that the pair $ij$ remains resolved during an evolution starting from the scale $Q_h^2$ down to the scale $Q^2$. After that, the pair $ij$ might appear as coming either from the pair $i'j$ (first term in (\ref{eq:dPij})) or the pair $ij'$ (second term). In practice, this choice is made by selecting a scale for each one of the two channels. The highest scale determines which channel actually happens. This method is referred to as the ``competing veto algorithm'' \cite{Kleiss:2016esx}.

The form of \Eq (\ref{eq:dPij}) can be used to motivate the fact that $\pTs$ was chosen to be the argument of $\as$ in the case of DPS too. Indeed, \Eq (\ref{eq:dPij}) can be seen as the sum of two usual ISR branching probabilities as used in the SPS case, the main differences being the presence of dPDFs instead of sPDFs and the different upper boundaries for the integrals. For high-energy collisions, one expects the momentum fractions to be rather small (typically $x_{1,2}\sim10^{-3}$). Therefore, in practice, one has $1-x_{1,2}\simeq1$ and the kinematic conditions are hence similar to the SPS case ones. One can thus expect that most of the arguments presented in \cite{Amati:1980ch,Ciafaloni:1981nm,Catani:1989ne,Catani:1990rr} should hold for the DPS case too. As a reminder, the choice of the argument of the strong coupling leads to contributions which are beyond the accuracy of a LO shower so this choice should not have a significant impact on the numerical results. 

\Eq (\ref{eq:dPij}) also motivates the fact that the whole shower evolution in the case of DPS is gauge invariant. Indeed, the fact that the DPS ISR evolution is similar to the sum of two usual SPS ISR evolutions leads us to divide the radiation pattern into FSR and ISR in the same way as in the SPS case, where both components make use of the gauge-invariant Altarelli-Parisi splitting functions for their calculation. Moreover, the dPDFs have a gauge-invariant operator definition \cite{Diehl:2011yj} and evolve according to gauge-invariant renormalisation group equations (recall \Eq (\ref{eq:homdDGLAP})). Finally, the ``initial conditions'' described in \Sec \ref{sec:ReviewDPS} reduce at small $y$ and adequate scales into the appropriate gauge-invariant perturbative splitting expressions (recall \Eq (\ref{eq:SplPT})).

\begin{figure}[t!] 
\centering
\includegraphics[width=0.7\textwidth]{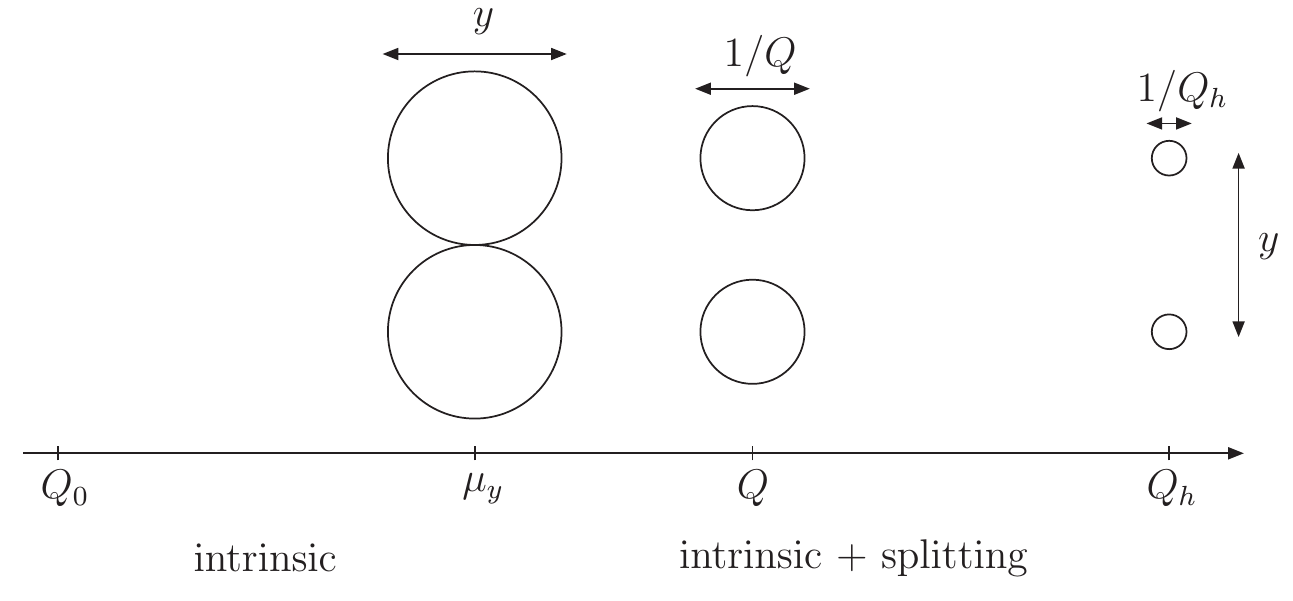}
\caption{Sketch of the backward evolution of a DPS. The axis shows the evolution scale. At the starting scale $Q_h$, the two partons belonging to the same proton have a size $1/Q_h$. This size increases during the backward evolution. At the scale $\mu_y$, the two partons have a size $y$ which is equal to the distance which separates them and they may be resolved into a single parton. The components of the dPDFs which are used are specified.}
\label{FigySize}
\end{figure}

Let us now come back to the DPS ISR evolution. Two regimes are defined. The evolution from the hard scale $Q_h^2$ down to the scale $\mu_y^2$ can be done using the two components of the dPDFs i.e. \mbox{$F_{ij}=F_{ij}^\mathrm{int}+F_{ij}^\mathrm{spl}$}. Since the splitting component is not defined for $Q^2\leq\mu_y^2$ according to the procedure prescribed by \cite{Diehl:2017kgu}, the evolution from the scale $\mu_y^2$ down to the scale $Q_0^2$ is performed using the intrinsic part only i.e. $F_{ij}=F_{ij}^\mathrm{int}$. The philosophy of the backward evolution of DPS is then the following. At the starting scale $Q_h^2$ of the evolution, the two partons of flavour $i$ and $j$ belonging to the same protons have a size of order $1/Q_h\leq y$ so one can talk about DPS. However, after an evolution downwards in $Q^2$ which leads to $Q^2=\mu_y^2$, the partons now have a size of order $1/Q=y$ and the pair $ij$ might be resolved into a single parton of flavour $k$. In the following, such a phenomenon will be referred to as ``merging'' since, from a backward-evolution point of view, it seems that the two partons of flavours $i$ and $j$ merge into a single parton of flavour $k$. The merging happens with a probability given by $p_\mathrm{Mrg}=F_{ij}^\mathrm{spl}(x_1,x_2,\y,\mu_y^2)/F_{ij}(x_1,x_2,\y,\mu_y^2)$. In the case where a merging happens, then a simple backward evolution of the single parton is performed. If no merging occurs then the two partons remain resolved as a pair and the evolution is carried on with the intrinsic part of the dPDFs only. More details about the merging procedure will be given in a dedicated section. An illustration of the backward evolution is given in \Fig \ref{FigySize}. The algorithm has the following structure:

\begin{enumerate}
\item Define two hard processes with their common scale $Q_h^2$. Those two hard processes are initiated by four partons of flavour $i$, $j$, $k$ and $l$.
\item Select the momentum fractions $x_1$, $x_2$, $x_3$, and $x_4$ of the four initial partons with \Eq (\ref{eq:FactDPSDiff}) (see the previous section). A value for $y$ is also selected within the range $b_0/Q_h<y<y_\mathrm{cut}$. 
\item \label{enu:step3} Evolve the two pairs $ik$ and $jl$ downwards in $Q^2$ from the scale $Q_h^2$ down to $\mu_y^2$ according to the branching probabilities $\dP_{ik}^\mathrm{ISR}$ and $\dP_{jl}^\mathrm{ISR}$ respectively. The two components of the dPDFs are used i.e. $F_{ik}=F_{ik}^\mathrm{int}+F_{ik}^\mathrm{spl}$. For each emission, the channel which wins is the one with the highest scale.
\item After the parton shower has been  performed (i.e. Step (\ref{enu:step3})), the scale $Q^2$ is now equal to $\mu_y^2$. Consider a proton in which a pair of partons with flavours $i'$ and $k'$ and with fractions $x'_1$ and $x'_3$ is resolved. If there exists a parton of flavour $h$ such that the branching $h\to i' +k'$ exists then take a random number $R$ uniformly distributed between 0 and 1. If $R<p_\mathrm{Mrg}=F_{i'k'}^\mathrm{spl}(x'_1,x'_3,\y,\mu_y^2)/F_{i'k'}(x'_1,x'_3,\y,\mu_y^2)$ then merge the two partons into a single one with momentum fraction $x'_1+x'_3$. After the merging, a usual backward evolution of the single parton $h$ is performed from the scale $\mu_y^2$ until the minimum scale $Q_0^2$ is reached. In the case where such a flavour $h$ does not exist or if $R>p_\mathrm{Mrg}$ then proceed with Step (\ref{enu:step5}). Do the same for the other proton.
\item \label{enu:step5} Evolve each remaining pair of partons downwards in $Q^2$ from the scale $\mu_y^2$ down to some infrared cut-off $Q_0^2\sim 1\,\mathrm{GeV}^2$. In the expression of $\dP_{ik}^\mathrm{ISR}$, only the intrinsic part of the dPDFs is now used i.e. $F_{ik}=F_{ik}^\mathrm{int}$.
\end{enumerate}

In the absence of merging, the construction of the kinematics at the end of the shower follows the exact same procedure as the one sketched in \Sec \ref{Sec:PSkin}. More precisely, the kinematics is constructed for each one of the two hard processes so that the respective invariant mass and rapidity of each subsystem are conserved. Since the two hard processes are separate, no complications appear. The kinematic constraints, imposing that the partons inside the same proton cannot carry more energy than what is available, are implemented within the evolution (recall the boundaries of the integral inside \Eq (\ref{eq:dPij})). The generation of FSR is the same as in the SPS case, since no PDFs are involved.

The main difference with the SPS case is the choice of the starting scale $Q_h^2$ for the ISR evolution. As mentioned in \Sec \ref{Sec:PSkin}, this choice depends on the types of dipoles involved and the partons initiating the hard process might have different starting scales. The problem with DPS is that the two partons extracted from the same proton must start with the same scale, since only a set of equal-scale dPDFs is available. Let us take the example of same-sign WW production. Here, one starts with two II dipoles, since the W bosons are colour singlets. If the first hard process is initiated by partons $i$ and $j$ and the second by partons $k$ and $l$, then the conditions on the starting scales in an angular-ordered shower ($Q^2=\qtis$) read

\begin{equation}
\label{eq:consStart}
\qtis_{h,i}\,\qtis_{h,j}=\hat{s}^2_A,
\hspace{80pt}
\qtis_{h,k}\,\qtis_{h,l}=\hat{s}^2_B,
\end{equation}

\noindent where $\hat{s}_A$ and $\hat{s}_B$ are the invariant masses squared of the two hard processes. The fact that the dPDFs use the same scale imposes $\qtis_{h,i}=\qtis_{h,k}$ and $\qtis_{h,j}=\qtis_{h,l}$. With the constraints above, this implies that $\hat{s}_A=\hat{s}_B$, which is not satisfied in general. Therefore, it is not possible to satisfy at the same time both of the constraints given in \Eq (\ref{eq:consStart}). The choice that is made in this instance is to set all the starting scales equal to $\min(\hat{s}_A,\hat{s}_B)$. This breaks one of the conditions of \Eq (\ref{eq:consStart}) but, in the case of WW production, one expects that $\hat{s}_A\simeq\hat{s}_B\simeq m^2_\W$ so the violation is not too large. Some complications arise when one wants to study W + 2 jets or 4-jet production. Indeed, in those cases, there is no reason why $\hat{s}_A$ should be comparable to $\hat{s}_B$. Moreover, the colour configuration might lead to some IF/FI dipoles. In these cases, the strategy which is adopted is to set the starting scales of the four incoming partons equal to a common scale $\qtis_{h}$. For the IF/FI dipoles, the starting scale for the FSR evolution of the final-state parton which is linked to the initial-state parton is then set to $\hat{t}^2/\qtis_{h}$, instead of simply $-\hat{t}$ as in the SPS case. This will ensure that, for any IF/FI dipole, the condition $\qtis_{h,i}\,\qtis_{h,j}=\hat{t}^2$ is satisfied. The choice of the common scale $\qtis_{h}$ depends on the dipole configuration. In the case where there are two II dipoles among the list of dipoles, then one should choose $\qtis_{h}=\min(\hat{s}_A,\hat{s}_B)$, as mentioned before. In contrast, if there is only one II dipole belonging, for example, to the hard process $A$, then the most reasonable choice seems to be $\qtis_{h}=\hat{s}_A$. Finally, if there are only IF/FI dipoles, then one should set $\qtis_{h}=\min(\{-\hat{t}_a\})$, with the index $a$ going through the list of IF/FI dipoles which are present. This is the most straightforward solution but more sophisticated strategies will be investigated in the future. Unfortunately, there is not really a solution for the case of two II dipoles with $\hat{s}_A$ and $\hat{s}_B$ very different. This is because a realistic description of W + 2 jets and 4-jet production via DPS requires a set of dPDFs with two different scales. Note that 4-jet production via DPS was studied in \cite{Blok:2015rka, Blok:2015afa}, since the approach of the authors was using unequal-scale dPDFs.

\subsection{Parton showering with merging}
\label{Sec:Mrg}

\subsubsection{Kinematics}

Some difficulties appear when one allows mergings to happen. Let us consider two partons $i$ and $j$ with momentum fractions $x_1$ and $x_2$ initiating two different hard processes. At the scale $Q^2=\mu_y^2$, they merge into a single parton $k$. The two partons $i$ and $j$ now get a space-like virtuality. However, the only scale which is present is the scale $\mu_y^2$ and there is some arbitrariness in how the virtualities should be related to that scale. Moreover, the construction of the kinematics is now troublesome. As explained in \Sec \ref{Sec:PSkin}, momentum conservation gets broken by the fact that the two partons that initiate the hard process obtain a transverse momentum and a space-like virtuality because of the ISR evolution. This is solved by applying a longitudinal boost on each side, so that the invariant mass squared $\hat{s}$ and the rapidity $Y$ of the system are conserved. In the case of SPS, this works since one has two degrees of freedom (two boosts) and two constraints ($\hat{s}$ and $Y$). In the case of DPS, the situation is different. For example, if a merging happens inside one beam but not within the other one, then the whole system is initiated by three partons. Therefore, one is left with only three degrees of freedom (one longitudinal boost for each parton). This is not enough to satisfy the fact that the invariant mass and the rapidity of each subsystem must be conserved, which in total gives four constraints. The situation becomes worse if two mergings happen, which then reduces the number of degrees of freedom to two. The system is thus overconstrained in the case of merging. This issue has already been mentioned and discussed for the transverse-momentum-ordered shower of \PyEight\, in \cite{Sjostrand:2004ef}. There, the merging is referred to as ``joined interaction''. The authors proposed two solutions to such a problem. The first one is to drop the statement that parton $k$ must have a momentum fraction equal to $x_1+x_2$. This removes a constraint and allows the kinematics to be established in the case of a transverse-momentum-ordered shower. Nevertheless, the momentum fractions used as arguments for the dPDFs must be adapted in order to account for such a change. The corrections are then of order $\mathcal{O}(\mu_y^2/Q_h^2)$. With the second solution, this constraint is kept but no transverse momentum is given to the virtual partons $i$ and $j$, which also allows the construction of the kinematics. Those prescriptions were proposed for a shower with a local-recoil strategy. For a global-recoil strategy as in angular-ordered showers, one needs to adapt them.

No ultimate solution has been found yet. However, the procedure presented in the following seems to give reasonable results, although there is room for improvement. At the scale $Q^2=\mu_y^2$, the evolution of the two hard processes gets frozen and the probability $p_\mathrm{Mrg}$ is evaluated for each proton. If no mergings happen, then the evolution is carried on and the construction of the kinematics is the same as in the SPS case. If at least one merging happens, then the first step is to construct at the scale $Q^2=\mu_y^2$ the individual kinematics of each hard process using the procedure described in \Sec \ref{Sec:PSkin}. This is done before implementing the mergings. The two hard processes are thus still separated. After that, the idea is to define a new hard process which absorbs the two hard processes and all the emissions that have occurred so far. This new hard process is characterised by a squared invariant mass $\hat{s}'$ and a rapidity $Y'$, which can be calculated. Before actually implementing the mergings, this new hard process is initiated by four partons: $i$ and $k$ on one side, and $j$ and $l$ on the other side. It is convenient to come back to a hard process initiated by only two partons. Therefore, one can define some pseudo-initiators with four-momenta $p_i+p_k$ and $p_j+p_l$ respectively. Those momenta are actually the ones which are assigned to the mother partons after merging. Let us take the example where partons $i$ and $k$ merge into a single parton of flavour $h$ and partons $j$ and $l$ do not merge. The new hard process is then physically initiated by the three partons $h$, $j$ and $l$. The momentum of $h$ is defined as $p_h=p_i+p_k$, which leads to $x_h=x_i+x_k$. Here, it is important to emphasise that the momentum fractions $x_i$ and $x_k$ are not exactly the same as the momentum fractions  which were generated by the shower and used to evaluate the probability $p_\mathrm{Mrg}$. These latter ones will be referred to as $\xi_i$ and $\xi_k$. The fact that $x_{i,k}\neq \xi_{i,k}$ is the price to pay to allow emissions before the merging phase and to be able to conserve the invariant mass and the rapidity of each hard system. More precisely, the momentum fractions $x_{i,k}$ and $\xi_{i,k}$ are related by the longitudinal boosts that are applied to the incoming partons before actually implementing the mergings. The longitudinal boosts have the following form

\newcommand{\widest}{\ensuremath{ccc}}
\newcommand{\newWidth}[1]{\makebox[\widthof{\widest}]{\ensuremath{#1}}}

\begin{equation}
\Lambda(\lambda)=\left(\begin{array}{cccc}\mathrm{ch}(\lambda) & 0 & 0 & \mathrm{sh}(\lambda) \\ 0 & \newWidth{1} & 0 & 0 \\ 0 & 0 & \newWidth{1} & 0 \\ \mathrm{sh}(\lambda) & 0 & 0 & \mathrm{ch}(\lambda) \end{array}\right),
\end{equation}

\noindent with 

\begin{equation}
\mathrm{ch}(\lambda) = \frac{\lambda^2+1}{2\lambda}, 
\hspace{80pt}
\mathrm{sh}(\lambda) = \frac{\lambda^2-1}{2\lambda}.
\end{equation}

\noindent In practice, $\lambda\simeq 1$ since the shower should not alter too much the initial kinematics of the hard systems. Before applying the boosts, partons $i$ and $k$ have momenta $(\sqrt{s}/2)\,\xi_{i,k}(1;0,0,1)$ in the laboratory frame. After applying the longitudinal boosts, the two momenta are $p_{i,k}=(\sqrt{s}/2)\,\lambda_{i,k}\,\xi_{i,k}(1;0,0,1)$. Thus, after implementing the merging, parton $h$ has a momentum fraction given by

\begin{equation}
x_h=x_i+x_k=\lambda_i\,\xi_i+\lambda_k\,\xi_k\simeq \xi_i+\xi_k.
\end{equation}

In this work, no transverse momentum is given to partons $i$ and $k$. However, a variant of this procedure that generates a transverse momentum for partons $i$ and $k$ will be investigated in future works. Since $p_i$ and $p_k$ are light-like momenta along the beam pipe and pointing in the same direction, their sum $p_h$ is necessarily a light-like momentum along the beam pipe too. Adding emissions to parton $h$ can thus only turn its light-like momentum into a space-like momentum. For the calculations, a pseudo-initiator $\tilde{g}$ with momentum $p_{\tilde{g}}=p_j+p_l$ is defined. This pseudo-initiator is simply a mathematical tool and is not physically implemented. The new hard process thus has a squared invariant mass and a rapidity given by 

\begin{equation}
\hat{s}'=(p_h+p_{\tilde{g}})^2=(x_i+x_k)(x_j+x_l)s,
\hspace{50pt}
Y'=\frac{1}{2}\ln\left(\frac{x_i+x_k}{x_j+x_l}\right).
\end{equation}

\begin{figure}[t!] 
\centering
\includegraphics[width=0.45\textwidth]{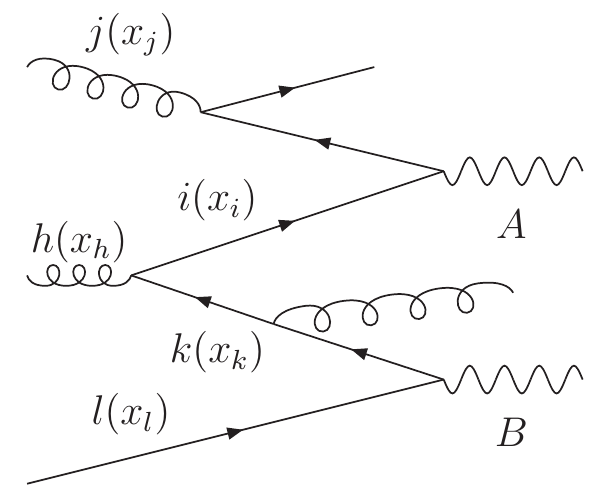}
\includegraphics[width=0.5\textwidth]{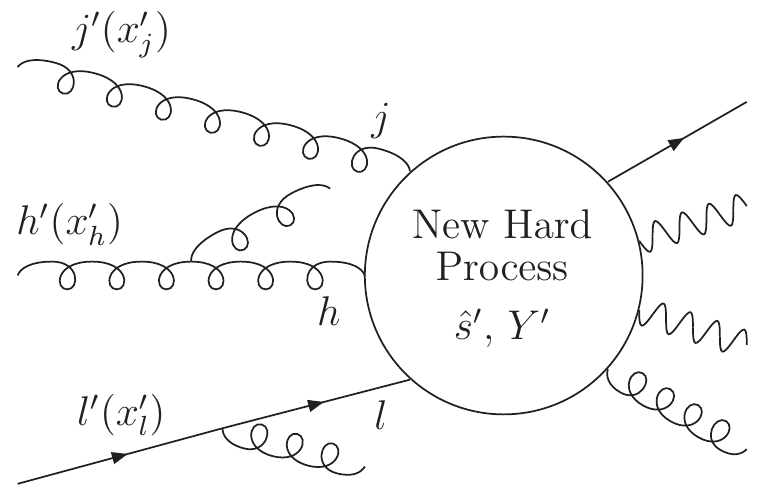} \\
(a) \hspace{200pt} (b)
\caption{Sketch of the merging phase. In (a), $Q^2=\mu_y^2$ and partons $i$ and $k$ merge into parton $h$. A new hard process initiated by partons $h$, $j$ and $l$ is defined. In (b), $Q^2=Q_0^2$ and the incoming partons extracted from the beams are now $h'$, $j'$ and $l'$.}
\label{Fig:Mrg}
\end{figure}

\noindent An illustration of this example is given in \Fig \ref{Fig:Mrg}. After this has been done, the evolution is carried on from the scale $\mu_y^2$ down to $Q_0^2$. In this example, a simple backward evolution of parton $h$ is performed, whereas the evolution of the pair $jl$ is carried on as described in \Sec \ref{Sec:dShowerSh}. At the end of the shower (i.e. $Q^2=Q_0^2$), the three partons that are extracted from the beams are $h'$, $j'$ and $l'$. During the evolution, subsequent emissions were attached to the two partons $h$ and $\tilde{g}$ that initiate the new hard process defined previously. Because of these emissions, partons $h$ and $\tilde{g}$ are now virtual particles and their momenta are not the light-like momenta $p_h$ and $p_{\tilde{g}}$ anymore. Instead, partons $h$ and $\tilde{g}$ have obtained a transverse momentum and a space-like virtuality, which break momentum conservation. In order to recover momentum conservation, one can now use the same strategy as the one used in the usual SPS case. In particular, the kinematics can be constructed. One has two degrees of freedom (a rescaling factor for parton $h$ and one for parton $\tilde{g}$) and two constraints ($\hat{s}'$ and $Y'$). The kinematics is thus constructed so that the invariant mass and the rapidity of the new hard process are conserved. The rescaling factor for parton $h$ defines the longitudinal boost that must be applied to parton $h'$ and its shower, whereas the one for the pseudo-initiator $\tilde{g}$ gives the longitudinal boost which is applied to both partons $j'$ and $l'$, as well as their respective shower.

Regarding the evolution after the mergings, one needs to redefine the dipoles since the colour flow might have been modified, as it will be seen in the next section. This means that new starting scales must be assigned to each parton. The strategy is similar to the one proposed in \Sec \ref{Sec:dShowerSh}. All the partons initiating the new hard process start with the initial scale $\mu_y^2$, regardless of whether they belong to an II or an IF/FI dipole. In the case of an IF/FI dipole, this means that the final-state parton linked to the initial-state parton starts with a scale $\min(\mu_y^2,\hat{t}^2/\mu_y^2)$. Here, it is made sure that the maximum starting scale for any parton is $\mu_y^2$, since the evolution stopped at this scale. With this prescription, each parton gets a unique chance to radiate within an interval of scales. This should avoid double-counting issues. Indeed, if the final-state parton were starting its evolution with the scale $\hat{t}^2/\mu_y^2$, then, in the case where $\hat{t}^2/\mu_y^2>\mu_y^2$, the parton could radiate again in the interval $[\mu_y^2,\hat{t}^2/\mu_y^2]$ despite this possibility already being covered during the first evolution down to $\mu_y^2$, before the merging phase. For the same reason, the starting scale for a parton belonging to a final-final (FF) dipole is set to $\min(\mu_y^2,m_\mathrm{dip}^2)$, where $m_\mathrm{dip}$ is the dipole mass. 

Although the strategy mentioned above seems to remove the double-counting issues, it is not clear whether it produces under-counting issues or not. To answer this question, one would need a full matching between the emission patterns of the system before and after the merging phase, which is beyond the scope of this work. In the case of $\W^+\W^+$ production, it will be seen later\footnote{See \Sec \ref{Sec:setup}.} that the cross section is dominated by the intrinsic $\times$ intrinsic part, which leads to a sampling of the variable $y$ biased towards the large values. Therefore, $\mu_y$ is usually close to the infrared cut-off $Q_0$ of the shower, leaving little room for extra emissions after the merging phase. In contrast, the merging phase may happen much earlier during the shower evolution for other processes such as $\Z^0\Z^0$ production. For these processes, further work is needed regarding the choice of the starting scales after the merging phase.

The kinematics for FSR also needs to be discussed in the case of merging. After the mergings have happened, the new dipole configuration might generate further FSR. Those emissions come either from the decay products of the two hard subsystems or from the final-state partons generated previously by ISR. When those particles radiate, they obtain a virtuality, which breaks momentum conservation. As described in \Sec \ref{Sec:PSkin}, momentum conservation is recovered by boosting the resulting jets along the direction of their respective progenitor. Here, the invariant mass which is conserved is $\sqrt{\hat{s}'}$. Unfortunately, this procedure may in general alter the individual kinematics of each hard system. In particular, there is no guarantee that the invariant mass of each hard system will be preserved, since it is the invariant mass of the new hard process which is conserved instead. This might be an issue since one expects the invariant masses to be distributed according to the cross section formula. Nevertheless, the situation becomes better in the case where the decay products of the hard systems are colour singlets. In this specific case no QCD radiation is attached to those particles so they do not get any virtuality. Thus, the invariant mass can be preserved. For instance, in the case of WW pair production, each W boson may decay into a pair of leptons. In the absence of an electroweak shower, no extra emissions are attached to those leptons. However, after merging, those leptons must be considered as FSR progenitors, even if they do not radiate. This is because they must balance the momenta of the jets in the equations that state global momentum conservation. The crucial point here is to consider the sum of the two leptons as a single FSR progenitor, and not each lepton on its own. It is this sum which is then used within the equations. The same boost is thus applied to both leptons. This ensures that the invariant mass of the lepton pair (which is the W mass in our instance) is preserved. A similar strategy unfortunately does not work in the case where the decay products are colour charged (e.g. $\Z^0$ boson decaying into jets), since their virtuality will be modified due to additional QCD radiation.

\subsubsection{Colour flow}

\begin{figure}[t!] 
\centering
\includegraphics[width=0.45\textwidth]{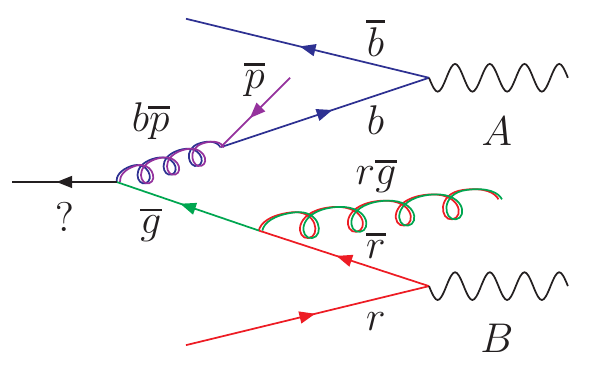}
\includegraphics[width=0.45\textwidth]{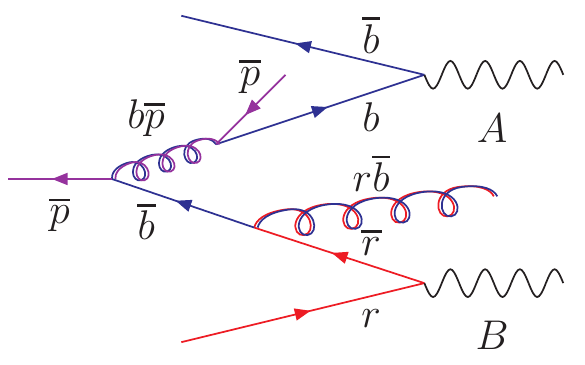} \\
(a) \hspace{200pt} (b)
\caption{Treatment of the colour flow in the case of merging: (a) colour flow before correction, (b) colour flow after correction. Here, an infinite number of colours are available (leading-colour approximation) so the label $p$ stands for the new colour purple.}
\label{Fig:Col}
\end{figure}

In the case of merging, the colour flow needs to be corrected. This is due to the fact that ISR is performed as a backward evolution. Under the leading-colour approximation \cite{tHooft:1973alw}, which is the framework of current parton showers \cite{Buckley:2011ms}, each time a new parton is emitted, a new colour is generated. Thus, the colours of the partons that are meant to merge do not match. This would mean that the merging cannot occur, at least from a colour point of view. Therefore, the colours need to be matched. The main idea is illustrated in \Fig \ref{Fig:Col}. In this example, the backward evolution leads to a green antiquark which should be merged with a blue-purple gluon. This implies that the colours blue and green should be set to be equal. The strategy is to change the most recent colour, which is green in this instance. This aims to disturb the colour flow as little as possible. In the case of a double merging, the colour flow needs to be corrected twice. In this case, one needs to make sure that a colour is not modified more than once. Otherwise, this might lead to some final-state gluon with a colour equal to its anticolour (referred sometimes as ``singlet gluons''), which should not be included under the leading-colour approximation. If such a situation appears, despite the precautions implemented, then the configuration must be vetoed, since such a gluon would cause problems during the hadronisation phase.

The treatment of colour flow in the case of merging is not fundamental in the context of a parton-level simulation. However, the modification of the colour flow due to a merging might affect the hadronisation phase significantly \cite{Buckley:2011ms}. This is the reason why this issue has been addressed in this work.

\subsection{Parton showering with different scales}
\label{Sec:unequal}

As mentioned in \Sec \ref{Sec:dShowerSh}, a realistic description of DPS with hard processes characterised by two different scales requires unequal-scale dPDFs. Such a set is not available yet. However, one can already extend the algorithm which has been proposed previously. The evolution of the dPDFs with two different factorisation scales has been discussed for instance in \cite{Gaunt:2009re, Ceccopieri:2010kg}. 

Let us consider two hard processes characterised by the scales $\mu_A^2$ and $\mu_B^2$ and initiated by four partons: $i(x_1)$ and $k(x_3)$ on one side, and $j(x_2)$ and $l(x_4)$ on the other side (recall \Fig \ref{FigFactDPS}). The kinematics of the two hard processes can be selected using \mbox{\Eq (\ref{eq:FactDPS})}, with the unequal-scale dPDFs. The pairs $ik$ and $jl$ need to be evolved. The starting scales are now allowed to be different. Let us assume that $\qtis_{h,k}<\qtis_{h,i}$. The strategy is to evolve parton $i$ from the scale $\qtis_{h,i}$ down to $\qtis_{h,k}$ by using the following branching probability

\begin{equation}
\dP_{i/k}= \frac{\d \qtis_i}{\qtis_i}\sum_{i'}\int_{x_1}^{1-x_3}\frac{\d x'_1}{x'_1}\,\frac{\as(\pTs)}{2\pi}\,P_{i'\to i}\left(\frac{x_1}{x'_1}\right)\,\frac{F_{i'k}(x'_1,x_3,\y,\qtis_i,\qtis_{h,k})}{F_{ik}(x_1,x_3,\y,\qtis_i,\qtis_{h,k})},
\end{equation}

\noindent which needs to be corrected by the appropriate Sudakov factor, as usual. This is nothing but the usual DGLAP equations, except that the sPDFs are replaced by the unequal-scale dPDFs and the upper boundary of the integral is now $1-x_3$ instead of unity. Thus, the evolution affects parton $i$ only and parton $k$ acts simply as a spectator. Once the scale $\qtis_i$ has been brought to the scale $\qtis_{h,k}$, the scales are now equal. The evolution can then be carried on using the equal-scale dPDFs and the procedure described in \Sec \ref{Sec:dShowerSh}.

\subsection{Defining valence and sea components for the dPDFs}
\label{Sec:Scheme}

In event generators, one wants to be able to decompose the dPDFs involving u and d quarks into valence and sea components which have a literal interpretation as probabilities to find valence and sea quarks inside the proton. This is particularly important for some hadronisation models. For example, a bookkeeping of valence and sea quarks is required in order to establish the structure of the beam remnants in \PyEight\, \cite{Sjostrand:2004pf}. Moreover, such a separation allows us to enforce some physical requirements inside the shower evolution of a pair of partons. For instance, the fact that the dPDF $\dv\dv$ is identically zero at all scales ensures that the pair of partons cannot be evolved back to a pair of valence d quarks. This latter property is the reason why the valence and sea components are separated in the shower evolution\footnote{In the shower evolution, $\u_{\mathrm{v}}$ and $\u_{\mathrm{s}}$ are treated like two different partons, with their own evolution equation.} described in \Sec \ref{Sec:dShowerSh}. Note that such a separation is not strictly necessary for the parton-level simulation that is presented here. However, we intend ultimately to incorporate this model into an existing event generator with a hadronisation model, and so make this valence-sea separation with this in mind.

The conventional definition of the valence-valence distribution is

\begin{equation}
\uv\uv = uu -u\ubar - \ubar u +\ubar\ubar.
\label{eq:convSch}
\end{equation}

\noindent However, such a definition leads to negative values. This is an issue since one wants to be able to interpret $\uv\uv$ as a probability density, which therefore must be positive-definite. Moreover, applying the conventional scheme given by \Eq (\ref{eq:convSch}) assigns non-zero values to the splitting part $[\uv\uv]_\mathrm{spl}$ of the dPDF whereas one would expect $[\uv\uv]_\mathrm{spl}$ to be identically zero at all scales. Indeed, this parton configuration does not involve any sea quark so no terms due to a $1\to 2$ perturbative splitting can contribute to the dPDF $\uv\uv$. Therefore, there is a need to define a new scheme. This scheme should satisfy the following properties:

\begin{enumerate}
\item \label{enu:prop1} The valence and sea components must be positive-definite.
\item \label{enu:prop2} They must sum up to the full dPDFs. For example, one should be able to write
\begin{equation}
uu = \uv\uv + \us\uv + \uv\us + \us\us,
\hspace{30pt}
ud = \uv\dv + \us\dv + \uv\ds + \us\ds,
\end{equation}
\noindent and so forth.
\item \label{enu:prop3} They must satisfy the intuitively expected dDGLAP evolution equations. For example, in the case of the $\uv\ds$ distribution, the associated evolution equation should not contain any term that involves the distribution $g\ds$. This is because the u quark is a valence quark so it cannot come from the branching of a gluon at a lower scale.
\item \label{enu:prop4} The valence-valence distributions must satisfy the intuitively expected number sum rules. More precisely, the valence-valence distribution $q_{\mathrm{v}}q_{\mathrm{v}}$ must satisfy the relation
\begin{equation}
\int_0^{1-x_2}\left(\int\d^2\y\,\Phi(y\nu)\,q_{\mathrm{v}}q_{\mathrm{v}}(x_1,x_2,\y,\mu^2)\right)\d x_1=(N_{\mathrm{q}_{\mathrm{v}}}-1)\,q_{\mathrm{v}}(x_2,\mu^2),
\end{equation}
\noindent with $N_{\mathrm{q}_{\mathrm{v}}}$ the number of valence quarks.
\end{enumerate}

\noindent Such a scheme can be derived. The valence-valence, valence-sea and sea-sea components are defined as follows:

\begin{subequations}
\begin{align}
[\uv\uv]_\mathrm{int}&= [uu  - \ubar u  - u\ubar + \ubar\ubar]_\mathrm{int}, \\
[\uv\us]_\mathrm{int}&=[\uv\ubar]_\mathrm{int}=[u\ubar - \ubar\ubar]_\mathrm{int}, \\
[\us\us]_\mathrm{int}&=[\us\ubar]_\mathrm{int} =[\ubar\us]_\mathrm{int} =[\ubar\ubar]_\mathrm{int},  
\end{align}
\end{subequations}

\noindent for the intrinsic part, and

\begin{subequations}
\begin{align}
[\uv\uv]_\mathrm{spl}&= 0, \\
[\uv\us]_\mathrm{spl}&=[\uv\ubar]_\mathrm{spl}=\frac{1}{2}[u\ubar-\ubar u-\ubar\ubar+uu]_\mathrm{spl}, \\
[\us\ubar]_\mathrm{spl}&=\frac{1}{2}[u\ubar+\ubar u+\ubar\ubar-uu]_\mathrm{spl}, \\
[\us\us]_\mathrm{spl}&=[\ubar\ubar]_\mathrm{spl},
\end{align}
\end{subequations}

\noindent for the splitting part. The scheme for the intrinsic part is the conventional one given by \Eq (\ref{eq:convSch}). However, the scheme for the splitting part is more complicated. The idea is to set $[\uv\uv]_\mathrm{spl}$ to the desired value i.e. zero for all scales. More generally, the splitting part of any valence-valence component must be set to zero for all scales since these components do not get any contributions from $1\to2$ perturbative splittings. The other components are then constructed so that \Props (\ref{enu:prop2}) and (\ref{enu:prop3}) are satisfied. Also, the correct initial conditions must be applied. For example, the $[\us\ubar]_\mathrm{spl}$ distribution must be initialised by the $1\to2$ splitting term since a gluon is allowed to split into a $\u_\mathrm{s}\bar{\u}$ pair. In contrast, the distributions $[\us\us]_\mathrm{spl}$ and $[\ubar\ubar]_\mathrm{spl}$ are initialised to zero since these configurations cannot originate from the splitting of a gluon. These initial conditions are guaranteed by the way the distributions are defined. This has the important consequence of breaking the symmetry between the distributions $\us\us$, $\ubar\ubar$ and $\us\ubar$. Note that this symmetry is maintained for the intrinsic part, as long as it is satisfied at the initial scale of the evolution. The same scheme is used for the d sector. There is no need to define such a scheme for the s, c and b sectors since the proton does not contain any s, c or b valence quark. However, one needs to take care of the distributions that couple u and d. In particular, the constraint $[\uv\dv]_\mathrm{spl} = 0$ must be imposed since $\uv\dv$ is a valence-valence component. The scheme for the intrinsic part is the conventional one. For the splitting part, the following one will be used:

\begin{subequations}
\begin{align}
[\uv\dv]_\mathrm{spl}&= 0, \\
\label{eq:uvds}
[\uv\ds]_\mathrm{spl}&=\frac{1}{2}[u\dbar-\ubar d-\ubar\dbar+ud]_\mathrm{spl}, \\
[\us\dv]_\mathrm{spl}&=\frac{1}{2}[\ubar d-u \dbar-\ubar\dbar+ud]_\mathrm{spl}, \\
[\uv\dbar]_\mathrm{spl}&=[u\dbar - \ubar\dbar]_\mathrm{spl}, \\
[\ubar\dv]_\mathrm{spl}&=[\ubar d - \ubar\dbar]_\mathrm{spl}, \\
[\us\dbar]_\mathrm{spl}&=[\us\ds]_\mathrm{spl} = [\ubar\ds]_\mathrm{spl} =  [\ubar\dbar]_\mathrm{spl}. 
\end{align}
\end{subequations}

\noindent The last configurations which have not been specified yet can be defined according to the conventional scheme for both the intrinsic and the splitting parts. For example, one can write $\uv g= ug - \ubar g$ and $\us g = \ubar g$ for a configuration involving a u quark and a gluon.

It can be checked using the equations above that the scheme satisfies \Prop (\ref{enu:prop2}). The fact that the distribution $\uv\ds$ verifies \Prop (\ref{enu:prop3}) is explicitly shown in \App \ref{app:evol}. Similar steps to those given in \App \ref{app:evol} can be used to show that the other distributions also satisfy \Prop (\ref{enu:prop3}); for brevity, we do not present these explicitly here. \Prop (\ref{enu:prop4}) is strongly dependent on the inputs used for the dPDFs at the initial scale of the evolution (recall \Eqs (\ref{eq:StartInt}) and (\ref{eq:StartSpl})). In particular, the DGS set of $\y$-dependent dPDFs generated in \cite{Diehl:2017kgu} does not satisfy \Prop (\ref{enu:prop4}). In order to approximately recover the number sum rules in the valence-valence sector, the initial conditions of the DGS set were modified. The modifications which were made as well as their impact on the sum rules are presented in \App \ref{app:inputs}. Note that no specific modifications were made to improve the way the sum rules are verified in the other sectors.

A general argument that \Prop (\ref{enu:prop1}) holds can be given following a similar logic as the one presented in \Sec 5.3 of \cite{Diehl:2013mla}. The distributions defined in the scheme  satisfy the expected evolution equations (see \Prop (\ref{enu:prop3})) and are all initialised with a value which is positive (or zero). The evolution equations state that the derivative of a dPDF with respect to the scale is equal to the convolution of some dPDFs with the regularised splitting kernels. If the splitting kernels are positive-definite, then the derivative will start with a positive value, which means that the dPDF will increase with the scale and therefore remain positive at higher values of the scale. At LO, the splitting kernels can be divided into a positive-definite part and a negative part\footnote{This negative part is contained inside the plus-prescription.} which is the virtual contribution to the kernel at $z=1$. From the structure of the DGLAP equations, one can notice that the negative contribution to the evolution of a dPDF is proportional to the dPDF itself. Therefore, the negative contribution cannot change the sign of the dPDF, which thus remains positive.

\section{Results}
\label{Sec:results}

\subsection{Setups of the simulations}
\label{Sec:setup}

The simulation introduced in the previous section has been used to generate parton-level events for $\W^+\W^+$ pair production via DPS only. The results will be compared with \mbox{\PyEight} and with \Herwig. The two hard processes are $\u\bar{\d}\to\W^+\to\mathrm{e}^+\nu_\mathrm{e}$ and \mbox{$\u\bar{\d}\to\W^+\to\mu^+\nu_\mu$}.\footnote{The symmetry factor is therefore unity here.} For all simulations, the factorisation scale and the argument of the couplings used in the cross-section calculations are set to be equal to $\min(\sqrt{\hat{s}_A},\sqrt{\hat{s}_B})$. The produced leptons are constrained to have a transverse momentum $\hat{p}_\perp> 20$ GeV and a rapidity $|\eta|<5$ in the centre-of-mass frame of the collision. It is assumed that the neutrinos can be exactly reconstructed. The set of LO sPDFs that is used is the 3-flavour MSTW2008 one \cite{Martin:2009iq, Martin:2010db}. In order to be consistent with the sPDFs, the scheme for the strong coupling $\as$ will be the 3-flavour scheme developed by the same authors \cite{Martin:2009bu}. With this scheme, the value of $\as$ at the $\Z^0$ mass is $\as(m_\mathrm{Z})=0.126$. For the production of the W bosons, only the channel $\u\bar{\d}\to\W^+$ will be considered. However, the strange quark is included within the shower, as well as the u and d quarks. The shower cut-off of the simulation for both ISR and FSR is set to \mbox{$p_{\perp,\mathrm{0}}=1$ GeV} (which means that  $\tilde{q} > 2$ GeV). For \PyEight\, and \Herwig, the hadronisation phase, MPI and matrix-element corrections are switched off. 

In the plots presented in the next section, the names refer to the following setups. The curves named ``Pythia'' and ``Herwig'' refer to the results obtained with \textsc{Pythia 8.2.40} and \textsf{Herwig 7.1.4}  respectively. The implementation of the algorithm described in \Sec \ref{Sec:dShower} will be referred to as ``dShower''. This simulation uses the set of $\y$-dependent dPDFs generated in \cite{Diehl:2017kgu}, with the scheme defined in \Sec \ref{Sec:Scheme}. The setup ``dSh-NoSpl'' refers to the same setup as ``dShower'', but with the splitting part of the dPDFs set equal to zero i.e. $F=F_\mathrm{int}$. This implies that the possiblity for merging is switched off (i.e. $p_\mathrm{Mrg}=0$). A comparison between those two setups will allow us to estimate the size of the contribution of the splitting part of the dPDFs. In order to measure the effect of the dPDFs $\y$-dependence,\footnote{And not the cumulative effect of the dPDFs in addition to the parton shower.} two other setups are defined. The idea is to use dPDFs that do not depend on $\y$, unlike the DGS set. In these setups, the $\y$-dependence is removed using a factorisation ansatz for the dPDFs. More precisely, the $\y$-dependent part of the dPDFs is factorised into a function $F(\y)$. The first setup, ``Fact'', uses a product of sPDFs as in \Eq (\ref{eq:AnsatzProd}). For this setup, a usual angular-ordered shower is added. The second setup, ``GS09'', instead employs the GS09 set (limited to three flavours) developed in \cite{Gaunt:2009re}. For this last setup, no shower is added, since this would require dedicated work.\footnote{The GS09 set includes contributions from $1\to2$ splittings and the shower needs to be consistent with this aspect.}

For the setups that are based on a factorisation ansatz, the value of the effective cross section $\sigma_\mathrm{eff}$ needs to be consistent with the inputs for the dPDFs given in \Eqs (\ref{eq:StartInt}) and (\ref{eq:StartSpl}). More specifically, the value of the effective cross section is directly linked to the values of the widths $h_{ij}$ which are given in \Eq (\ref{eq:hVal}). In the case of $\W^+\W^+$ production, the main contribution comes from the width $h_{\u\bar{\d}}$ of the $u\bar{d}$ distribution. One can now find out which $\sigma_\mathrm{eff}$ the value of $h_{\u\bar{\d}}$ corresponds to. The effective cross section is given by \Eq (\ref{eq:sigmaEff}). For the function $F(\y)$, one can use the same Gaussian form factor as the one used in \Eqs (\ref{eq:StartInt}) and (\ref{eq:StartSpl}). Note that this construction does not take into account the fact that the effective width of the $u\bar{d}$ distribution will gradually change during the evolution due to the mixing of this distribution with $ug$, $g\bar{d}$ and $gg$ which have different widths. The value obtained for $\sigma_\mathrm{eff}$ will thus be a rough estimate. The function $F(\y)$ can be decomposed into intrinsic and splitting parts, as for the dPDFs. Therefore, the effective cross section gets contributions from intrinsic $\times$ intrinsic, intrinsic $\times$ splitting and splitting $\times$ splitting pieces which are referred to in the literature as 2v2, 2v1 and 1v1 contributions respectively. In the case of $\W^+\W^+$ production, the 2v2 contribution is the dominant one since one needs at least two pertubative splittings to reach the configuration $\u\bar{\d}$ (e.g. $\u\to\u\g$ followed by $\g\to\d\bar{\d}$) \cite{Diehl:2017kgu}. The contributions involving splitting parts will then be neglected for this calculation. In this context, the effective cross section can be approximated to be 

\begin{equation}
\label{eq:hud}
\sigma_\mathrm{eff}^{-1}\simeq\int F^2_\mathrm{int} (\y)\,\d^2\y=\frac{\pi}{(4\pi h_{\u\bar{\d}})^2}\int_0^{+\infty} \exp\left(-\frac{2y^2}{4 h_{\u\bar{\d}}}\right) \d y^2=\frac{1}{8\pi h_{\u\bar{\d}}}.
\end{equation} 

\noindent This leads to $\sigma_\mathrm{eff}\simeq8\pi h_{\u\bar{\d}}= 69.09$ mb, which is the value that will be used in the following. This is larger than the values measured by CDF and D0. However, the value of $h_{\u\bar{\d}}$ which has been used corresponds to fits to data for GPDs \cite{Diehl:2004cx}.

\Eq (\ref{eq:hud}) can be used to justify the value of the cut-off $y_\mathrm{cut}= 8\,\mathrm{GeV}^{-1}$ defined in \Sec \ref{Sec:2hard}. Indeed, if one limits the integral in \Eq (\ref{eq:hud}) to the region defined by $0<y < y _\mathrm{cut}$, then one gets

\begin{equation}
\frac{\pi}{(4\pi h_{\u\bar{\d}})^2}\int_0^{y_\mathrm{cut}^2} \exp\left(-\frac{2y^2}{4 h_{\u\bar{\d}}}\right) \d y^2= \frac{1}{8\pi h_{\u\bar{\d}}}\,\left(1-\exp\left(-\frac{2y_\mathrm{cut}^2}{4 h_{\u\bar{\d}}}\right)\right)\simeq0.989\times\frac{1}{8\pi h_{\u\bar{\d}}}.
\end{equation}

\noindent Thus, in the case of WW pair production, $99\%$ of the full integral to infinity is taken into account with this value of the cut-off. One may worry about the 2v1 and 1v1 contributions. However, these contributions get at least one factor $1/y^2$ from the splitting part, which makes the integral converge faster. The approximation $y< y_\mathrm{cut}=8\,\mathrm{GeV}^{-1}$ is therefore valid in the context of WW pair production.

\subsection{Total DPS cross section and partonic luminosity}

The first observable which will be studied is the total DPS cross section. The contribution from SPS is here omitted. For the setups using the $\y$-dependent dPDFs, this means that the cross section is calculated using \Eq (\ref{eq:FactDPS}) only. The subtracting term $\sigma_\mathrm{sub}$ and the SPS cross section $\sigma_\mathrm{SPS}$ introduced in \Eq (\ref{eq:sub}) are thus neglected. In the case of same-sign WW pair production, $\sigma_\mathrm{sub}$ is irrelevant as it has been numerically shown in \cite{Diehl:2017kgu}. This is again because there is no direct LO splitting that leads to the configuration $\u\bar{\d}$. The fact that $\sigma_\mathrm{sub}$  can be neglected also implies that the dependence of $\sigma_\mathrm{DPS}$ on the unphysical scale $\nu$ should be negligible.\footnote{See \App \ref{app:validation}.} In contrast, $\sigma_\mathrm{SPS}$ may not be negligible. The results are presented in \Tab \ref{table:DPSxs}. 

\begin{table}[h!]
\centering
\begin{tabular}{|c|c|c|} 
\hline \multirow{3}{*}{Setup} 
& \multirow{3}{*}{\begin{tabular}{c} $\sqrt{s}=7$ TeV \end{tabular}} 
& \multirow{3}{*}{\begin{tabular}{c} $\sqrt{s}=14$ TeV \end{tabular}}
\\ & &  \\ & &  \\ \hline 
\multirow{3}{*}{\begin{tabular}{c} dShower \end{tabular}} &
\multirow{3}{*}{$0.170\pm0.002$} &
\multirow{3}{*}{$0.718\pm0.007$} 
\\ & &  \\ & &  \\ \hline 
\multirow{3}{*}{\begin{tabular}{c} dSh-NoSpl \end{tabular}} &
\multirow{3}{*}{$0.102\pm0.001$} &
\multirow{3}{*}{$0.451\pm0.004$} 
\\ & &  \\ & &  \\ \hline 
\multirow{3}{*}{\begin{tabular}{c} Fact \end{tabular}} &
\multirow{3}{*}{$0.1571\pm0.0001$} &
\multirow{3}{*}{$0.6558\pm0.0006$} 
\\ & &  \\ & &  \\ \hline 
\multirow{3}{*}{\begin{tabular}{c} GS09 \end{tabular}} &
\multirow{3}{*}{$0.1364\pm0.0001$} &
\multirow{3}{*}{$0.6001\pm0.0005$} 
\\ & &  \\ & &  \\ \hline 
\multirow{3}{*}{\begin{tabular}{c} \textsc{\PyEight} \end{tabular}} &
\multirow{3}{*}{$0.1349\pm0.0004$} &
\multirow{3}{*}{$0.584\pm0.002$} 
\\ & &  \\ & &  \\ \hline 
\multirow{3}{*}{\begin{tabular}{c} DPS pocket formula \end{tabular}} &
\multirow{3}{*}{$0.1585\pm0.0004$} &
\multirow{3}{*}{$0.660\pm0.002$} 
\\ & &  \\ & &  \\ \hline 
\end{tabular}
\caption{Total DPS cross section in femtobarns [fb] for different setups and for different centre-of-mass energies $\sqrt{s}$. The statistical error is given.}
\label{table:DPSxs}
\end{table}

The first aspect which should be noticed is that the GS09 set, \PyEight\, and the setup dSh-NoSpl all lead to a smaller cross section than the one which could be predicted from the DPS pocket formula. This is due to the fact that these three setups include number effects and a suppression of the dPDFs near the kinematic limit. More precisely, GS09 and dSh-NoSpl use phase-space factors whereas \PyEight\, uses a rescaling of the PDFs, see \Sec \ref{Sec:currentMPI}. Both number effects and the kinematic suppressions result in a decrease of the cross section. The observation made for the GS09 set is consistent with the results presented in \cite{Gaunt:2010pi}. One may wonder why the setup Fact gives a slightly lower cross section than the one obtained with \Eq (\ref{eq:pocket}). This is because the DPS pocket formula uses the total cross section for single $\W^+$ production, which does not include the kinematic constraint $x_1+x_2\leq 1$. This constraint is in contrast implemented for the setup Fact.

The dShower algorithm gives a higher cross section than that predicted with the DPS pocket formula. This is because the splitting part of the dPDFs is now included and the 2v1 and 1v1 terms enhance the cross section. The GS09 also includes the 2v1 and 1v1 contributions, albeit not correctly taking into account the $\y$-dependence. However, the cross section for GS09 remains smaller than the one found with \Eq (\ref{eq:pocket}). The main difference is that the 2v1 and 1v1 terms get geometrical enhancements when one uses \mbox{$\y$-dependent} dPDFs \cite{Blok:2011bu, Gaunt:2012dd}. This explains why the cross section found for dShower is higher than the one obtained with the GS09 set.

In order to better understand the differences between all the setups, one needs to consider a less inclusive quantity than the total cross section. Here, the partonic luminosity will be used. It is closely related to the total cross section, but offers the possibility to study a PDF set in different regions of phase space. In the case of DPS, it is defined as

\begin{equation}
\label{eq:lumi}
\mathcal{L} = \sum_{i,j,k,l}\int\d^2\y\,\Phi^2(y\nu)\,F_{ik}(x_1,x_3,\y,\mu^2)\,F_{jl}(x_2,x_4,\y,\mu^2),
\end{equation}

\noindent where here we use $\mu=\nu=m_\W$, $x_1=x_2=m_\W/\sqrt{s}$ and $x_{3,4}=(m_\W/\sqrt{s})\exp(\pm Y)$ i.e. one hard system is produced at zero rapidity and the other one is at $Y$. This expression will be used for the setups dShower and dSh-NoSpl. If the setups involve dPDFs $f_{ij}$ that do not depend on $\y$ (e.g. GS09 and \PyEight), then one can use \Eq (\ref{eq:sigmaEff}) to simplify the expression of the luminosity:

\begin{equation}
\label{eq:lumiNoy}
\mathcal{L} \simeq \frac{1}{\sigma_\mathrm{eff}}\sum_{i,j,k,l} f_{ik}(x_1,x_3,\mu^2)\,f_{jl}(x_2,x_4,\mu^2).
\end{equation}

\noindent If one now uses the ansatz given by \Eq (\ref{eq:AnsatzProd}) (e.g. for the setup Fact), then one gets the following formula

\begin{equation}
\label{eq:lumiFact}
\mathcal{L} \simeq \frac{1}{\sigma_\mathrm{eff}}\sum_{i,j,k,l} f_{i}(x_1,\mu^2)\,f_{k}(x_3,\mu^2)\,f_{j}(x_2,\mu^2)\,f_{l}(x_4,\mu^2). 
\end{equation}

\begin{figure}[t!] 
\centering
\includegraphics[width=0.95\textwidth]{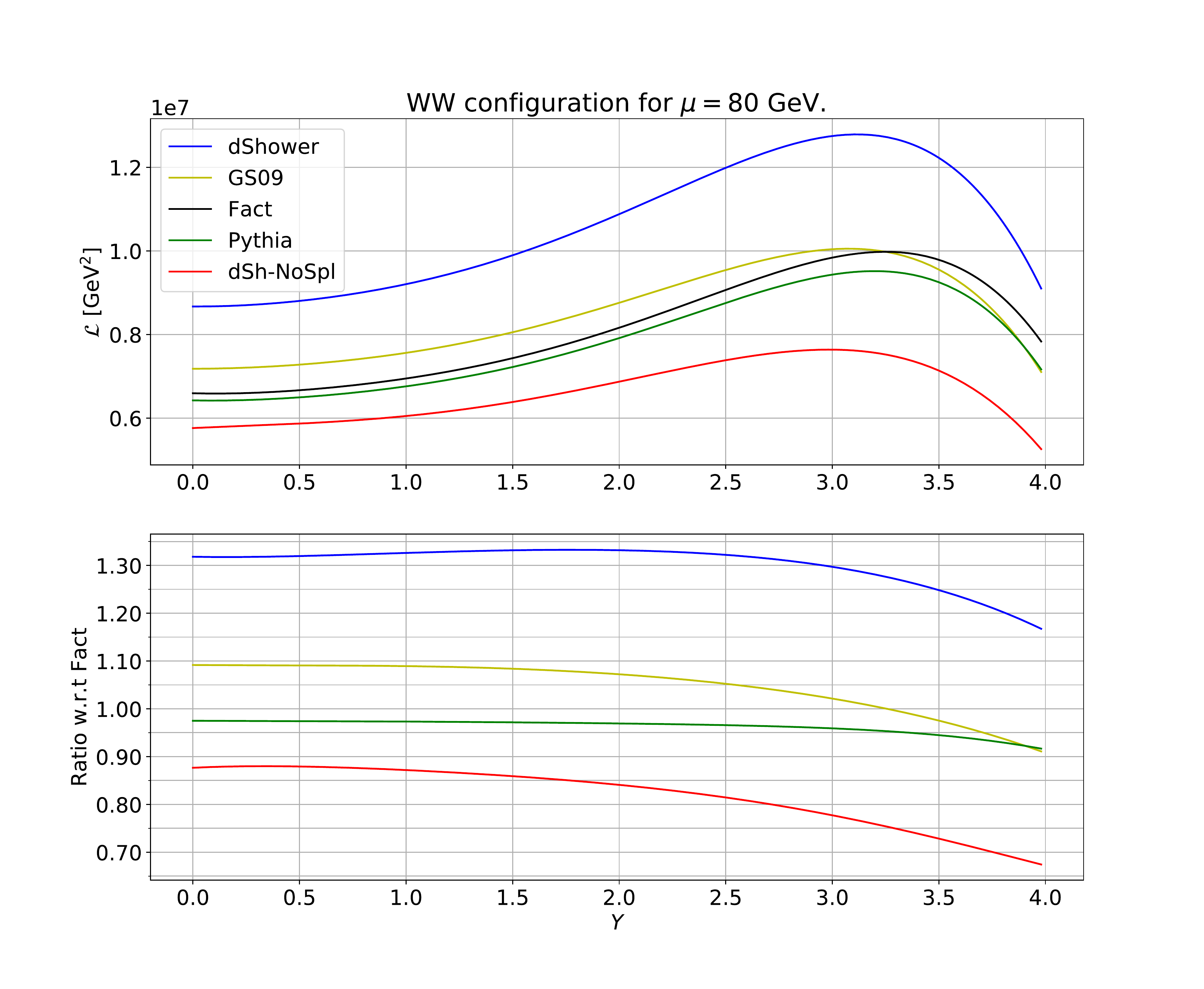} 
\caption{$\mathcal{L}(Y)$ for $\sqrt{s}=14$ TeV and for different setups. The partonic configuration WW refers to \mbox{$u\dbar \dbar u+\dbar uu\dbar+uu\dbar\dbar+\dbar\dbar uu$}. The setup Fact is used as a reference in the ratio plot. Note the suppressed zero on the vertical axis.}
\label{Fig:lumiWW}
\end{figure}

\noindent The luminosity $\mathcal{L}$ is plotted as a function of the rapidity $Y$ of the second hard process in \Fig \ref{Fig:lumiWW}.

An important feature that appears in the ratio plot is the fact that all the ratios with respect to the setup Fact are roughly constant over a large band of rapidities, and then start to decrease around $Y=3$. This is an effect of the kinematic suppression of the dPDFs. Indeed, the higher the value of the rapidity is, the closer to the kinematic limit the system is and the dPDFs get suppressed. This suppression is not present for the setup Fact.

The shape of the luminosity for the GS09 set requires some explanations. For low values of the rapidity, the kinematic suppression is not relevant so the 2v1 and 1v1 contributions raise the luminosity above the one obtained with a simple factorisation ansatz. However, for higher values of the rapidity, the phase-space factor of the GS09 dPDFs starts to suppress the dPDFs since the system is tending towards the kinematical limit. This results in a drop of the luminosity, which goes below the Fact luminosity.

The setups Fact, Pythia\footnote{The rescaling of the PDFs in \PyEight\, might recreate some 2v1 or 1v1 features, however.} and dSh-NoSpl only involve the 2v2 contributions. Therefore, one would expect similar shapes for the luminosity for these three setups. On one hand, \PyEight\, leads to a luminosity which is very close to the one obtained with the factorisation ansatz; the only difference being the drop at large $Y$ for the Pythia setup which is caused by the kinematic suppression. On the other hand, the luminosity for dSh-NoSpl is significantly smaller than the two other ones. This is a direct effect of the evolution of the $\y$-dependent dPDFs of the dSh-NoSpl setup. With our choice for the value of $\sigma_\mathrm{eff}$, one can consider that the three dPDF sets start at low scale with the same $y$-dependence for the dPDFs that involve two quarks. More precisely, this $y$-dependence is Gaussian and reads

\begin{equation}
\label{eq:Gaussian}
F(\y)=\frac{1}{4\pi h_{\u\bar{\d}}}\exp\left(-\frac{y^2}{4 h_{\u\bar{\d}}}\right).
\end{equation}

\begin{figure}[t!] 
\centering
\includegraphics[width=0.95\textwidth]{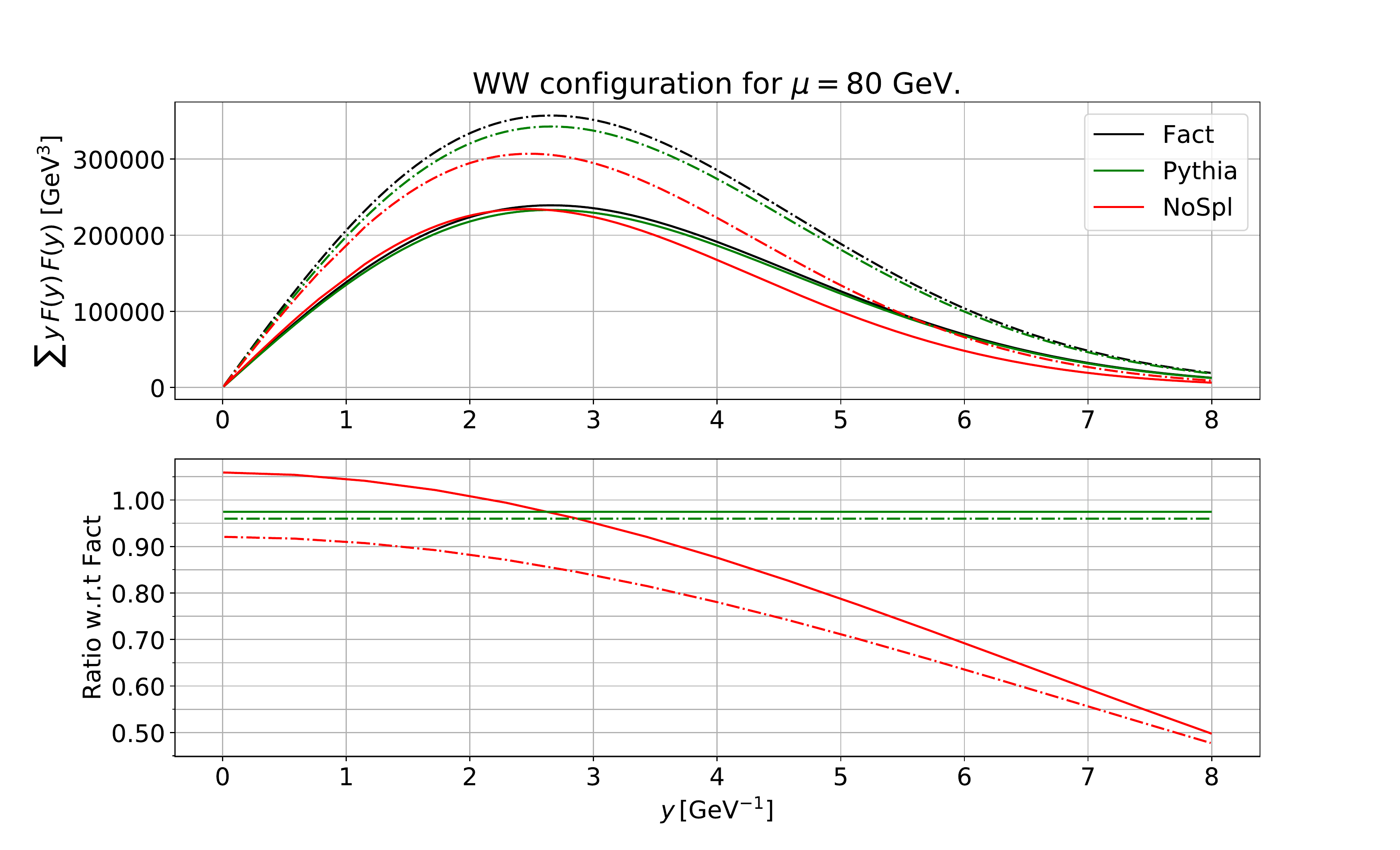} 
\caption{Integrand of the luminosity (divided by $2\pi$) given by \Eq (\ref{eq:lumi}) as a function of $y$ for $\sqrt{s}=14$ TeV and for the WW configuration. The solid lines correspond to the value $Y=0$, whereas the dashed-dotted lines give the results for $Y=3$. The $y$-dependence of the dPDFs for the setups Fact and Pythia is simply the Gaussian factor given by \mbox{\Eq (\ref{eq:Gaussian})}. The setup Fact is the reference in the ratio plot.}
\label{Fig:yDep}
\end{figure}

\noindent In the case of the setups Fact and Pythia, this Gaussian factor is integrated inside the luminosity expression and leaves behind the factor $1/\sigma_\mathrm{eff}$. In particular, the $y$-dependence is fixed and equal to that at the starting scale. In the case of the dSh-NoSpl setup, the evolution alters the $y$-dependence of the intrinsic part of the dPDFs due to mixing with other dPDFs which have different widths. Thus, the shape is no longer Gaussian at high scales. As it can be seen in \Fig \ref{Fig:yDep}, the tail of the $y$-distribution of the dPDFs has been dampened by the evolution. This effect has already been observed and discussed in \cite{Diehl:2014vaa}. The dampening is even stronger for large values of the rapidity. Since the luminosity at a given value of $Y$ is the area under the curve represented in \Fig \ref{Fig:yDep}, the dampening of the $y$-distribution results in a lower value of the luminosity. Thus, the fact that the dSh-NoSpl setup shows a smaller luminosity is a consequence of the evolution of the intrinsic part of the $\y$-dependent dPDFs.

\subsection{Asymmetry}

A relevant observable for same-sign WW pair production via DPS is the lepton pseudorapidity asymmetry. It is defined as \cite{Gaunt:2010pi}

\begin{equation}
\mathcal{A}=\frac{\sigma(\eta_{\ell,1}\times\eta_{\ell,2}<0)-\sigma(\eta_{\ell,1}\times\eta_{\ell,2}>0)}{\sigma(\eta_{\ell,1}\times\eta_{\ell,2}<0)+\sigma(\eta_{\ell,1}\times\eta_{\ell,2}>0)},
\end{equation}

\noindent where $\sigma$ is the inclusive DPS cross section and $\eta_{\ell,1}$ and $\eta_{\ell,2}$ are the pseudorapidities of the two charged leptons in the centre-of-mass frame of the pp collision. A lot of attention has been brought to this observable during the last few years \cite{Gaunt:2010pi, Gaunt:2012, Cotogno:2018mfv}. The asymmetry is sensitive to the correlations between the two hard systems. Indeed, if the two hard processes are completely independent, then there is no reason why the probability for the leptons to be emitted in the same hemisphere would differ from that to be emitted in different hemispheres. The resulting asymmetry is therefore equal to zero. In contrast, with parton correlations, the fact that a lepton is produced in one hemisphere affects the probability for the second lepton to be located in the same hemisphere.

\begin{figure}[t!] 
\centering
\includegraphics[width=0.7\textwidth]{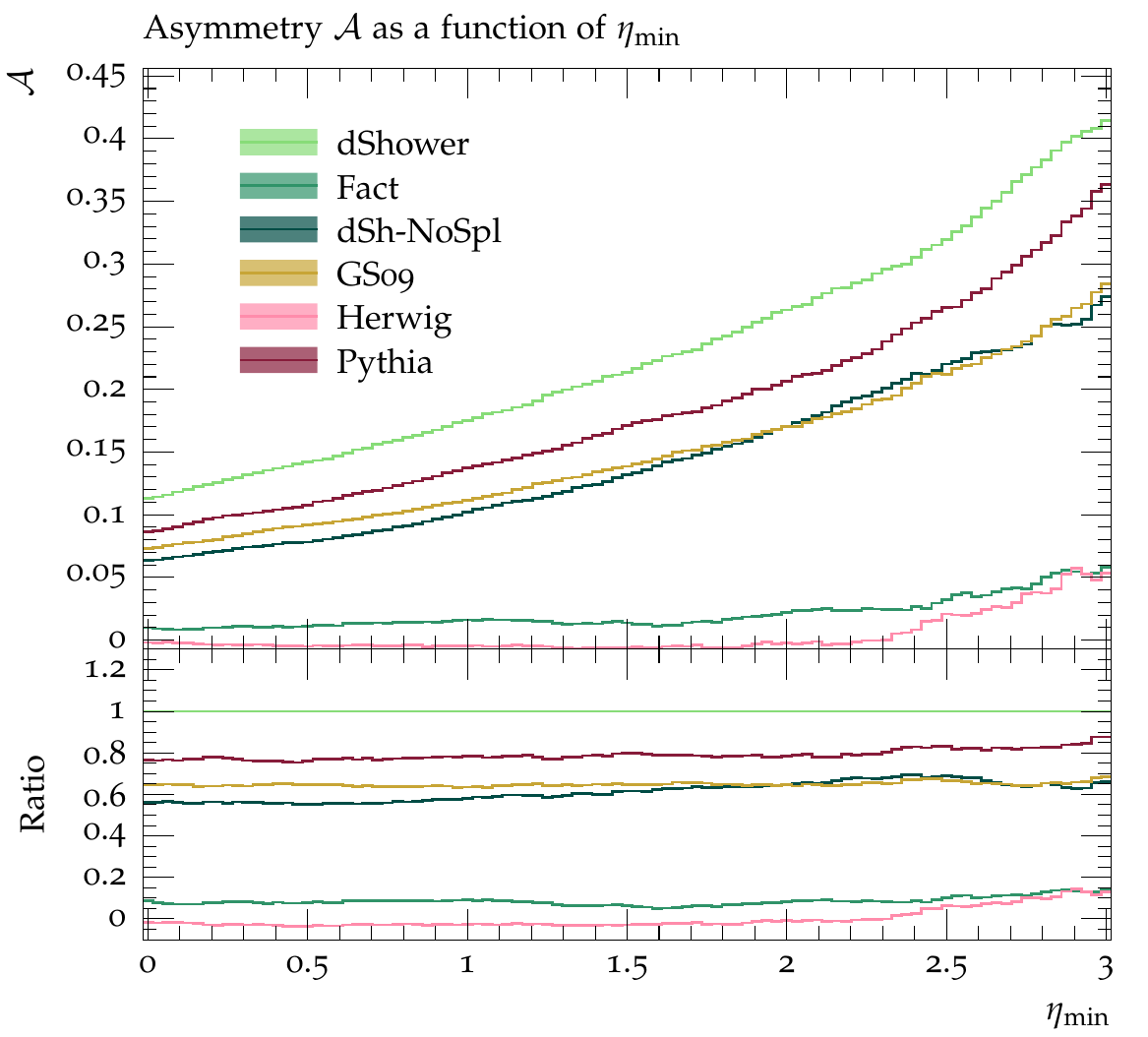} 
\caption{$\mathcal{A}(\eta_\mathrm{min})$ for $\sqrt{s}=14$ TeV and for different setups. The setup dShower is the reference in the ratio plot. Recall that there is no shower for the GS09 setup.}
\label{Fig:asym}
\end{figure}

In order to probe large momentum fractions, it is useful to define a minimum value for the rapidity $\eta_\mathrm{min}$ such that $|\eta_{\ell,1}|,|\eta_{\ell,2}|>\eta_\mathrm{min}$. For high values of  $\eta_\mathrm{min}$, the parton correlations should contribute significantly and a large asymmetry should be observed. The asymmetry is given as a function of  $\eta_\mathrm{min}$ in \Fig \ref{Fig:asym} for all the different setups. It can be noticed that all the setups which include a kinematic suppression of the dPDFs as well as number effects lead to a significant asymmetry which increases with $\eta_\mathrm{min}$. In contrast, the setups Fact and Herwig give an asymmetry which is roughly constant and close to zero. For the setup Fact, the two hard systems are completely independent so this result is expected. The DPS model implemented within \Herwig\,  also uses two independent hard processes. However, the backward evolution of the second hard process is performed using a modified version of the sPDFs, as mentioned in \Sec \ref{Sec:currentMPI}. This might explain the rise of the asymmetry for large values of $\eta_\mathrm{min}$.

The dShower setup gives the largest asymmetry so including $\y$-dependent 2v1 and 1v1 contributions seem to enhance parton correlations. As explained in the previous section, the GS09 setup does not lead to the same asymmetry since this setup does not employ $\y$-dependent dPDFs. The effects of the 2v1 and 1v1 contributions are thus less significant for this setup. The dSh-NoSpl setup gives a smaller asymmetry than the one obtained with dShower most likely because it only includes the 2v2 terms. It is somewhat surprising that \PyEight\, leads to a higher asymmetry than the ones generated with the GS09 and the dSh-NoSpl setups. This is perhaps linked to the way the dPDFs are implemented inside \PyEight, see \Sec \ref{Sec:currentMPI}. 

The effects of including the $1\to2$ splittings with the $\y$-dependence has already been investigated in \cite{Azzi:2019yne}. It was found there that the $\y$-dependent 2v1 and 1v1 contributions increase the asymmetry, which is consistent with our observation.

\subsection{Event shapes}

\begin{figure}[t!] 
\centering
\includegraphics[width=0.6\textwidth]{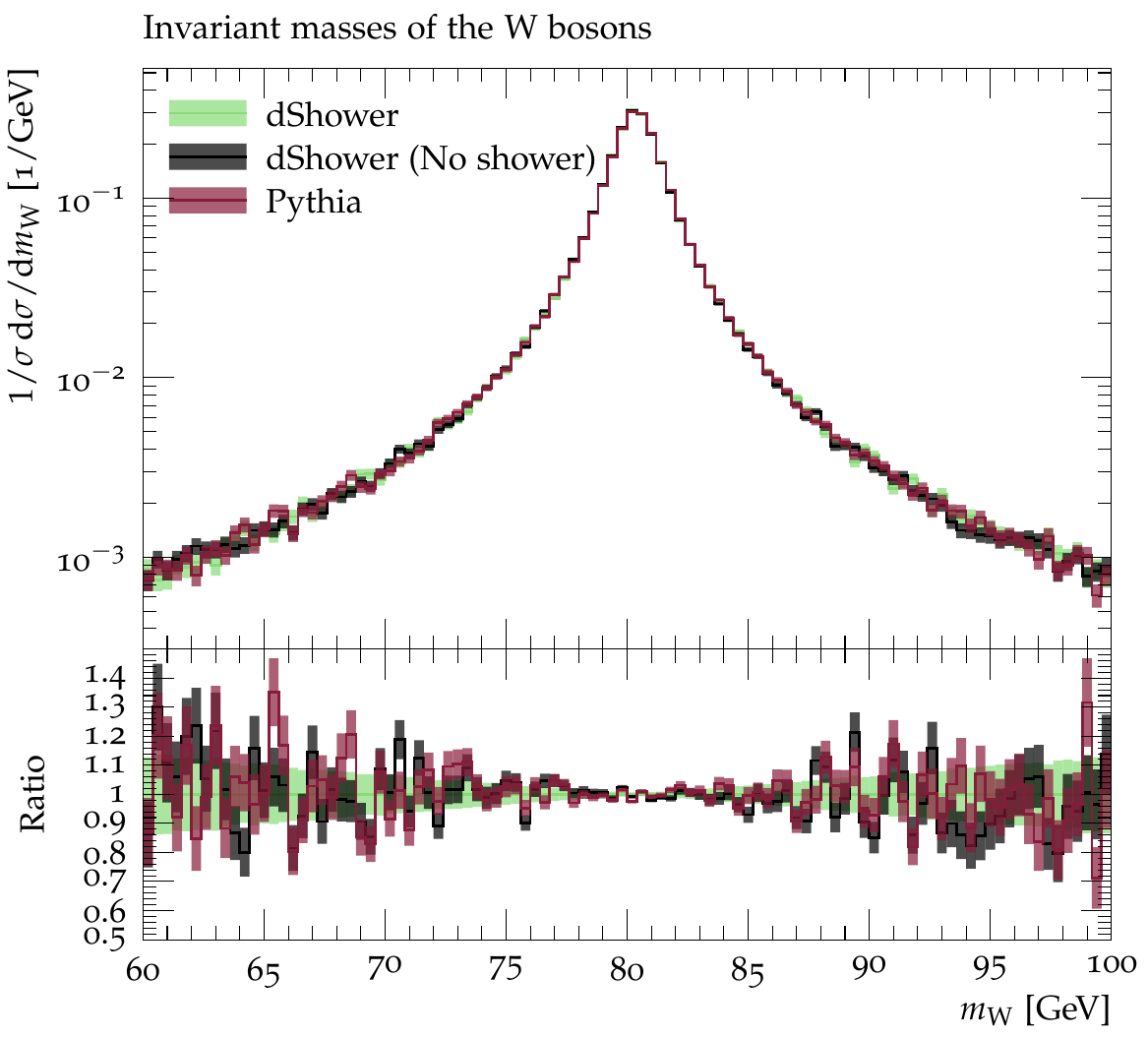} 
\caption{Mass spectrum of the W bosons. This histogram has been made by recording the values of the masses of the two W bosons produced in each event.}
\label{Fig:mW}
\end{figure}

This section will end with some event shapes for $\sqrt{s}=14$ TeV. In all the histograms which will be presented, the error bands represent the statistical errors due to the use of Monte-Carlo techniques. The first one is the mass spectrum of the W bosons represented in \Fig \ref{Fig:mW}. This plot is introduced for validation purposes. The mass spectrum is given for the dShower setup with and without shower. One can see that the two mass spectra exactly match. This is an important result since it means that the parton shower with mergings does not alter the mass distribution of the W bosons, whose shape is determined by the cross section of the hard process, recall \Eq (\ref{eq:FactDPS}). This has been achieved with the procedure described in \mbox{\Sec \ref{Sec:Mrg}}. The mass spectrum obtained with \PyEight\, is also represented for comparison. More validation plots are given in \App \ref{app:validation}.

\begin{figure}[t!] 
\centering
\includegraphics[width=0.49\textwidth]{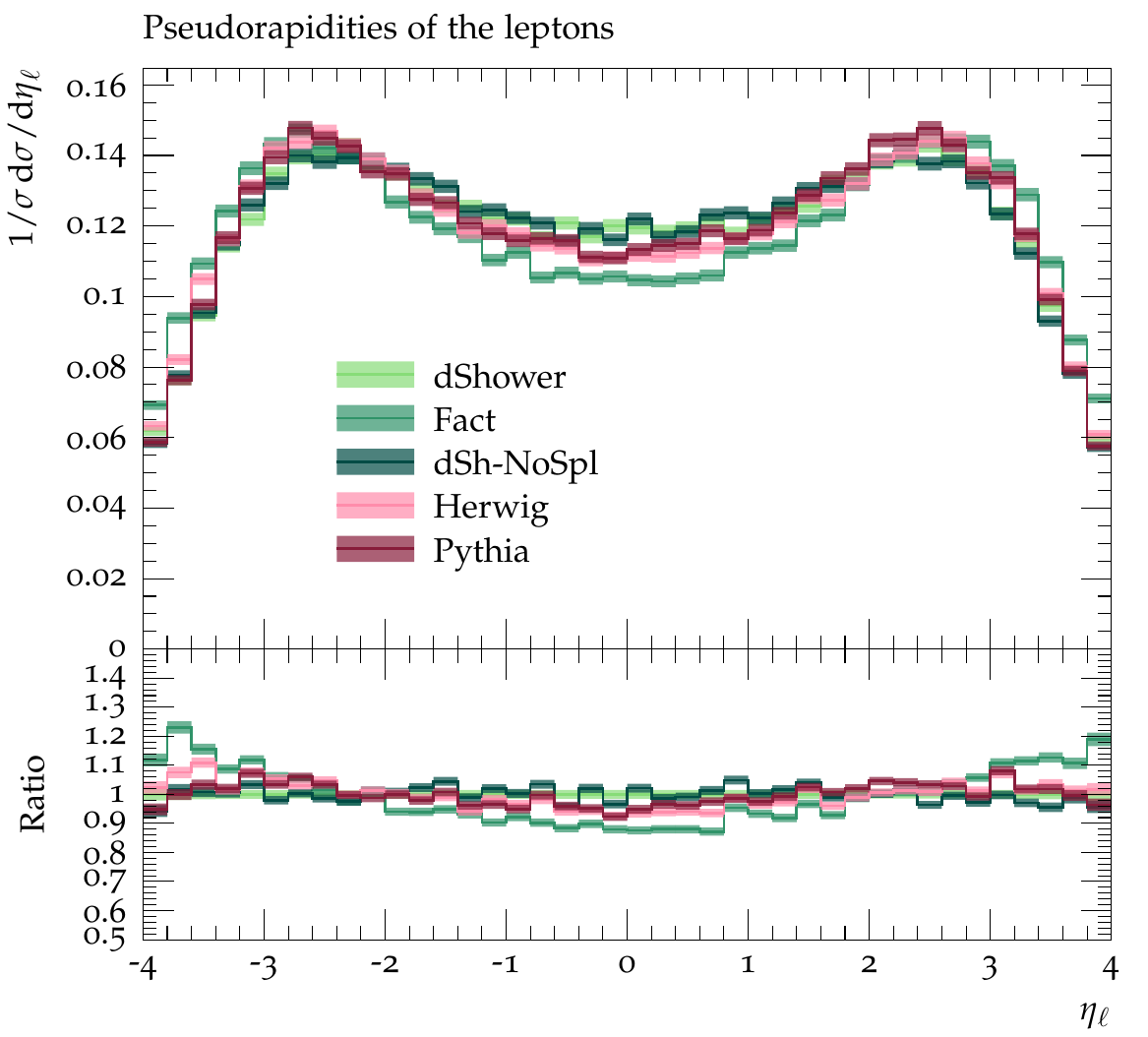} 
\includegraphics[width=0.49\textwidth]{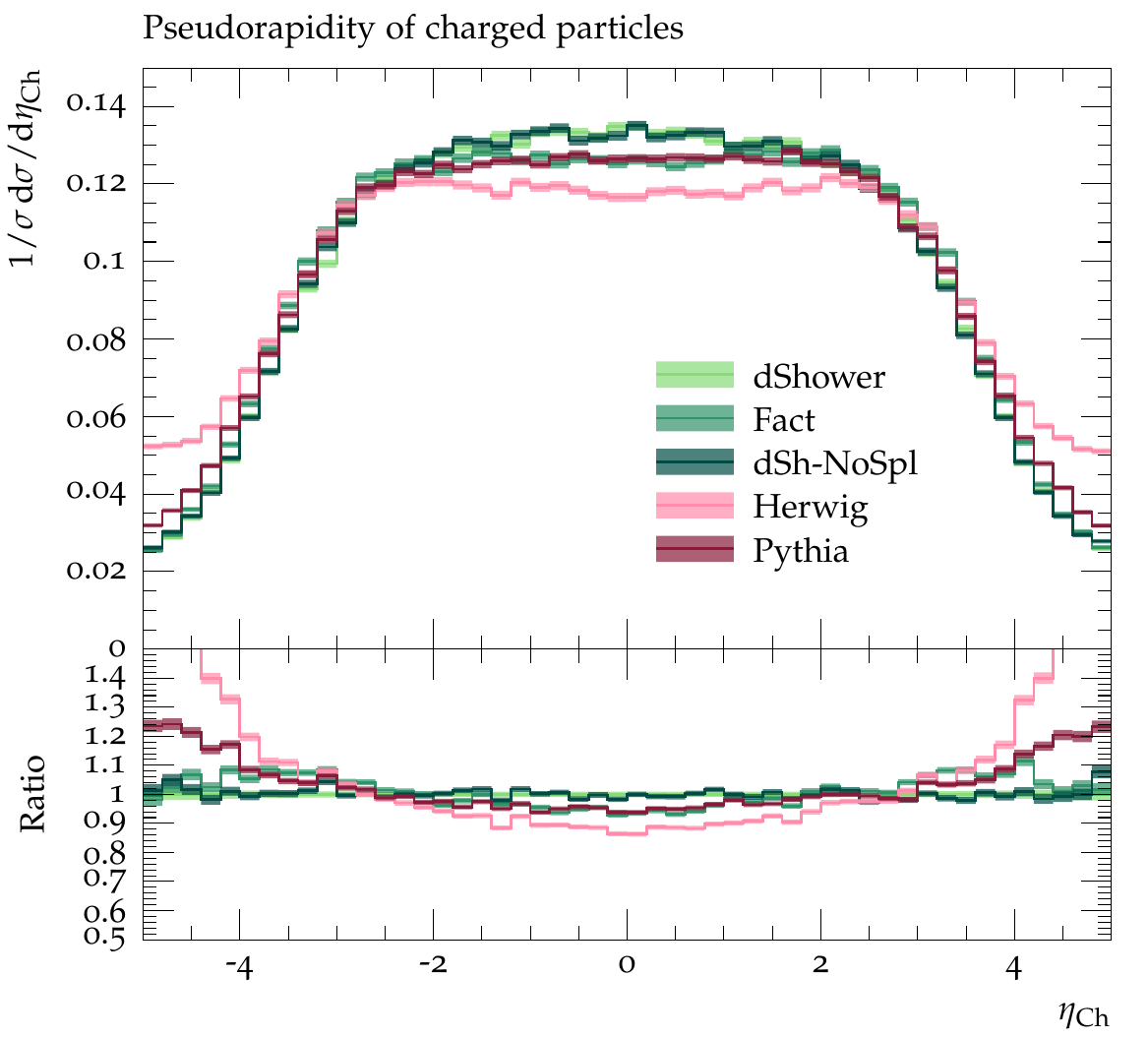} \\
(a) \hspace{200pt} (b) 
\caption{(a) Pseudorapidities of the two charged leptons. (b) Pseudorapidities of all the charged particles produced in the event. The setup dShower is the reference in the ratio plot.}
\label{Fig:Eta}
\end{figure}

Let us now study some pseudorapidity distributions. In \Fig \ref{Fig:Eta}a, the pseudorapidity distribution of the two charged leptons is given. This distribution is sensitive to the suppression of the dPDFs near the kinematic boundaries. Indeed, leptons with high rapidities imply that the system reaches the kinematic boundaries and suppressions occur. Therefore, a setup which includes kinematic suppressions of the dPDFs should have fewer events that contain leptons with high rapidities. This is what is observed on the plot. The setup Fact, which does not include any kinematic suppression, is above the other setups in the high-rapidity ranges, whereas it is below in the central region. This distribution has already been discussed in \cite{Gaunt:2010pi, Gaunt:2012}. 

The pseudorapidity distribution of the charged particles produced in the event is given in \Fig \ref{Fig:Eta}b. This observable is less sensitive to the kinematic suppression of the dPDFs. Indeed, it can be noticed that the setups Fact and Pythia give similar results in the central region, even if the Pythia setup includes some suppression effects whereas the setup Fact does not. Moreover, it seems that using a set of $\y$-dependent dPDFs makes a modest difference. However, as for the previous rapidity distribution, including the splitting part of the dPDFs does not have a strong effect. Indeed, on both plots of \Fig \ref{Fig:Eta}, the setups dShower and dSh-NoSpl overlap.

\begin{figure}[t!] 
\centering
\includegraphics[width=0.6\textwidth]{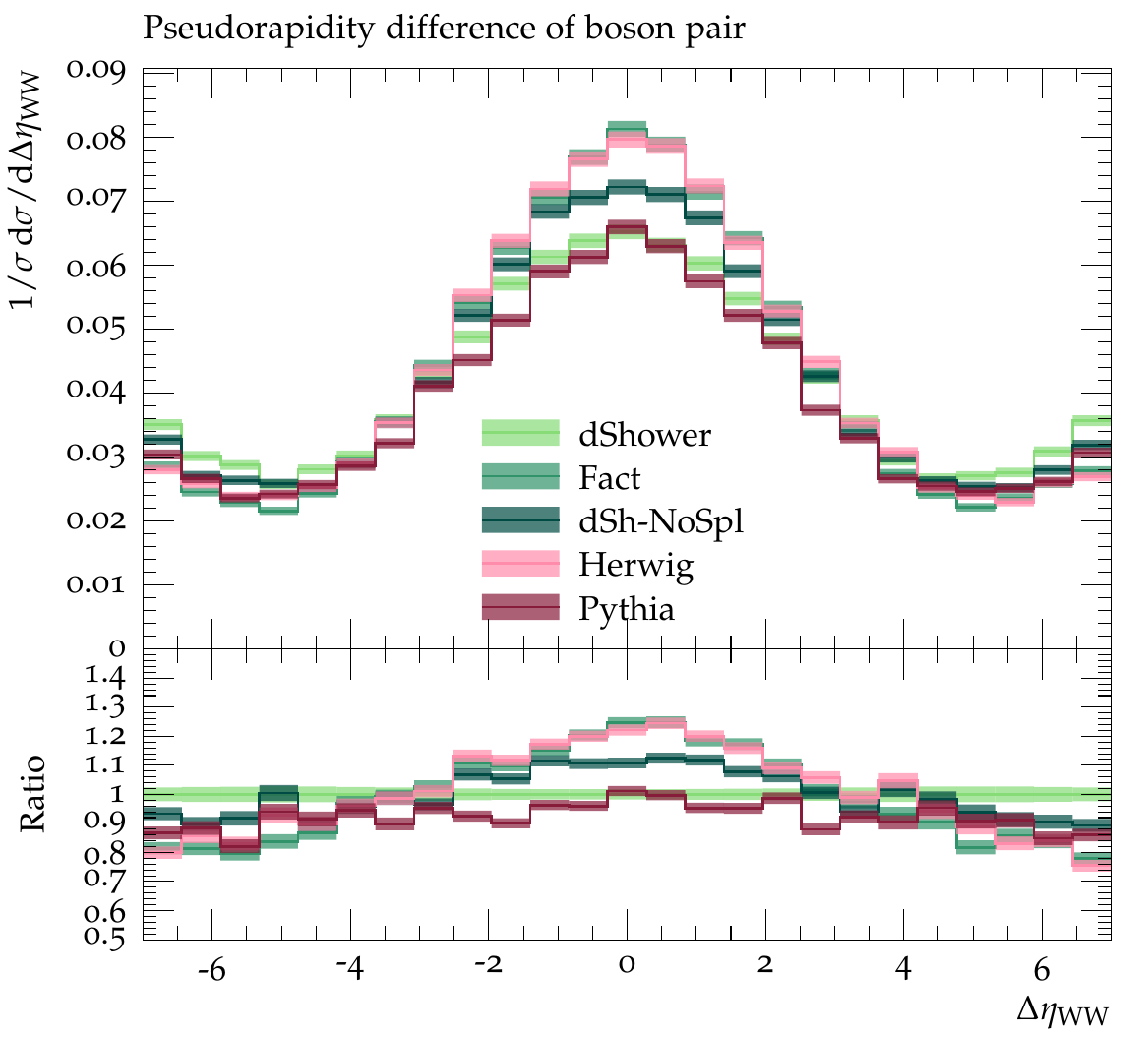} 
\caption{Difference between the pseudorapidities of the two W bosons. The setup dShower is the reference in the ratio plot.}
\label{Fig:PropWW1}
\end{figure}

\begin{figure}[t!] 
\centering
\includegraphics[width=0.49\textwidth]{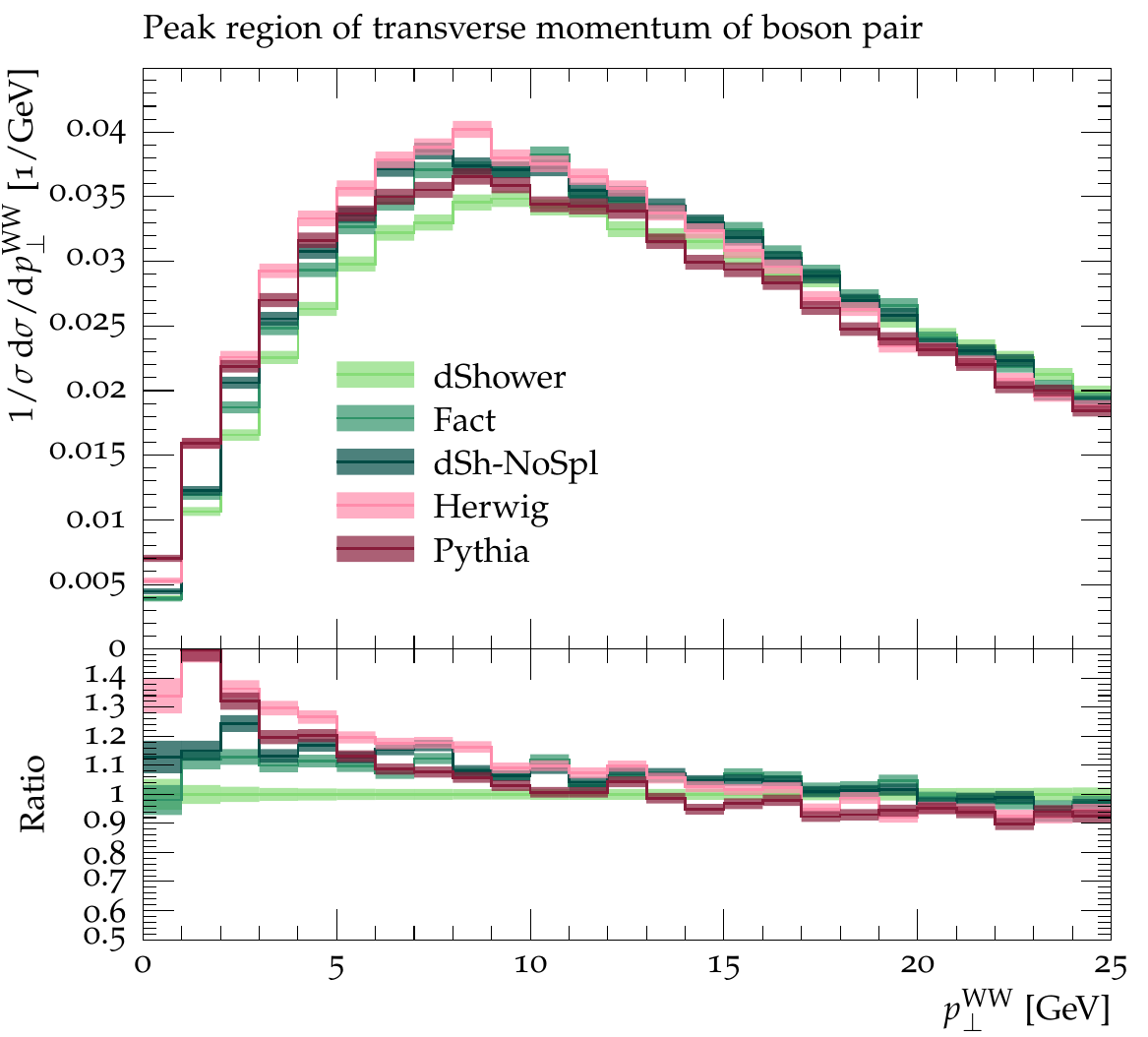} 
\includegraphics[width=0.49\textwidth]{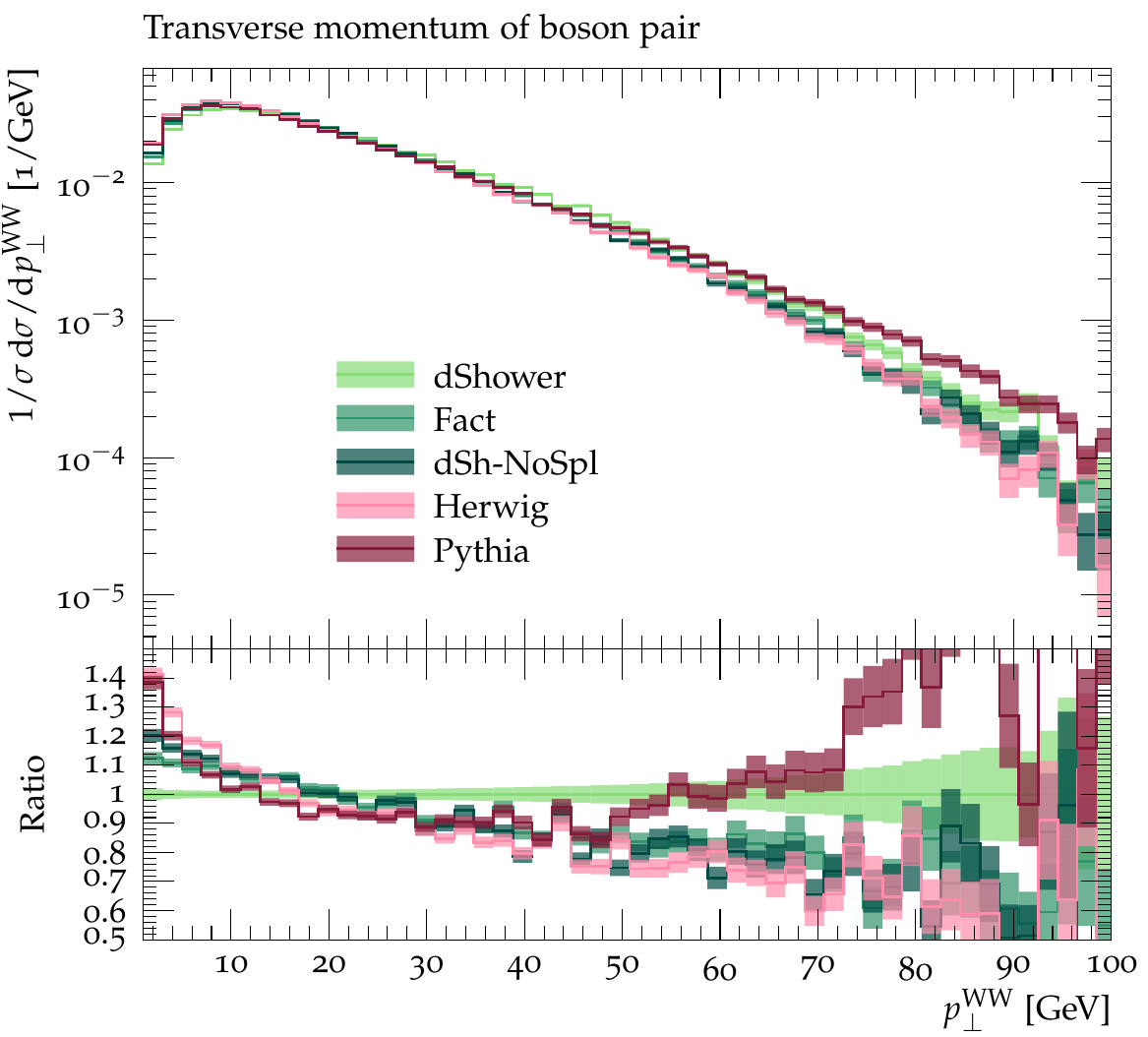} \\
(a) \hspace{225pt} (b)
\caption{(a) and (b) Transverse momentum of the WW pair. The setup dShower is the reference in the ratio plot.}
\label{Fig:PropWW2}
\end{figure}

In \Figs \ref{Fig:PropWW1} and \ref{Fig:PropWW2}, some properties of the WW pair are presented. First, let us discuss the pseudorapidity difference of the two W bosons given in \Fig \ref{Fig:PropWW1}. This observable is closely related to the asymmetry discussed earlier and is thus sensitive to parton correlations. The main idea is that in presence of suppression effects, the system will try to avoid the kinematic boundaries by producing the W bosons far apart i.e. with a large rapidity difference. This is what can be observed on the plot where the setups dShower, dSh-NoSpl and Pythia generate more events with large rapidity differences than the Fact setup that does not have any suppression effect. It is interesting to notice that including the splitting part of the dPDFs makes a difference for this distribution. This difference was already present on the asymmetry plot.

\begin{figure}[t!] 
\centering
\includegraphics[width=0.95\textwidth]{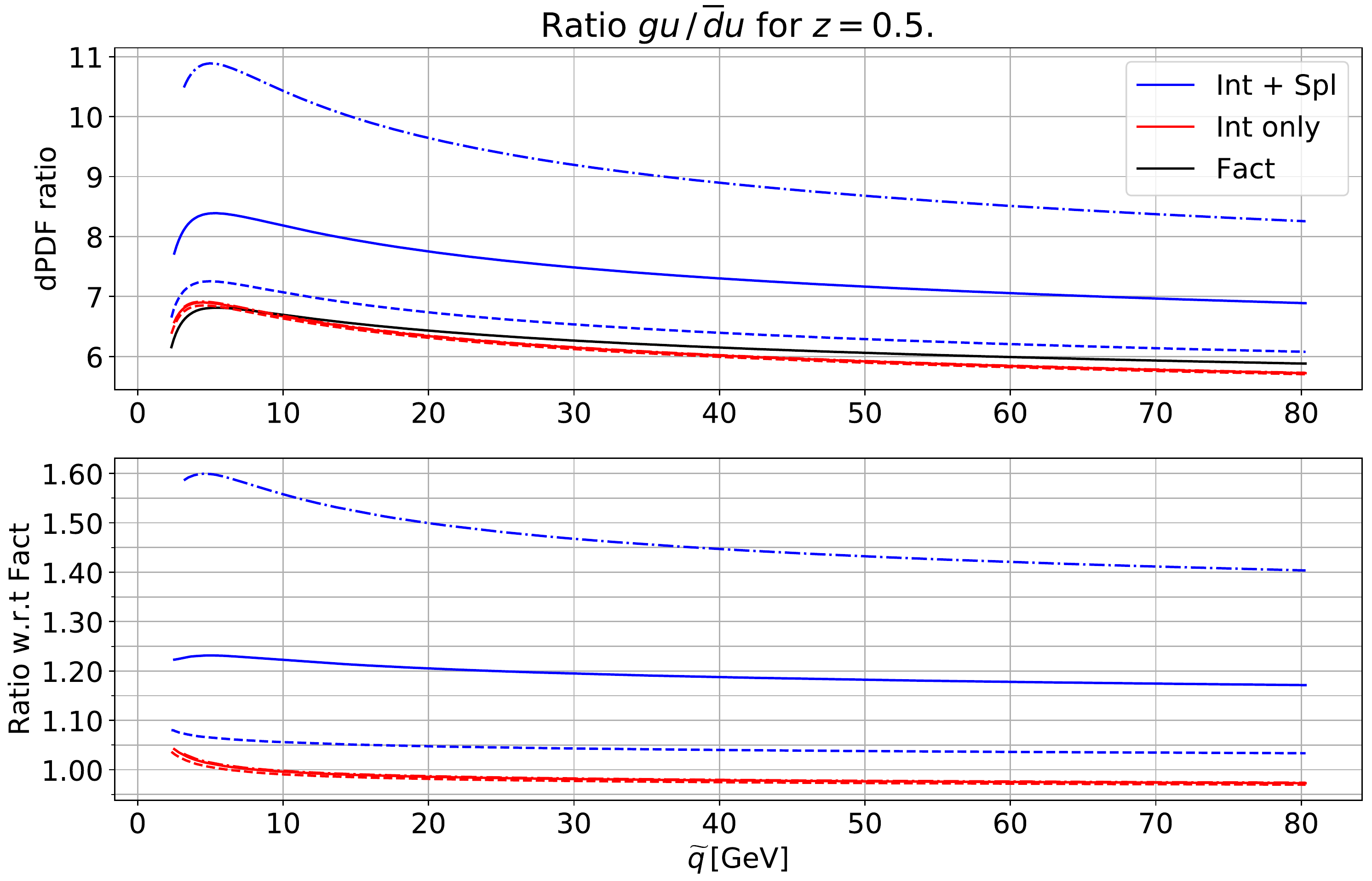} 
\caption{Ratio of dPDFs as a function of the evolution variable $\tilde{q}$ for different setups. The ratio corresponds to the backward branching $\g\to\bar{\d}$ with a u quark acting as a spectator. The values of the momentum fractions are $z=0.5$ and $x_{1,2}=m_{\W} / \sqrt{s}$. The dashed, solid and dashed-dotted lines correspond to the values $y=2\,\mathrm{GeV}^{-1}$,  $y=1\,\mathrm{GeV}^{-1}$ and  $y=0.5\,\mathrm{GeV}^{-1}$ respectively. The setup Fact is the reference in the ratio plot.}
\label{Fig:ratioPDF}
\end{figure}

The last observable which will be discussed is the transverse momentum of the WW pair, see \Fig \ref{Fig:PropWW2}. This observable is interesting since it is directly linked to the characteristics of the shower. Indeed, at LO, the DPS cross section for WW pair production predicts that the W bosons are created with zero transverse momenta. The two W bosons actually get a transverse momentum by recoiling against the partons which have been generated by ISR. Their transverse momenta is thus a pure product of the shower. The most important piece of information that should be extracted from \Fig \ref{Fig:PropWW2} is that the dShower setup leads to fewer events with a small $\pT^{\W\W}$ than the dSh-NoSpl setup. The most sensible explanation is that including the splitting part of the dPDFs results in a larger Sudakov factor and thus in a stronger suppression at small $\pT$. To see that, one needs to recall the branching probability for ISR defined in the case of DPS by \Eq (\ref{eqdPijSud}). The strength of the Sudakov factor is determined by the following quantity

\begin{equation}
\frac{F_{i'j}(x_1/z,x_2,\y,\qtis)}{F_{ij}(x_1,x_2,\y,\qtis)}\,\frac{\as}{2\pi}\,P_{i'\to i}(z),
\label{eq:ratioPDF}
\end{equation}

\noindent where a pair of partons $ij$ with momentum fractions $x_1$ and $x_2$ evolves backwards into a pair $i'j$ with fractions $x_1/z$ and $x_2$ via the QCD branching $i'\to i$. For the case of $\W^+\W^+$ production, the most relevant branchings  are $\g\to\bar{\d}$ and $\g\to\u$. In the case of the setups dShower and dSh-NoSpl, the splitting kernels are exactly the same so the differences are necessarily due to the ratio of dPDFs.  In order to investigate the differences between the two setups, the dPDF ratio $gu/\dbar u$, corresponding to the backward branching $\g\to\bar{\d}$ in the presence of a u quark, is plotted as a function of $\tilde{q}$ in \Fig \ref{Fig:ratioPDF}. It can be seen that including the splitting part of the dPDFs may considerably increase the dPDF ratio. More precisely, the lower the value of $y$ is, the larger the ratio is. This might have been expected since the splitting part of the $gu$ distribution behaves like $1/y^2$. In the case of the dSh-NoSpl setup, where only the intrinsic part of the dPDFs is taken into account, the ratio is close to the one obtained with a factorisation ansatz. Moreover, this ratio does not seem to be too sensitive to the value of $y$. Similar behaviours are seen for the ratio $g\dbar/\dbar\dbar$ with a $\bar{\d}$ quark, as well as for the ratios $gu/uu$ and $g\dbar/u\dbar$ corresponding to the backward branching $\g\to\u$ in the presence of a u quark and a $\bar{\d}$ quark respectively. These observations are consistent with what can be observed on the $\pT$ spectrum. Indeed, the setups dSh-NoSpl and Fact have similar dPDF ratios, which results in a similar Sudakov suppression at small $\pT$. In contrast, the dShower setup includes the splitting part which leads to a larger ratio and thus to a stronger suppression at small $\pT$.

A similar difference between the setups dShower and dSh-NoSpl can be observed for high values of $\pT^{\W\W}$. More precisely, the setup dShower generates more events with a large $\pT^{\W\W}$ than the dSh-NoSpl setup. The reason is the same. For high $\pT$ values, the Sudakov factor is close to unity and is thus irrelevant in that region. The probability to have an emission at high $\pT$ is therefore driven by the terms which are present in \Eq (\ref{eq:ratioPDF}). Since including the splitting part of the dPDFs considerably enhances the dPDF ratio, it results that the dShower setup has a higher probability to emit in the high $\pT$ region than the dSh-NoSpl setup.

\section{Summary}

In this work, a parton-level simulation of DPS has been introduced. This simulation is based on the general DGS framework and includes the $1\to2$ perturbative splittings as well as the $\y$-dependence. Some first numerical results have been obtained by combining the set of $\y$-dependent dPDFs developed in \cite{Diehl:2017kgu} with an angular-ordered shower. Several issues relative to parton showering in the context of DPS have been addressed. More specifically, a lot of work has been dedicated to the kinematics and the treatment of the colour flow in the case of merging. As a by-product of this work, a new scheme for defining the valence and sea components of the dPDFs has been proposed. This scheme is consistent with the evolution equations and ensures that each component is initialised with a positive value.

The simulation has been used to study same-sign WW pair production. The results show some differences with the conventional approaches to dealing with DPS. In particular, using a $\y$-dependent set of dPDFs clearly enhances the 2v1 and 1v1 contributions to the cross section, as has been predicted in \cite{Blok:2011bu, Gaunt:2012dd}. This results in a larger cross section and in a larger asymmetry. From a parton-shower point of view, the $\y$-dependence of the splitting part seems to have a sizeable effect on the dPDF ratios involved in the branching probabilities. In the case of $\W^+\W^+$ production, this leads to a stronger Sudakov suppression at small $\pT$ values as well as a higher probability to emit at large $\pT$ values. Aside from these interesting effects, the impact of including the mergings is rather small. This is because of our choice of process. Indeed, in the case of WW pair production, a merging is possible only if at least one emission has happened beforehand during the backward evolution. In order to more clearly see the impact of the mergings, it would be more interesting in the future to study other processes such as $\Z^0\Z^0$ production which contains partonic configurations like $\u\bar{\u}$ or $\d\bar{\d}$ which can be more easily merged.   

Several directions can be indicated for future works. The most important one is the development of a shower evolution with two different hard scales. With the current state of the simulation, the only DPS processes which can be studied are the ones that involve two hard systems with comparable scales such as WW or $\Z^0\Z^0$ production. It would be very helpful to be able to study other processes such as W + 2 jets or 4-jet production, where the two scales are usually very different.

The kinematics of the shower in case of merging still requires some further work. Indeed, the one which has been established in this work preserves the invariant mass of the resonance only if the decay products of that resonance are all colour singlets. Therefore, it is necessary to develop in the future a procedure that preserves the invariant mass whatever the colour charge of the decay products is. This aspect is relevant if one wants to study $\Z^0\Z^0$ production with at least one of the $\Z^0$ bosons decaying into jets. It is recalled here that it is the FSR kinematics which is creating problems, whereas the ISR one seems to be performing as expected.

The simulation should also be improved in order to allow more realistic studies. A first upgrade would be for example to extend the quark content from three to five flavours. The DGS set of $\y$-dependent dPDFs has already been extended to include five flavours \cite{Azzi:2019yne}. The remaining task would be to include the mass thresholds inside the shower evolution. A second upgrade would be to link the simulation to a hadronisation model so that it generates hadronic final states. This would allow to study the impact of the mergings on the colour flow.

Finally, some work is needed to implement \Eq (\ref{eq:sub}) inside the simulation in a consistent way. Naively, the simulation should be able to switch from a DPS description to an SPS description for $y\lesssim 1/Q_h$, with $Q_h$, the hard scale. Such a simulation would allow for example to model WW pair production with both DPS and SPS processes taken into account. Including the subtraction term from \Eq (\ref{eq:sub}) should also remove any dependence on the unphysical scale $\nu$. Some techniques used in event generators for the matching-and-merging procedure \cite{Bengtsson:1986hr, Seymour:1994we, Seymour:1994df, Miu:1998ju, Lonnblad:1995ex,Frixione:2002ik, Frixione:2007vw, Nason:2004rx, Catani:2001cc, Lonnblad:2001iq, Mrenna:2003if} might be helpful to achieve such a goal.

The ideas developed in the context of this simulation could be used to try to improve the current MPI models. For example, a system containing $n$ different interactions could be divided into $n(n-1)/2$ pairs of interactions. Each one of these pairs can be seen as a DPS so the approach introduced in this work could be applied to each pair separately. The shower evolution developed for DPS and the whole merging procedure could then be embedded inside the MPI evolution. This would hopefully include some of the parton correlations.


\acknowledgments

BC would like to thank Michael Seymour for his advice, comments and thoughtful remarks towards this work, as well as Torbj\"{o}rn Sj\"{o}strand and Peter Skands for useful and inspiring discussions. The help of Matthew De Angelis is also greatly acknowledged. This work has received funding from the European Union's Horizon 2020 research and innovation programme as part of the Marie Sk\l{}odowska-Curie Innovative Training Network MCnetITN3 (grant agreement no. 722104). The histograms have been produced with Rivet \cite{Buckley:2010ar} and the sketches with Axodraw \cite{Collins:2016aya}.

\appendix
\section{Evolution equations for valence and sea components}
\label{app:evol}

Let us take the example of the $\uv\ds$ distribution. For this specific dPDF, one expects the following dDGLAP equation

\begin{equation}
\mu^2\frac{\partial}{\partial\mu^2}\,\uv\ds = \hat{P}_{\q\to\q}\,\underset{1}{\otimes}\,\uv\ds + \hat{P}_{\q\to\q}\,\underset{2}{\otimes}\,\uv\ds + \hat{P}_{\g\to\q}\,\underset{2}{\otimes}\,\uv g, 
\label{eq:dDGLAPuvds}
\end{equation}

\noindent which is basically \Eq (\ref{eq:homdDGLAP}) with the following notation\footnote{The operator $\underset{2}{\otimes}$ is defined in a similar way.}

\begin{equation}
\hat{P}_{i'\to i}\,\underset{1}{\otimes}\,F_{i'j} = \int_{x_1}^{1-x_2}\frac{\d x_1'}{x_1'}\,\frac{\as(\mu^2)}{2\pi}\,\hat{P}_{i'\to i}\left(\frac{x_1}{x_1'}\right)\,F_{i'j}(x_1',x_2,\y,\mu^2).
\end{equation}

\noindent In order to show that the distribution $\uv\ds$ defined with the scheme given in \Sec \ref{Sec:Scheme} actually satisfies \Eq (\ref{eq:dDGLAPuvds}), the starting point is to write the dDGLAP equations for the $ud$, $u\dbar$, $\dbar u$ and $\ubar\dbar$ dPDFs. One has

\begin{subequations}
\begin{align}
\mu^2\frac{\partial}{\partial\mu^2}\,ud &= \hat{P}_{\q\to\q}\,\underset{1}{\otimes}\,ud + \hat{P}_{\q\to\q}\,\underset{2}{\otimes}\,ud + \hat{P}_{\g\to\q}\,\underset{1}{\otimes}\,gd + \hat{P}_{\g\to\q}\,\underset{2}{\otimes}\,ug, \\
\label{eq:udbar}
\mu^2\frac{\partial}{\partial\mu^2}\,u\dbar &= \hat{P}_{\q\to\q}\,\underset{1}{\otimes}\,u\dbar + \hat{P}_{\q\to\q}\,\underset{2}{\otimes}\,u\dbar + \hat{P}_{\g\to\q}\,\underset{1}{\otimes}\,g\dbar + \hat{P}_{\g\to\q}\,\underset{2}{\otimes}\,ug, \\
\mu^2\frac{\partial}{\partial\mu^2}\,\ubar d &= \hat{P}_{\q\to\q}\,\underset{1}{\otimes}\,\ubar d + \hat{P}_{\q\to\q}\,\underset{2}{\otimes}\,\ubar d + \hat{P}_{\g\to\q}\,\underset{1}{\otimes}\,gd + \hat{P}_{\g\to\q}\,\underset{2}{\otimes}\,\ubar g, \\
\mu^2\frac{\partial}{\partial\mu^2}\,\ubar\dbar &= \hat{P}_{\q\to\q}\,\underset{1}{\otimes}\,\ubar\dbar + \hat{P}_{\q\to\q}\,\underset{2}{\otimes}\,\ubar\dbar + \hat{P}_{\g\to\q}\,\underset{1}{\otimes}\,g\dbar + \hat{P}_{\g\to\q}\,\underset{2}{\otimes}\,\ubar g.
\label{eq:ubardbar}
\end{align}
\label{set:ud}
\end{subequations}

In the scheme, the intrinsic part of $\uv\ds$ is defined as $[\uv\ds]_\mathrm{int}=[\uv\dbar]_\mathrm{int}=[u \dbar-\ubar\dbar]_\mathrm{int}$. Therefore,  the difference between \Eqs (\ref{eq:udbar}) and (\ref{eq:ubardbar}) gives

\begin{equation}
\mu^2\frac{\partial}{\partial\mu^2}\,[\uv\ds]_\mathrm{int}=\hat{P}_{\q\to\q}\,\underset{1}{\otimes}\,[u\dbar-\ubar\dbar]_\mathrm{int}+\hat{P}_{\q\to\q}\,\underset{2}{\otimes}\,[u\dbar-\ubar\dbar]_\mathrm{int}+\hat{P}_{\g\to\q}\,\underset{2}{\otimes}\,[ug-\ubar g]_\mathrm{int}
\label{eq:uvdsint}
\end{equation}

\noindent For the splitting part, one needs to use the definition of $[\uv\ds]_\mathrm{spl}$ given by \Eq (\ref{eq:uvds}). Combining the equations in the set (\ref{set:ud}) according to \Eq (\ref{eq:uvds}) leads to

\begin{equation}
\begin{split}
\mu^2\frac{\partial}{\partial\mu^2}\,[\uv\ds]_\mathrm{spl}=&\hat{P}_{\q\to\q}\,\underset{1}{\otimes}\,\frac{1}{2}[u\dbar-\ubar d-\ubar\dbar+ud]_\mathrm{spl}+\hat{P}_{\q\to\q}\,\underset{2}{\otimes}\,\frac{1}{2}[u\dbar-\ubar d-\ubar\dbar+ud]_\mathrm{spl} \\
&+\hat{P}_{\g\to\q}\,\underset{2}{\otimes}\,[ug-\ubar g]_\mathrm{spl}.
\end{split}
\label{eq:uvdsspl}
\end{equation}

\noindent Finally, adding \Eqs (\ref{eq:uvdsint}) and (\ref{eq:uvdsspl}) gives back \Eq (\ref{eq:dDGLAPuvds}), as desired. It can be noticed that for both the intrinsic and the splitting parts, the terms associated to $g d$ and $g\dbar$ in the set of equations (\ref{set:ud}) cancel each other. This is exactly why the distributions $[\uv\ds]_\mathrm{int}$ and $[\uv\ds]_\mathrm{spl}$ are defined the way they are. The term related to $g\ds$ cannot be present in \Eq (\ref{eq:dDGLAPuvds}).

The proof for the other distributions follows a similar reasoning. One needs to write the dDGLAP equations for the plain dPDFs and combine them according to the definitions of the valence and sea distributions given in the scheme. 

\section{Initial conditions and number sum rules}
\label{app:inputs}

\begin{figure}[t!] 
\centering
\includegraphics[width=0.85\textwidth]{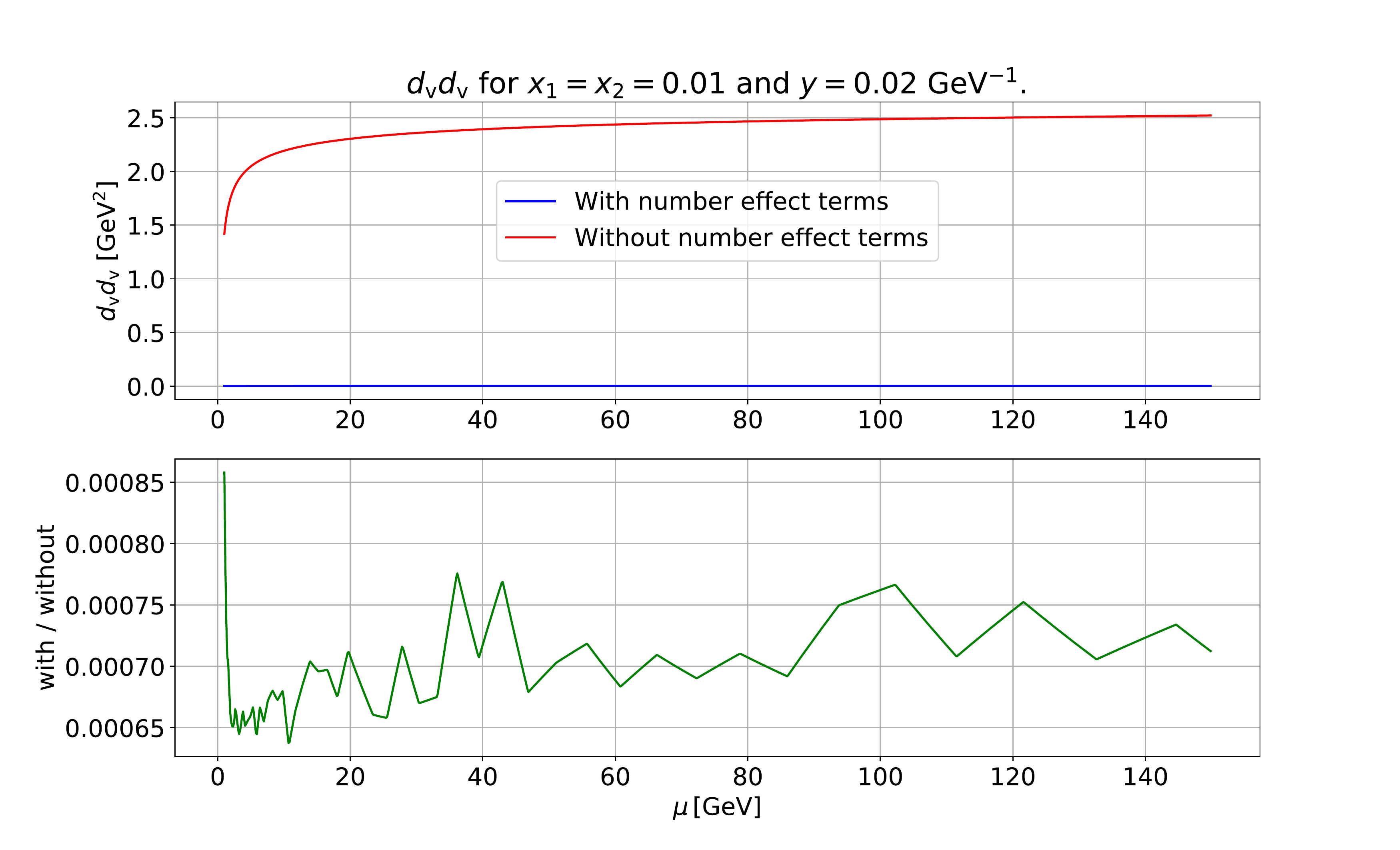}
\caption{Valence-valence component of $dd$ for two different kinds of initial conditions. The distribution in blue has been generated with initial conditions that include the number effect terms, whereas the one in red use initial conditions that do not include any number effect term.}
\label{Fig:valval}
\end{figure}

With the initial conditions given by \Eq (\ref{eq:StartInt}), the DGS set of $\y$-dependent dPDFs does not satisfy the number sum rules. To see that, let us focus on the $\dv\dv$ dPDF. In \mbox{\Fig \ref{Fig:valval}}, the evolution of $\dv\dv$ as a function of the scale $\mu$ is given. One can see that the quantity $\dv\dv$ generated with the input (\ref{eq:StartInt}) is not identically zero. As mentioned in \Sec \ref{Sec:currentMPI}, this is unphysical. Fortunately, this issue can be solved by modifying the initial conditions for the intrinsic part. Indeed, it is argued in \cite{Gaunt:2009re} that one should subtract from \Eq (\ref{eq:StartInt}) for the $dd$ distribution the following term

\begin{equation}
\frac{1}{4\pi h_{\d\d}(x_1,x_2)}\exp\left(-\frac{y^2}{4h_{\d\d}(x_1,x_2)}\right) \dv(x_1,\mu_0^2)\,\dv(x_2,\mu_0^2)\,\frac{(1-x_1-x_2)^2}{(1-x_1)^2\,(1-x_2)^2}.
\end{equation}

\noindent This has the effect of removing the valence-valence contribution from the full $dd$ dPDF at the starting scale $\mu_0$ and thus forbidding this configuration. Since the homogeneous dDGLAP equations preserve the sum rules \cite{Gaunt:2009re}, this constraint should be present at higher scales too. For the u sector, one can apply the same method. More specifically, the initial conditions should take into account the fact that extracting a valence u quark halves the probability of finding a second valence u quark. This is fulfilled by subtracting the following term from \Eq (\ref{eq:StartInt}) for the $uu$ distribution

\begin{equation}
\frac{1}{4\pi h_{\u\u}(x_1,x_2)}\exp\left(-\frac{y^2}{4h_{\u\u}(x_1,x_2)}\right) \frac{1}{2}\,\uv(x_1,\mu_0^2)\,\uv(x_2,\mu_0^2)\,\frac{(1-x_1-x_2)^2}{(1-x_1)^2\,(1-x_2)^2}.
\end{equation}

\begin{figure}[t!] 
\centering
\includegraphics[width=0.85\textwidth]{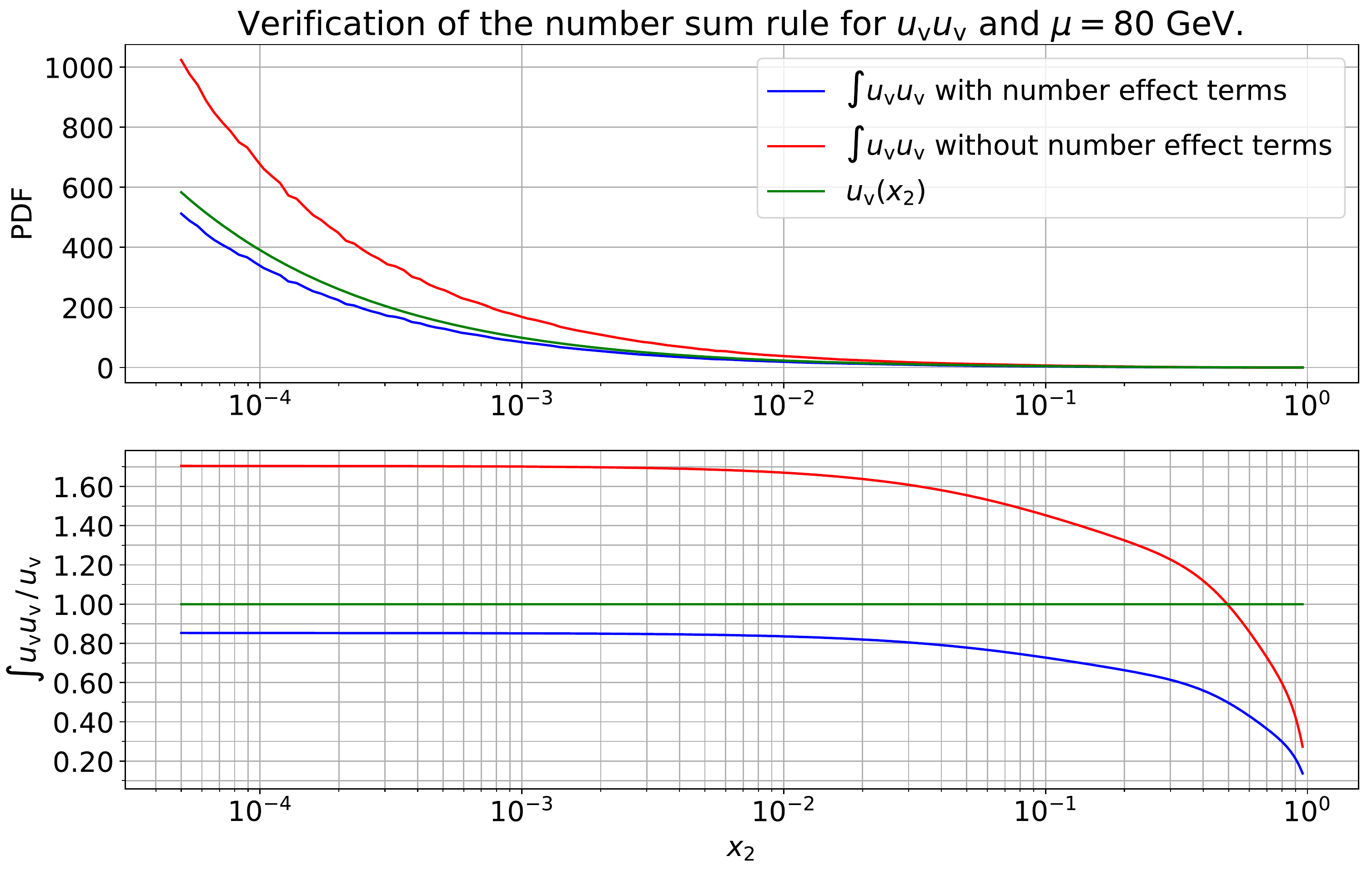}
\caption{Verification of the number sum rule for $\uv\uv$ for the two types of initial conditions. The blue and red curves give the evolution of the distribution defined in \Eq (\ref{eq:uvuv}) as a function of $x_2$ for $\mu=\nu=80\, \GeV$. The green curve gives the distribution $\uv(x_2,\mu^2)$. According to the number sum rule, the ratio should be equal to unity.}
\label{Fig:uuRule}
\end{figure}

\noindent It is shown in \cite{Gaunt:2009re} that including these subtraction terms, referred to as ``number effect'' terms, considerably improve the way the dPDFs describe the GS sum rules, at least for dPDFs that do not depend on $\y$. For $\y$-dependent dPDFs, this procedure seems to work too. Indeed, in \Fig \ref{Fig:valval}, it can be seen that the initial conditions which include the number effect terms lead to a $\dv\dv$ distribution which is zero for all scales, as wanted. This automatically implies that the number sum rule for the $\dv\dv$ distribution given in \Sec \ref{Sec:Scheme} (see \Prop (\ref{enu:prop4})) is satisfied. In \Fig \ref{Fig:uuRule}, the quantity

\begin{equation}
\int_0^{1-x_2}\left(\int\d^2\y\,\Phi(y\nu)\,\uv\uv(x_1,x_2,\y,\mu^2)\right)\d x_1
\label{eq:uvuv}
\end{equation}

\noindent is represented as a function of $x_2$ for the two types of initial conditions. According to the sum rule stated for the $\uv\uv$ distribution, this quantity should be equal to the distribution $\uv(x_2,\mu^2)$, which is also given in the figure. It appears that the initial conditions which include the number effect terms improve significantly the way the sum rule for $\uv\uv$ is verified.  

\section{Validation plots and scale variations}
\label{app:validation}

\begin{figure}[t!] 
\centering
\includegraphics[width=0.49\textwidth]{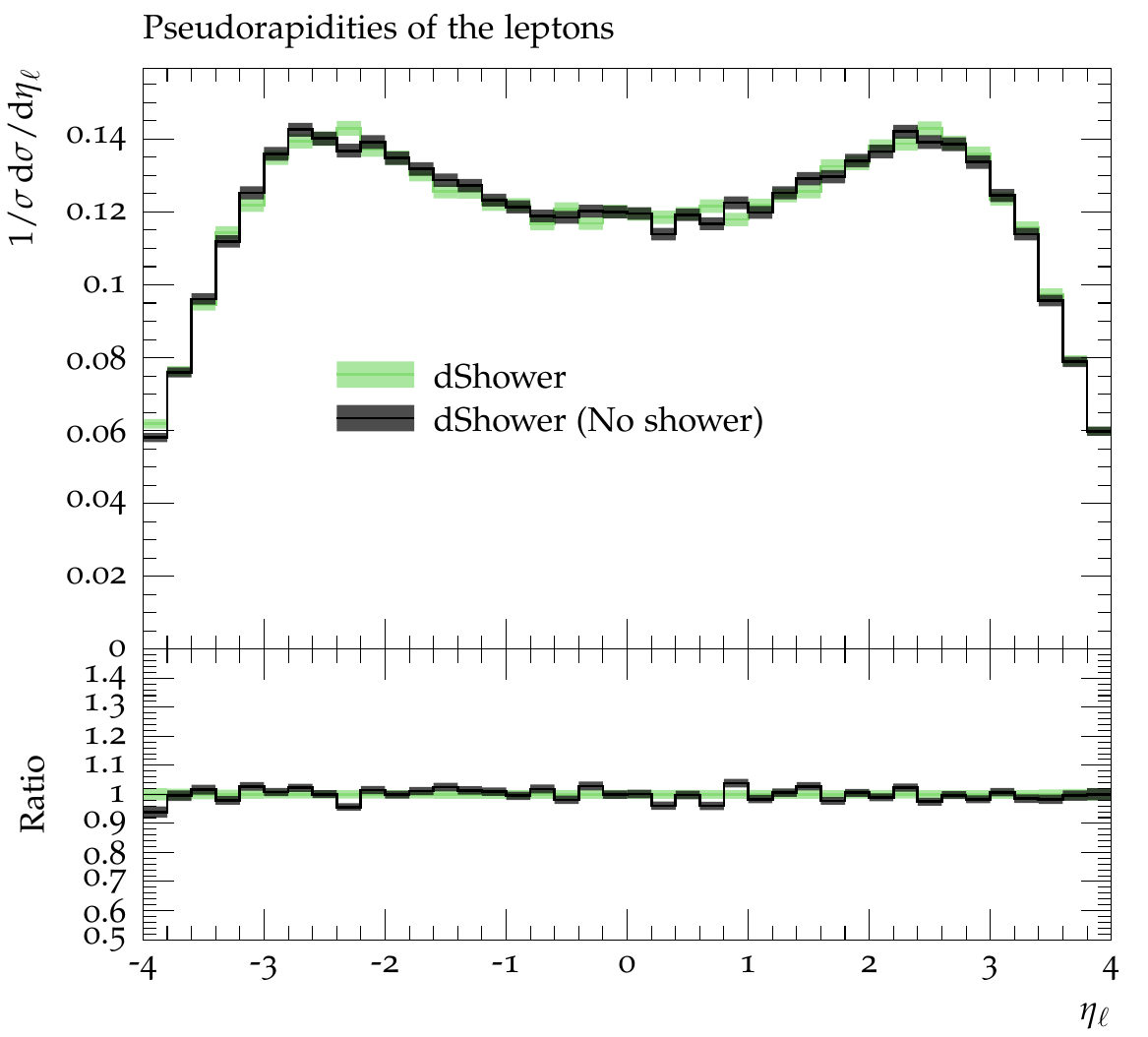} 
\includegraphics[width=0.49\textwidth]{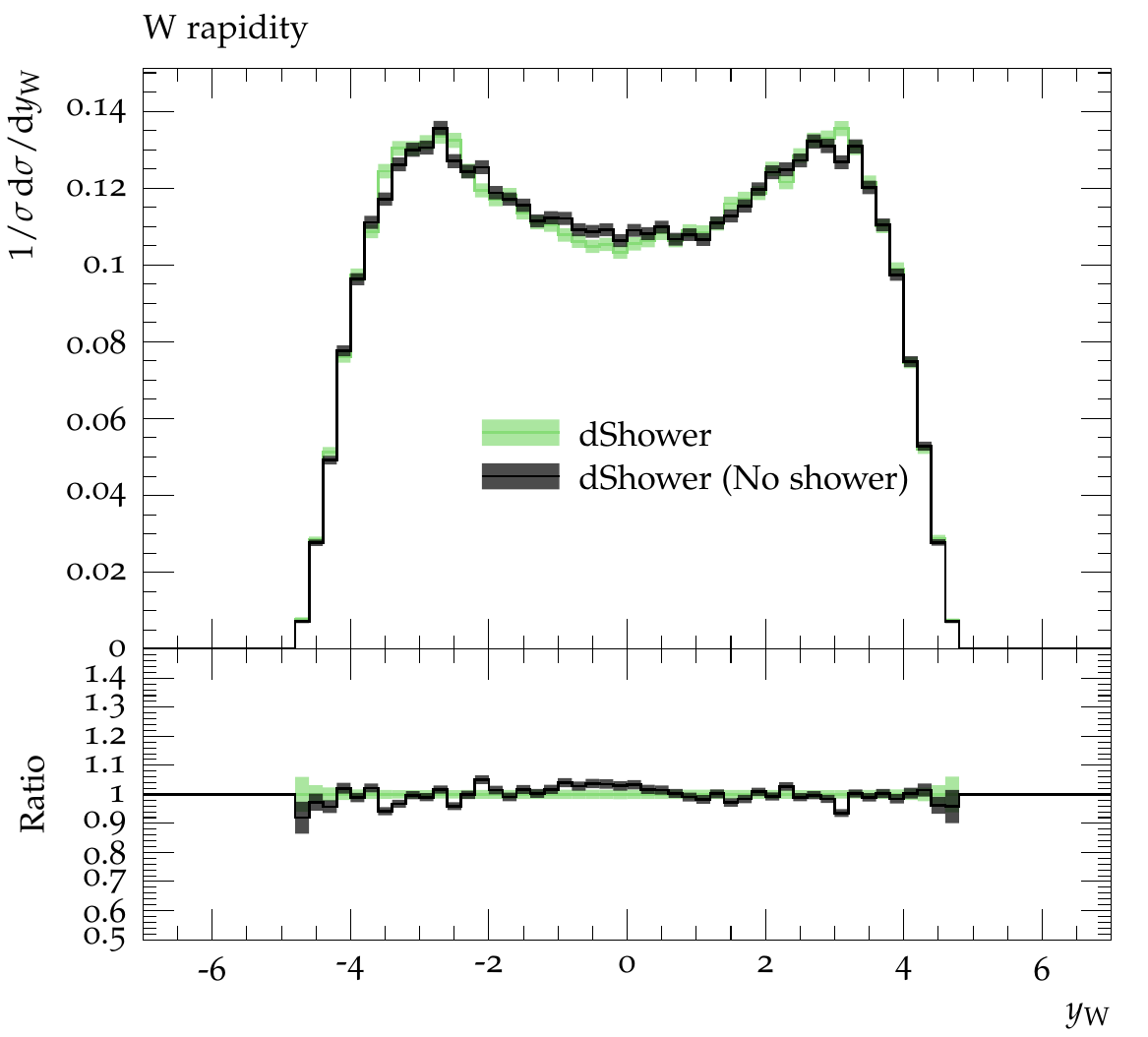} \\
(a) \hspace{225pt} (b) 
\caption{(a) Pseudorapidities of the two charged leptons. (b) Rapidities of the two W bosons.}
\label{Fig:validation}
\end{figure}

\begin{figure}[t!] 
\centering
\includegraphics[width=0.6\textwidth]{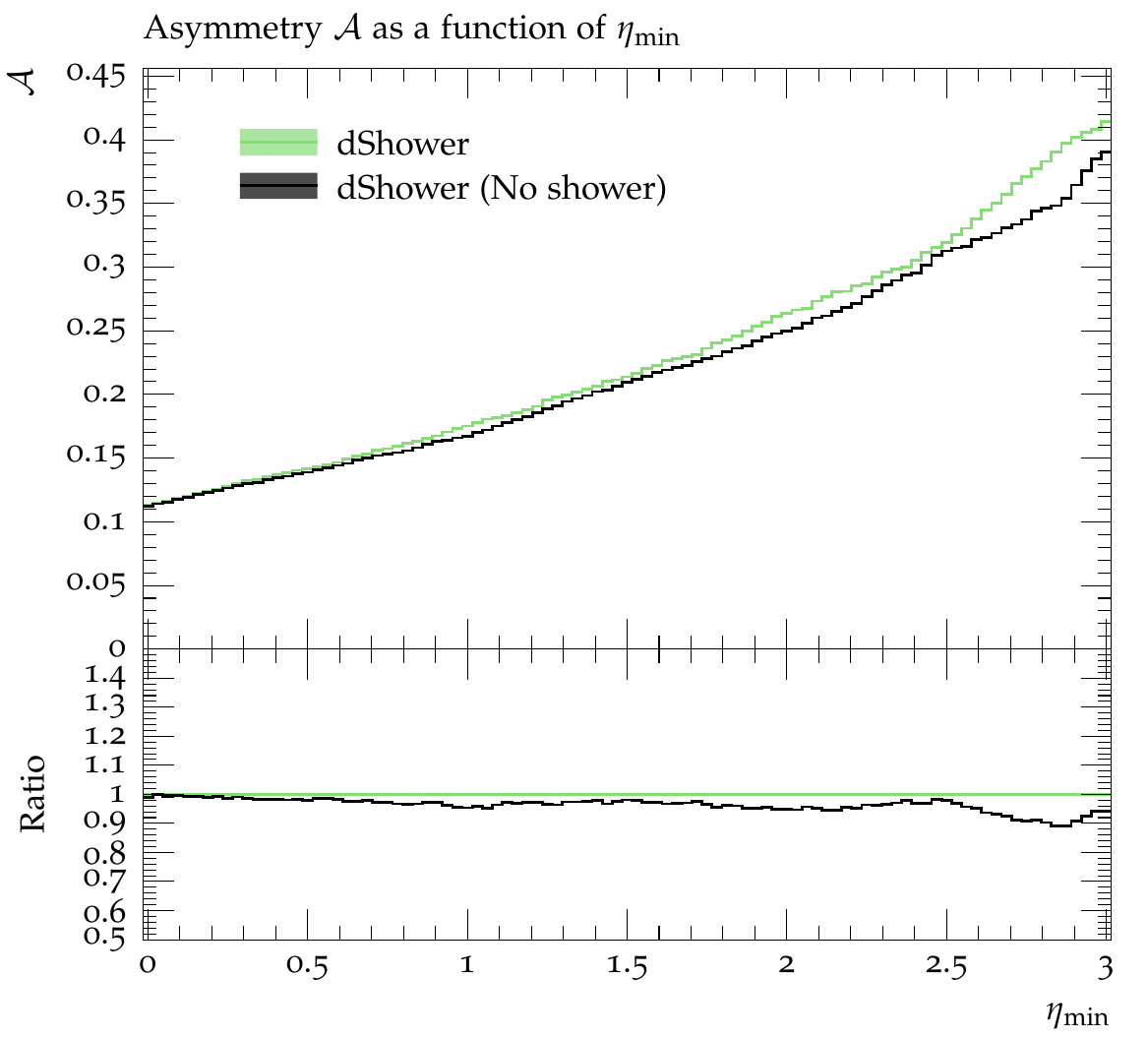}
\caption{Asymmetry $\mathcal{A}$ as a function of $\eta_\mathrm{min}$.}
\label{Fig:validation2}
\end{figure}

In \Figs \ref{Fig:validation} and \ref{Fig:validation2}, the pseudorapidities $\eta_\ell$ of the leptons, the rapidities $y_\W$ of the W bosons and the asymmetry are given for the dShower setup with and without shower. Those observables are determined by the DPS cross section (\ref{eq:FactDPS}) and the shower should not affect their shape too much. The issue is that the procedure introduced in \Sec \ref{Sec:Mrg} shuffles around the individual kinematics of each hard process. A solution has been found to preserve the invariant masses of the W bosons, recall \Fig \ref{Fig:mW}. In contrast, no solution has been implemented yet in order to conserve the values of the rapidities which hence might be altered by the shower. This can be seen in \Fig \ref{Fig:validation}, especially for the central region of the $y_\W$ distribution. Fortunately, the differences remain modest. The asymmetry seems to be slightly affected by the shower for large values of $\eta_\mathrm{min}$. It is important to specify here that the higher $\eta_\mathrm{min}$ is, the lower the statistics is since more events get vetoed.

\begin{figure}[t!] 
\centering
\includegraphics[width=0.49\textwidth]{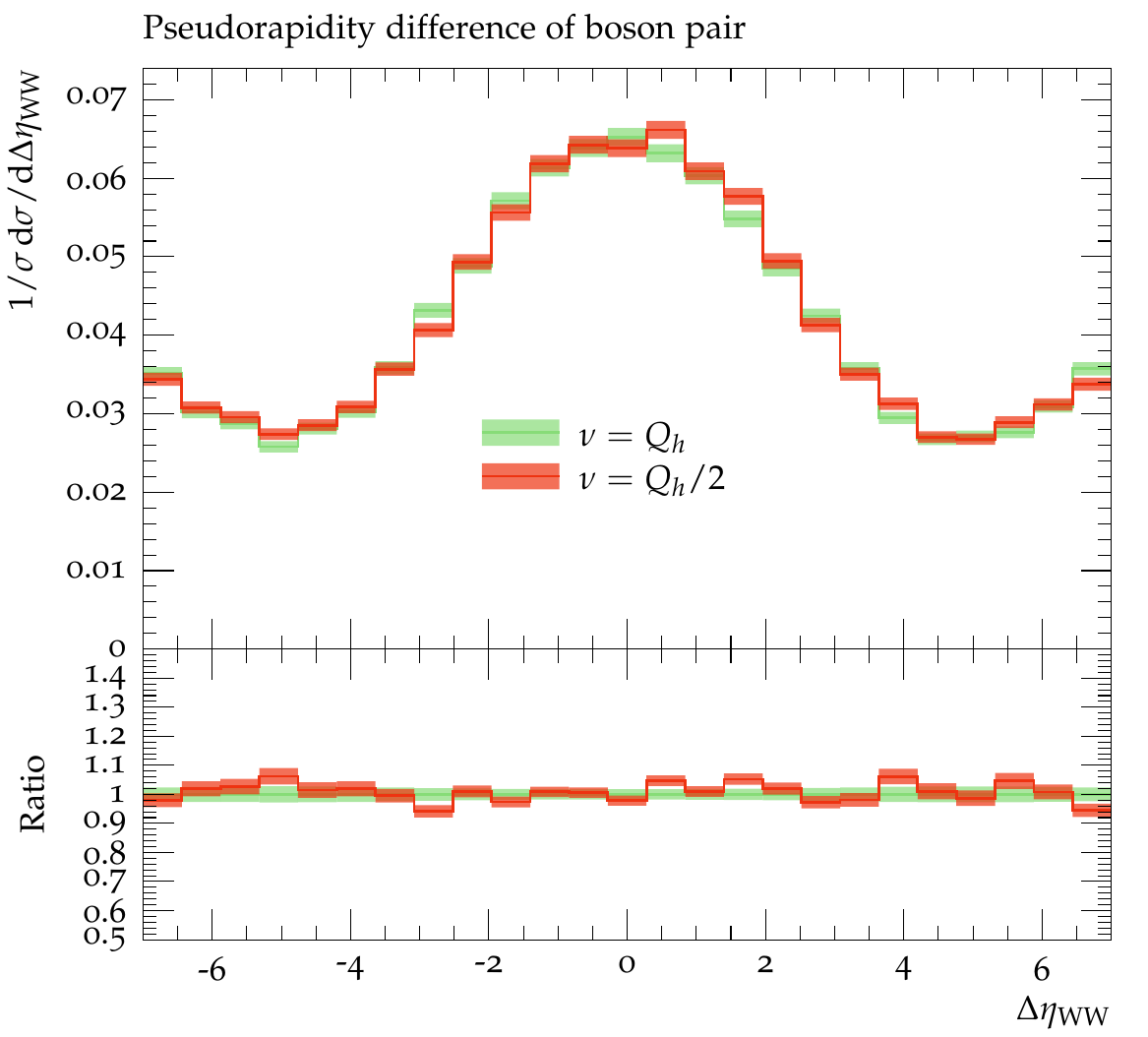} 
\includegraphics[width=0.49\textwidth]{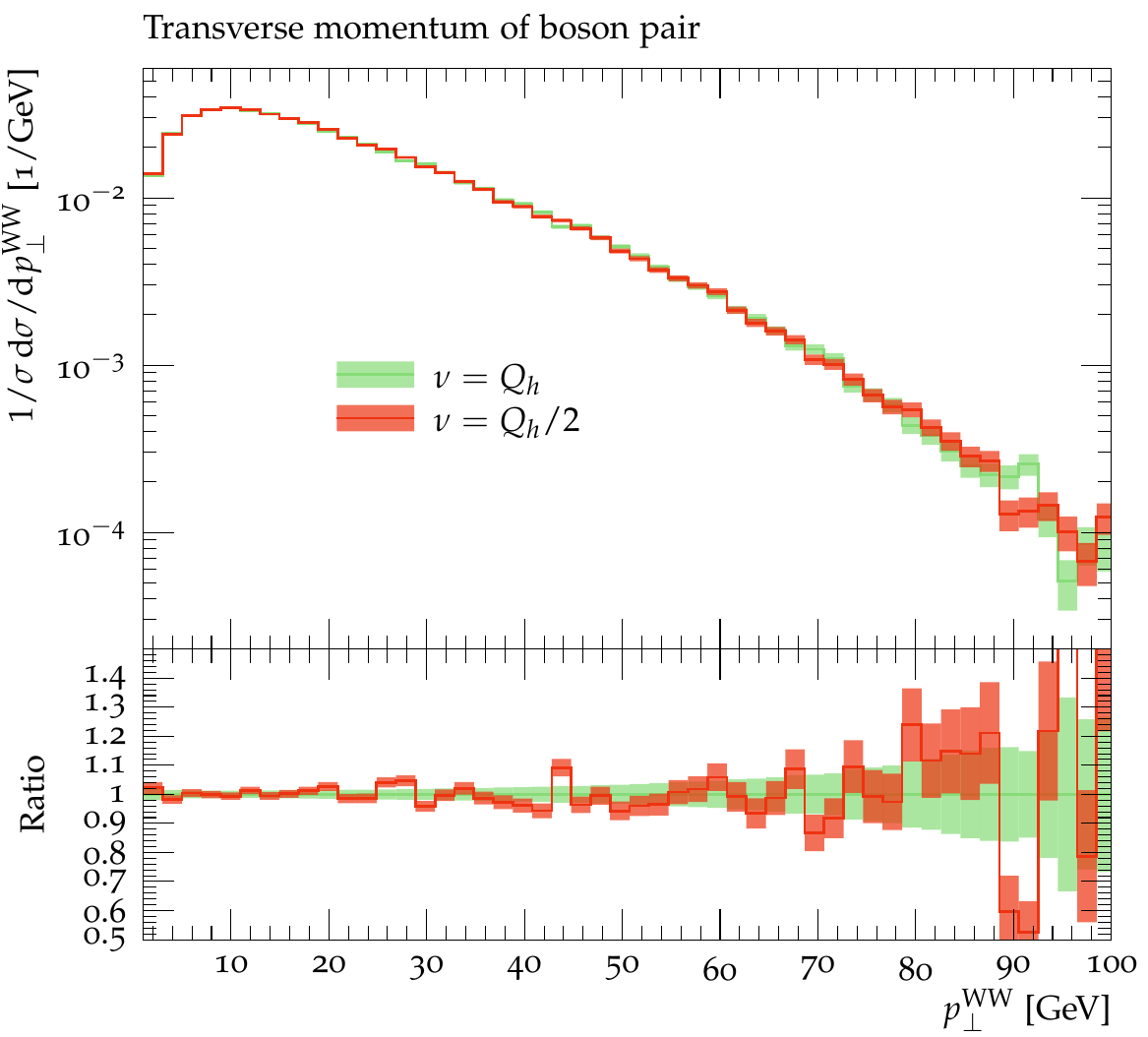} \\
(a) \hspace{225pt} (b)
\caption{(a) Difference between the pseudorapidities of the W bosons. (b) Transverse momentum of the WW pair.}
\label{Fig:scaleVar}
\end{figure}

\begin{figure}[t!] 
\centering
\includegraphics[width=0.6\textwidth]{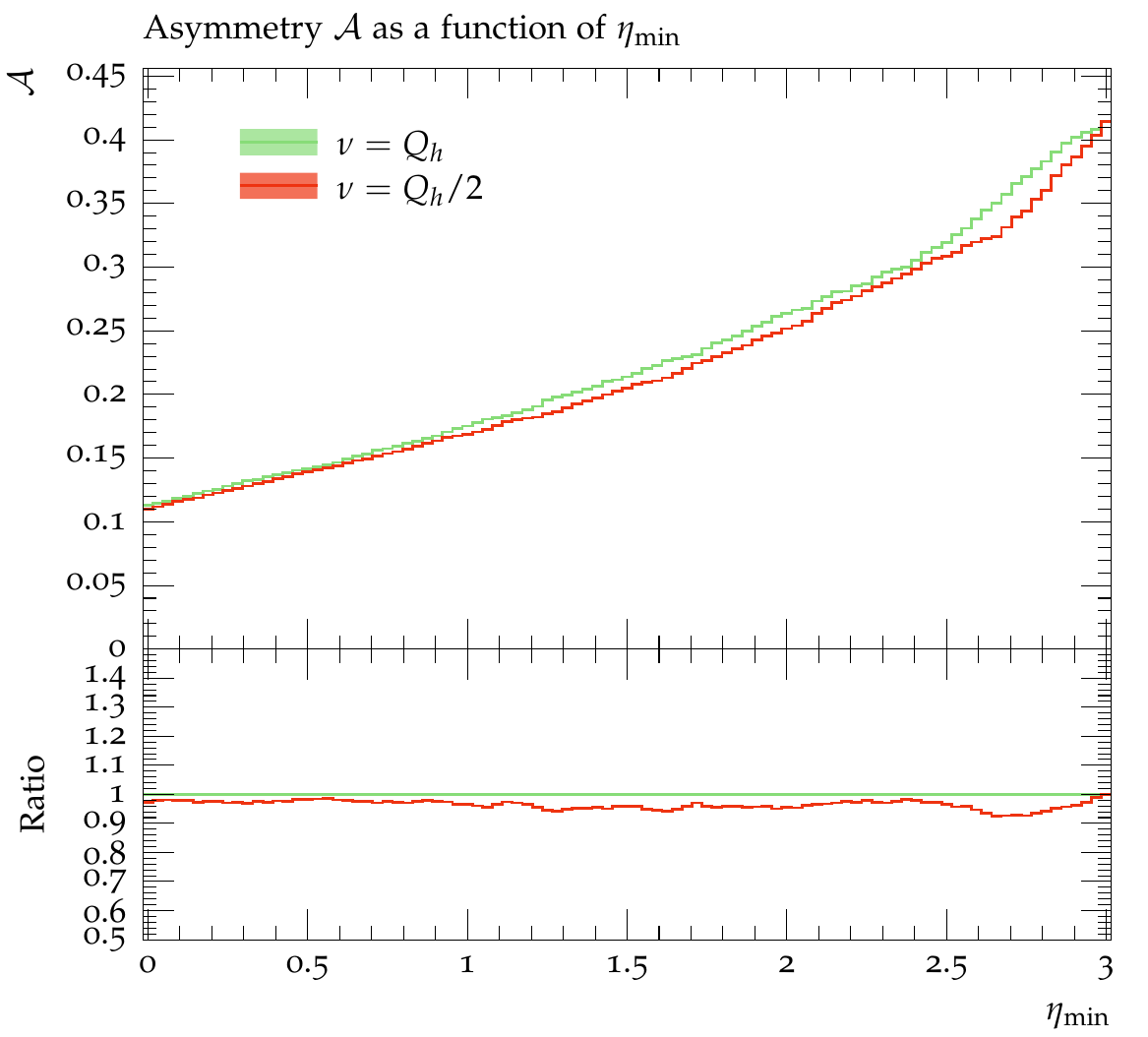}
\caption{Asymmetry $\mathcal{A}$ as a function of $\eta_\mathrm{min}$.}
\label{Fig:scaleVar2}
\end{figure}

It has been explained in \Sec \ref{sec:ReviewDPS} that the $\nu$-dependences of the DPS cross section and of the subtraction term in \Eq (\ref{eq:sub}) cancel each other, at least order by order. The problem is that the simulation of DPS introduced in this work does not include this subtraction term. Therefore, the results of the simulation are sensitive to the choice made for the scale $\nu$. This sensitivity can be assessed by varying the scale $\nu$. It is customary to vary the scale upwards ($\nu\leftarrow 2\nu$) and downwards ($\nu\leftarrow \nu/2$). In our case, the variation upwards is not possible. Indeed, the scale $\nu$ was originally chosen to be the hard scale $Q_h$, see \Sec \ref{Sec:2hard}. Hence, a variation upwards implies that the scale $\nu$ is now $2Q_h$ and it might happen that the value of $\mu_y$ is larger than the hard scale $Q_h$, since the lower limit for the range of $y$ values is now $b_0/(2Q_h)$. Unfortunately, the approach developed in \Sec \ref{Sec:dShowerSh} cannot handle such a configuration because it relies on the fact that $\mu_y < Q_h$. Nevertheless, a variation downwards is possible. The results of the scale variation are given in \Figs \ref{Fig:scaleVar} and \ref{Fig:scaleVar2}. It can be seen that the scale variation does not affect the event shapes. This is because the dominant contribution in the case of $\W^+\W^+$ production is the 2v2 part of the cross section, which is not too sensitive to the value of $\nu$. The asymmetry is nonetheless slightly reduced.

\clearpage

\bibliographystyle{JHEP}
\bibliography{Paper_dPDFs}

\end{document}